\documentclass[lettersize,journal]{IEEEtran}
\usepackage{amsmath,amsfonts}
\usepackage{algorithmic}
\usepackage{algorithm}
\usepackage{array}
\usepackage[caption=false,font=normalsize,labelfont=sf,textfont=sf]{subfig}
\usepackage{textcomp}
\usepackage{stfloats}
\usepackage{url}
\usepackage{xcolor}
\usepackage{verbatim}
\usepackage{graphicx}
\usepackage{cite}
\usepackage{threeparttable}
\usepackage{tabularx}
\usepackage{multirow, multicol}
\usepackage{caption}
\usepackage{slashbox}

\newcolumntype{L}[1]{>{\raggedright\let\newline\\\arraybackslash\hspace{0pt}}m{#1}}
\newcolumntype{C}[1]{>{\centering\let\newline\\\arraybackslash\hspace{0pt}}m{#1}}
\newcolumntype{R}[1]{>{\raggedleft\let\newline\\\arraybackslash\hspace{0pt}}m{#1}}

\begin{document}

%\sethlcolor{yellow}

\title{A Comprehensive Survey of 5G URLLC and Challenges in the 6G Era}

\author{Md. Emdadul Haque,~\IEEEmembership{Senior Member,~IEEE,}
        Faisal Tariq,~\IEEEmembership{Senior Member,~IEEE,}     
        Md. Sakir Hossain, ~\IEEEmembership{Senior Member,~IEEE,}
        Muhammad R. A. Khandaker,~\IEEEmembership{Senior Member,~IEEE,} 
        Muhammad A. Imran,~\IEEEmembership{Fellow,~IEEE}
        Kai-Kit Wong,~\IEEEmembership{Fellow,~IEEE}
        %MS,~\IEEEmembership{Fellow,~IEEE}
        %and~Jane~Doe,~\IEEEmembership{Life~Fellow,~IEEE}% <-this % stops a space
\thanks{Md. Haque is with Department of Information and Communication Engineering, University of Rajshahi, Bangladesh. e-mail: haque-ice@ru.ac.bd}% <-this % stops a space
%\thanks{Nurul H Mahmood is with Centre for Wireless Communication, University of Oulu, Finland. email:NurulHuda.Mahmood@oulu.fi}
\thanks{F. Tariq and M. A. Imran are with James Watt School of Engineering, University of Glasgow, United Kingdom. e-mail: \tt{Faisal.Tariq@glasgow.ac.uk, Muhammad.Imran@glasgow.ac.uk}}% <-this % stops a space
\thanks{Md. Hossain is with Department of Computer Science and Engineering, BRAC University, Bangladesh. e-mail: \tt{sakir.hossain@bracu.ac.bd}}%
\thanks{Muhammad Khandaker is with Main Roads, Western Australia. e-mail: \tt{M.Khandaker@hw.ac.uk}}%
\thanks{Kai-Kit Wong is with Department of Electronic and Electrical Engineering, University College London. He is also with  Yonsei Frontier Lab, Yonsei University, Seoul, Korea, e-mail: \tt{kai-kit.wong@ucl.ac.uk}}% <-this % stops a space
%\thanks{Manuscript received April 19, 2005; revised January 11, 2007.}
}

%\tableofcontents


\markboth{IEEE Communication Surveys and Tutorials}{February, 2020}



\maketitle

%\tableofcontents

\begin{abstract}
As the wireless communication paradigm is being transformed from human centered communication services towards machine centered communication services, the requirements of rate, latency and reliability for these services have also been transformed drastically. Thus the concept of \emph{Ultra Reliable and Low Latency Communication} (URLLC) has emerged as a dominant theme for 5G and 6G systems. %URLLC aims to support a vast set of 5G verticals such as tactile internet, industrial IoTs, intelligent transportation system, remote health care and so on. 
Though the latency and reliability requirement varies from one use case to another, URLLC services generally aim to achieve very high reliability in the range of 99.999\% while ensuring the latency of up to 1 ms. These two targets are however inherently opposed to one another. Significant amounts of work have been carried out to meet these ambitious but conflicting targets. In this article a comprehensive survey of the URLLC approaches in 5G systems are analysed in detail. Effort has been made to trace the history and evolution of latency and reliability issues in wireless communication. A layered approach is taken where physical layer, Medium Access Control (MAC) layer as well as cross layer techniques are discussed in detail. It also covers the design consideration for various 5G and beyond verticals. Finally the article concludes by providing a detailed discussion on challenges and future outlook with particular focus on the emerging 6G paradigm. 

\end{abstract}


%\tableofcontents

% \fi
\IEEEpeerreviewmaketitle

%%%%%%%%%%%%%%%%%%%%%%%%%%%%%%%
%%%%%%%%%%%%%%%%%%%%%%%%%%%%%%%
%%%                         %%%
%%%         SECTION         %%%
%%%                         %%%
%%%%%%%%%%%%%%%%%%%%%%%%%%%%%%%
%%%%%%%%%%%%%%%%%%%%%%%%%%%%%%%

\section{Introduction}
\label{sec:intro}

%{\color{red} Editor: FT \newline length: 1-2 pages}

% With the unprecedented growth of wireless communication the application areas are increasing surprisingly. 
 %The demanded services level quality and complexity of the applications are also increasing extraordinarily. 
 Wireless communication has experienced unprecedented growth in recent times which resulted in myriads of new applications and use-cases spanning almost every aspect of our life. While the 4G technology was dominated by human communication comprising voice and data services, emerging 5G and beyond services penetrated various industrial applications resulting in machine focused and highly agile communication. With the advent of technology capable of providing massive data rate and connecting any types of devices, diverse vertical industries such as smart health-care, manufacturing, transport, utilities and media are gradually embracing 5G systems and will be integral part of the emerging 6G communication system \cite{ftariq1}\cite{JAndrews2014}\cite{islam2025artificial}. This unprecedented proliferation leads to a set of Key Performance Indicators (KPI) such as data rate, latency and reliability requirement which was not fully experienced in previous generations \cite{AGupta15}\cite{wijethilaka2021survey}. For example, smart utility management system requires very low individual data rate but with very high reliability. Whereas multimedia applications such as ultra high density video requires very high data rate but it can cope with relatively low reliability.

 To accommodate these wide ranging demands from various use cases in different 5G verticals, International Telecommunication Union (ITU) defined following three main service classes for the emerging 5G systems: 
 \begin{itemize}
     \item \emph{enhanced Mobile Broadband (eMBB)}: This is aimed at supporting high data rate applications including video streaming, web access, video conferencing, and virtual reality \cite{fuad02}\cite{filali2022dynamic}.
     
     \item \emph{massive Machine-Type Communication (mMTC) }: This service class supports massive number of Internet of Things (IoT) devices, which are only sporadically activated and transmit data with small payloads \cite{CBockelmann}\cite{MShafi17}.
     \item \emph{Ultra Reliable and Low Latency Communication (URLLC)}:  This is conceivably the most innovative service class used for various mission-critical applications including tactile internet, autonomous vehicles, factory automation and remote action with robots \cite{popovski2}\cite{MsimsekJSAC}.

 \end{itemize} 
 
\begin{figure}[!t]
 \centering
 \includegraphics[width=0.5\textwidth]{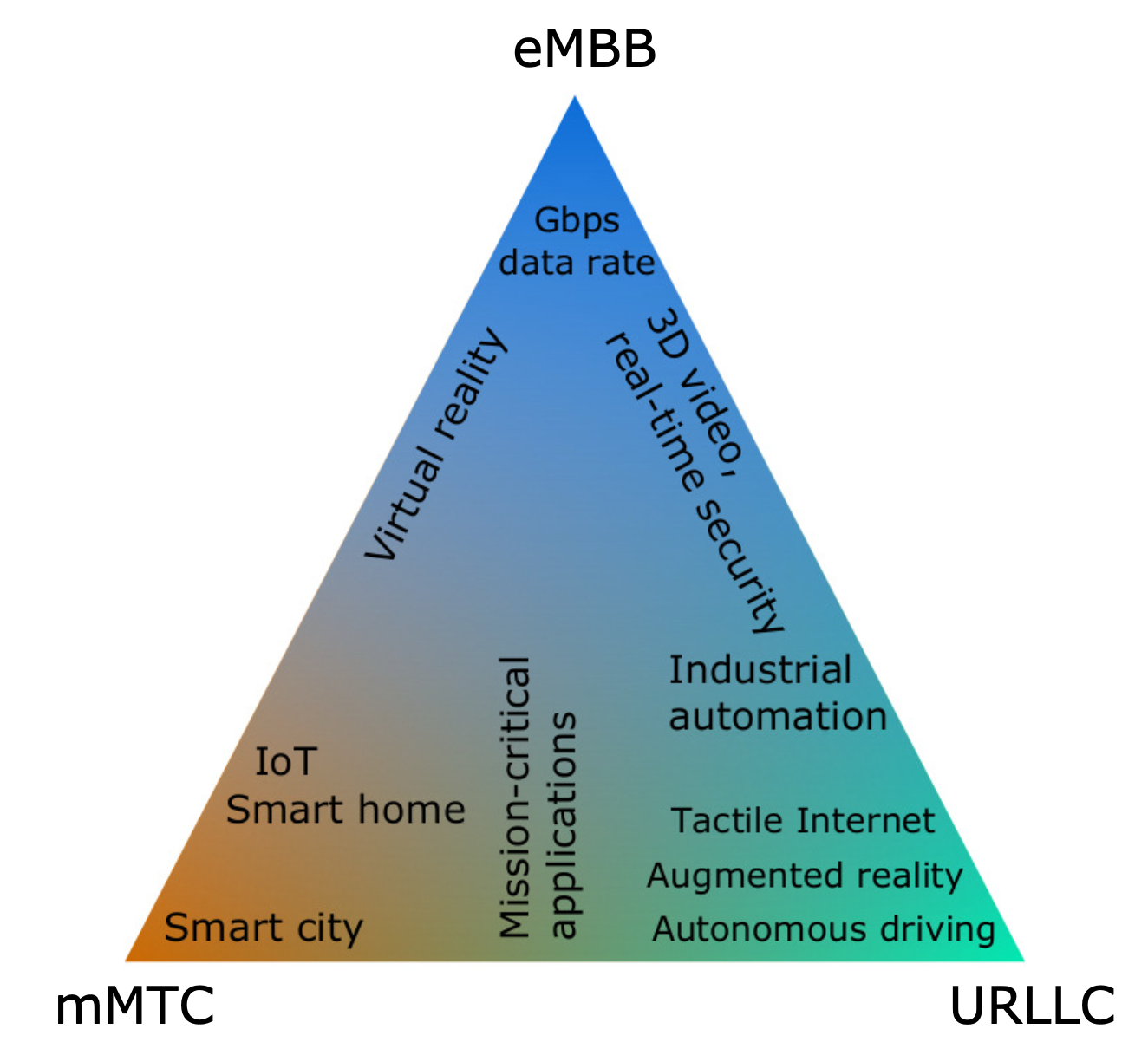}
 \caption{ITU IMT2020 use case depicting 3 different service classes for 5G.}
 \label{fig_1}
 \end{figure}

 The interrelation between various service classes and associated 5G applications based on ITU IMT2020 directives are demonstrated in  Fig. \ref{fig_1}. Though there are some overlaps in various service classes, URLLC systems have very specific requirements regarding latency and reliability.  For the URLLC, reliability is defined as \cite{alliance2019verticals}:
  
  \begin{quote}
  \emph{''The percentage value of the amount of sent packets/messages successfully delivered to a given node within the time constraint required by the targeted service, divided by the total number of sent packets/messages.''}
  \end{quote}
  
  and latency is defined as \cite{alliance2019verticals}: 
  \begin{quote}
      \emph{"The time that takes to transfer a given piece of information from a source endpoint device to a destination endpoint device, measured at the application service access points, from the moment it is transmitted by the source endpoint device to the moment it is successfully received at the destination endpoint device."}
  \end{quote}
  
 %It also displays a simplified version of the ITU IMT2020 triangle with eMBB, URLLC and mMTC as key 5G use case directions and the initial 5G focus.

  %To support very high data rate services for eMBB services, the network architecture is dominated by Ultra Dense Network (UDN) with relatively limited velocity of the mobile users . 

  The URLLC service is defined by ITU in IMT-2020 along with other requirements of the 5G network. The industrial standardisation group 3GPP has prepared a new standard, which is known as 5G New Radio (NR). Release 16 to 18 of 3GPP have the details of URLLC requirements and architectures. The initial 5G network have the features including eMBB, IoT, mission-critical control and fixed wireless access with very limited service of mMTC and URLLC.

 There are a number of characteristics which make URLLC fundamentally different from the traditional systems architectures. The major contrast for the packet structure between traditional systems and URLLC is the size of the packets. URLLC generally has very short packet size in order to meet the ultra low End-to-End (E2E) delay requirement typically in the range of $<1 ms$. It is also aimed to ensure that the packets are received correctly with a very high success probability in the range of $(1-10^{-5})$ to $(1-10^{-9})$. These stringent latency and reliability constraints are considered as the most challenging aspect of 5G network design. While achieving these stringent requirements in the link layer is relatively easy, particularly in a small area network, it is extremely difficult in the network layer where applications work over wide area networks such as remote surgery \cite{5gppp1}, \cite{challacombe1}. The issue of latency in wide area networks involves multiple factors including uplink and downlink delay, coding and processing delay, queuing delay and routing delay in backhaul and core network \cite{she1}.  While individual delays can be tackled with relative ease, the accumulation of multiple delays makes design the URLLC particularly challenging as it involves carefully optimising them to ensure that the combined delay remains below the required ultra low latency thresholds. %Although delay in the optical fiber core network is less than 1ms,

 The KPIs of Long Term Evolution (LTE) such as latency, reliability and throughput are inherently conflicting in nature and therefore requires the designer to reach an optimal trade-off \cite{soret1}. With this conflicting relationship, the typical error rate in 4G was $10^{-2}$. This  was manageable with conventional channel coding gain and re-transmission mechanism \cite{Abreu1}. However, these techniques are not sufficient on their own to meet the stringent latency and reliability requirements of URLLC as mentioned above. Therefore, new techniques need to be designed for URLLC. The latency requirement of URLLC is considered as the top priority in the 3GPP NR standard meeting \cite{ji1}. The reliability is considered as the secondary priority due to the fact that the existing techniques such as channel coding, space, antenna and frequency diversity techniques can be exploited to ensure reliability to a large extent \cite{johansson1}. %\cite{sybis2016channel}
 As mentioned previously, the most obvious way of achieving low latency is to shorten the packet size in the range of a few bytes (e.g. 20 bytes or even smaller) \cite{she1}. However, the short packet size severely restricts the traditional way of increasing reliability by means of coding gain \cite{durisi1}.%\cite{masaracchia2021uav} . 
 Conversely, to increase the reliability redundant information needs to be transmitted (either in the form of re-transmitting the same packet to increase the probability of correct reception or adding redundant bits for error detection and correction) which ultimately affects the latency performance. Thus it is obvious that a paradigm shift in design thinking is necessary to fully realise the potential of the  URLLC concept.

 To achieve the goal of radically new design thinking, a comprehensive understanding of what has been done so far for URLLC system design is very important. In this paper, a comprehensive taxonomy of the state-of-the-art research works on URLLC is provided. To the best of authors knowledge, there are some survey papers related to URRLLC that have been published recently \cite{Survey2019PhyMac}, \cite{siddiqi20195g}, \cite{shaik2024ai}, \cite{pradhan2024survey}, \cite{elgarhy2024energy}, etc. that investigate several aspects of URLLC. However, there is no survey that comprehensively covers all the aspects of URLLC and also provides an overall coverage of the issues related to URLLC. This paper aims to fill that gap by providing a comprehensive discussion on URLLC issues and associated approaches. It also provides an in-depth discussion on future challenges and possible ways of addressing them.
 
\begin{figure*}[!t]
 \centering
 \includegraphics[width=1\textwidth]{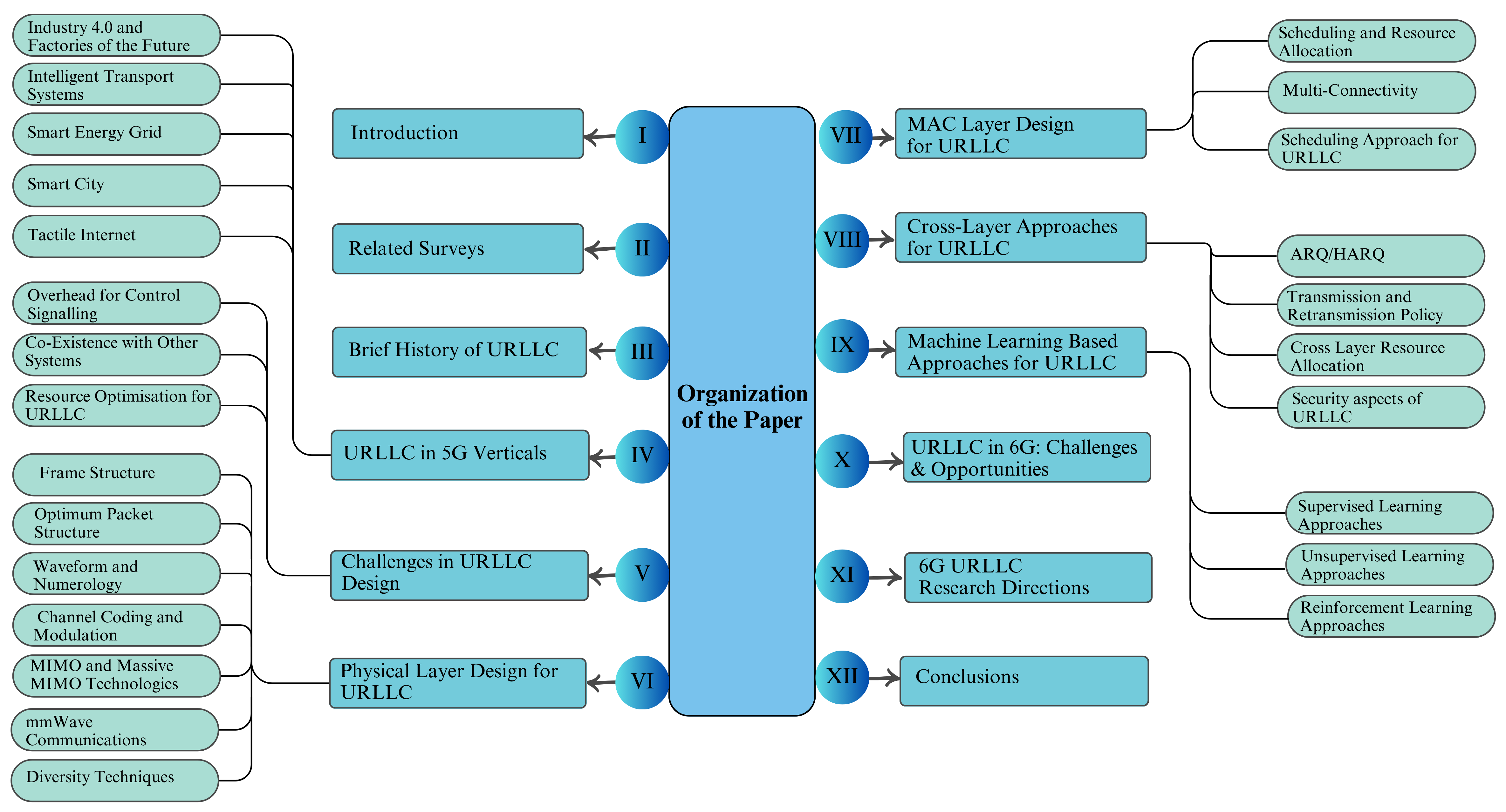}
 \caption{Organisation of the paper.}
 \label{fig_organization}
 \end{figure*}

The rest of the paper is organised as follows (see Fig. \ref{fig_organization}): Section \ref{sec:relatedSurvey} provides a comparison of this survey with the existing ones. Sections \ref{sec:history} and \ref{sec:5G_verticals} discuss the history and use cases of URLLC from 5G vertical aspects. Then, the challenges of the implementation of URLLC are discussed in Section \ref{sec:challenges_urllcDesign}. While Section \ref{sec:physicalLayer} gives the details of the physical layer design aspect of URLLC, the design of the MAC layer of URLLC is discussed in Section \ref{sec:MACLayerDesign}. The cross-layer design issues are outlined in Section \ref{sec:crossLayer}. Then, the URLLC solutions exploiting the machine learning algorithms are presented in Section \ref{sec:ML}. Sections \ref{sec:6GChallenges_Opportunities} and \ref{sec:6GResearchDirections} provide the 6G research challenges and research directions, respectively. Section \ref{sec:conclusions} concludes the article. 

\section{Related Surveys} \label{sec:relatedSurvey}

This section presents an overview of recent papers that investigate URLLC to establish the scope of this paper. Several URLLC survey papers exist in the literature (see Table~\ref{surveyComparison}). For example, in \cite{khan2022urllc}, the authors provide a comprehensive overview of URLLC and eMBB for Industrial IoT (IIoT), aiming to identify gaps between URLLC and eMBB services in 5G or beyond 5G (B5G) networks. As the stringent requirements of URLLC can degrade the performance of eMBB, the paper investigates the trade-off between the two. An overview of diverse applications of URLLC and eMBB is also presented. However, the paper mainly focuses on concerns related to URLLC and eMBB in IIoT without delving into a detailed investigation of URLLC. 

A brief investigation of URLLC and its importance in the industrial environment is presented in \cite{siddiqi20195g}. The IoT devices that require URLLC requirements are also addressed in the paper with technical details. Standardisation progress and implementation details are discussed, along with highlighting future research directions in the paper.

\begin{table*}[]
\centering
\caption{Comparison of the existing survey papers}
\label{surveyComparison}
\begin{tabular}{|l|c|c|c|c|c|c|c|}
\hline
Focus & Paper reference & PHY & MAC & URLLC & Cross layer & Machine learning & 6G overview \\ \hline
\multirow{3}{*}{IoT} & \cite{siddiqi20195g} & \checkmark & \checkmark &  & \checkmark &  &  \\ \cline{2-8} 
 & \cite{elgarhy2024energy} & \checkmark &  & \checkmark &  &  &  \\ \cline{2-8} 
 & \cite{khan2022urllc} & \checkmark & \checkmark & \checkmark &  &  &  \\ \hline
Capacity maximisation & \cite{shaik2024ai} & \checkmark &  & \checkmark &  & \checkmark &  \\ \hline
Security & \cite{pradhan2024survey} & \checkmark &  & \checkmark &  & \checkmark &  \\ \hline
Smart technologies & \cite{ho2019next} & \checkmark & \checkmark & \checkmark$^{1}$ &  & \checkmark & \checkmark \\ \hline
Interference management & \cite{siddiqui2023urllc} &  &  & \checkmark &  &  & \checkmark \\ \hline
\multirow{4}{*}{Comprehensive survey} & \cite{Survey2019PhyMac} & \checkmark & \checkmark & \checkmark & \checkmark &  & \checkmark \\ \cline{2-8} 
 & \cite{masaracchia2021uav} & \checkmark &  & \checkmark &  &  & \checkmark \\ \cline{2-8} 
 & \cite{ali2021urllc} & \checkmark & \checkmark & \checkmark &  & \checkmark$^{2}$ &  \\ \cline{2-8} 
 & This paper & \checkmark & \checkmark & \checkmark & \checkmark & \checkmark & \checkmark \\ \hline
\end{tabular}
\label{table:survey_review} % is used to refer this table in the text

\begin{tablenotes}
\centering
\item[1]  $^1$ Partial description of URLLC.   
\item[2]  $^2$ FRL only.
\end{tablenotes}
\end{table*}

In \cite{ho2019next}, the authors investigate the potential requirements of 5G communication to support automation in the vertical domain and highlight the enabling technologies to meet the requirements of 5G. %Specifically, the paper identifies smart factory, smart vehicle, smart grid, and smart city use cases and determines the service requirements, namely URLLC, eMBB, and mMTC. 
An in-depth survey of existing technologies that fulfill the service requirements is presented. The paper also provides a brief overview of 6G networks. However, it focuses more on discussing smart technologies within the context of 5G networks for major applications in our daily lives, without providing detailed descriptions of URLLC.

In \cite{Survey2019PhyMac}, the authors investigate the state-of-the-art physical and MAC layer protocols that cover both licensed and unlicensed bands. The paper investigates the Long-Term Evolution (LTE) as the coexistence technology for a potential enabler of URLLC in the unlicensed band. A brief overview of cross-layer mechanisms to achieve URLLC is included in the paper. It also presents potential applications with latency and reliability requirements, along with future research directions for URLLC. However, the paper focuses more on discussing coexistence technology and overlooks the recent progress of 5G networks.  

Unmanned Aerial Vehicle (UAV)-enabled URLLC networks with 5G and 6G technologies are presented in \cite{masaracchia2021uav}. The paper highlights the prominent features of UAV-aided communication, which is a promising solution for future networks. %Additionally, the paper provides an overview of network architectures and frameworks for URLLC. 
An analysis and classification of recent works on UAV-enabled URLLC are also presented in the paper to enhance the understanding of the UAV-enabled URLLC networks.

In \cite{siddiqui2023urllc}, the authors present an investigation on URLLC from the interference perspective, as the network design and modeling of access, along with all related technologies for B5G and 6G networks, pose challenges due to interference. %The paper discusses the comprehensive trade-off relationship among throughput, latency, and reliability, taking into account the impact of interference.
It also contains an in-depth discussion on recent wireless access technologies and their applications in optimising next-generation cellular systems. However, the paper covers only a narrow domain of URLLC.

In \cite{ali2021urllc}, the authors provide a detailed discussion on next-generation technologies and network intelligence in 5G NR for URLLC requirements. The paper presents a potential machine learning technique, namely federated reinforcement learning (FRL), for 5G NR URLLC. Furthermore, the paper includes a detailed discussion of the MAC layer channel access mechanism that enables URLLC in 5G NR for Tactile Internet (TI). %It also proposes seven FRL-based frameworks for 5G NR URLLC requirements. 
However, the paper mainly emphasises the FRL perspective as a probable technique for 5G NR URLLC requirements. 

The resource allocation strategies exploiting machine learning to maximise the capacity of URLLC wireless networks are summarised in \cite{shaik2024ai}. Among the machine learning models, only deep learning is included in the paper. %The resource allocation, beamforming, mmWave communications, modulation and coding issues are discussed in detail with a special emphasis on the resource allocation methods. 
A good number of future research challenges and directions are figured out in the paper. However, the scope of the survey is very limited covering only capacity maximisation of the network. Two more reductionistic surveys including  \cite{pradhan2024survey, elgarhy2024energy}, which discuss the physical layer security  of URLLC design \cite{pradhan2024survey}, and energy and latency optimisation methods of URLLC \cite{elgarhy2024energy}.

Although there have been some works focused on URLLC in 5G networks, they represent early efforts, and many recent works are ignored. The survey papers cover limited areas of the URLLC in 5G networks with limited number of papers while an in-depth discussion covering most of the areas with sufficient number of recent papers are absent in the survey papers. The existing surveys are summarised in Table \ref{surveyComparison}.  

%To the best of our knowledge, a comprehensive study of recent URLLC technologies is still lacking for both 5G and beyond systems. Therefore, this paper aims to fill this gap by providing a state-of-the-art survey of existing methodologies for 5G URLLC, while also discussing the challenges and requirements of future networks, with a particular focus on the 6G paradigm.

 \begin{figure}[!t]
 \centering
 \includegraphics[width=0.5\textwidth]{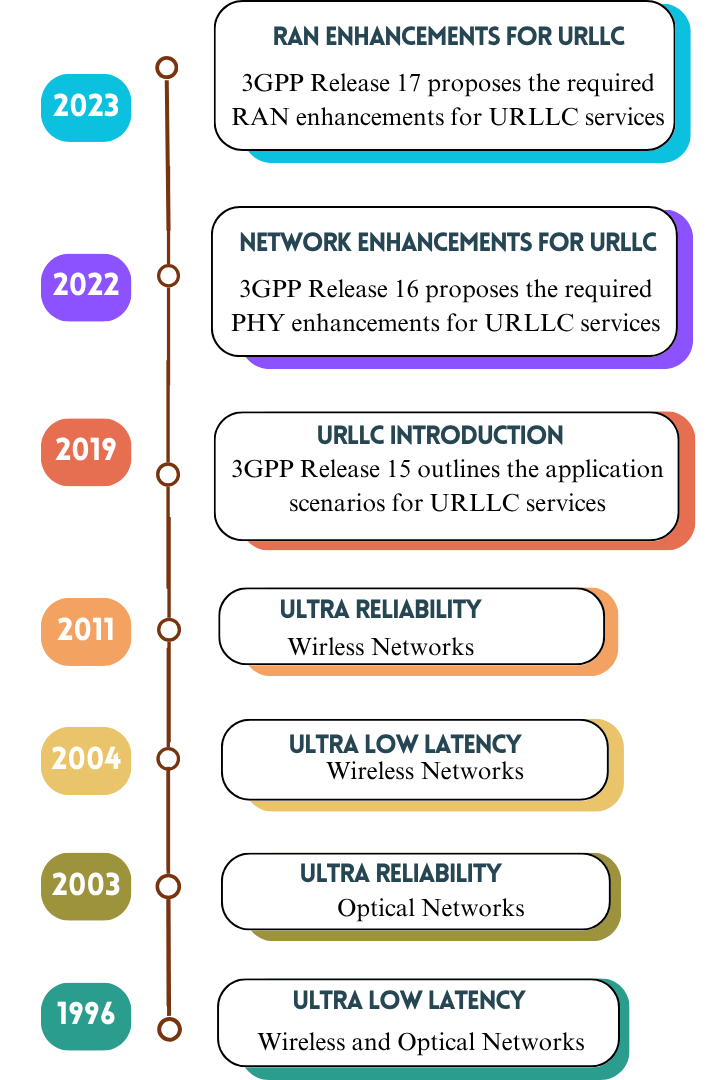}
 \caption{Evolution of URLLC}
 \label{fig_history}
 \end{figure}

\section{Brief History of URLLC}\label{sec:history}
\label{sec:brief}

During the last few decades wireless communication systems have penetrated all aspects of our life, particularly due to the advantage of mobility compared to any wired solution or systems that involve wired connectivity. Rapid technological development in other areas also played an important role in accelerating the pace of proliferation of wireless systems in our everyday life. Therefore, the uncertainty around seamless wireless access and reliability of wireless connectivity has significantly been reduced over time. Consequently, more and more mission critical applications are gradually opting for wireless connectivity despite the various challenges including latency and reliability. 

The application domains of the current times and recent past were dominated by human-centered applications for which latency and reliability requirements were not very stringent. For example,  Voice over Internet Protocol (VoIP) service can cope with up to 50 ms queuing delay and reliability of $2*10^{-2}$ with packet size is 1500 bytes \cite{ts3Gpp36814}. On the other hand, for machine control applications, the latency and reliability requirements are much higher than the aforementioned systems. For example, automation control in general requires 1ms of latency and and 99.9999\% reliability \cite{ts3Gpp22661}, \cite{attaran2021impact} with optimum packet size of 20 bytes or smaller \cite{ts3Gpp38913}.\\
%The maximum possible latency and reliability of the network are .... and ...... , respectively. 

In the 4G systems, the radio resources are allocated in every transmit time interval (TTI) which is set as 1 ms  \cite{capozzi2013downlink}. This means that each packet will have to wait in the queue of a base station for more than 1 ms. Since the latency calculation includes backhaul and core network delay, it is obvious that the required latency for URLLC can not be achieved with the existing 4G system architecture. 

The ultra-low latency communication is introduced in the mid-nineties in developing high-performance parallel interfaces for communication over copper wire and optical fiber. This standard allows point-to-point, full-duplex transmission at 6400 Mb/s \cite{hoffman1996hippi}. The discussion on ultra reliability can be traced back to the beginning of the century \cite{wen2003ultra, YWen05URC} where Ultra Reliable Communication (URC) was considered for optical communication networks. The timeline of the evolution of URLLC is depicted in Fig.~\ref{fig_history}. For optical networks, light-path diversity was used to provide ultra reliability and error rate of $10^{-16}$ was achieved when the number of lightpaths was 8 \cite{YWen05URC}. Initially, the reliability and latency requirements were mostly decoupled.

The discussion on URC in wireless communication can be traced back to as early as 2011 \cite{CDombrowski11}. The latency requirement was referred to as 'real time communication' and the requirement was not as stringent as of today. The work in \cite{CDombrowski11} was focused on theoretical analysis of using/combining various diversity techniques for reducing the outage probability to extremely low level (or achieving ultra high reliability in other words). Similar discussion could be found in a number of works in the subsequent years for various services and applications aiming at 5G systems \cite{AOsseiran13} including multimedia wireless system \cite{DSoldani1}, wireless sensor and actuator networks in harsh industrial environment \cite{JSilvo13}, automation and public safety communication \cite{RBaldemair13}. Authors in \cite{HSchotten14} proposed a new metric called \emph{Availability Indicator} as an enabler for URC. The authors in  \cite{popovski1} provided one of the early and comprehensive definitions of various service classes and applications for URC along with their specific requirements. The reliability requirement for URC was specified as $>$ 99.999\%.  For URC, services were classified into two types based on the latency requirement: long term URC with latency $\geq 10ms$ and short term URC with latency $\leq 10ms$. In 3GPP release 14, latency requirements for URC services were revised down to 4 ms. These requirements were later adopted in 5G wireless system specification to cover various applications including home automation and industrial automation. 

%\cite{fanibhare2021survey

Around similar time, in addition to reliability, ultra low latency (ULL) was also gradually started to be considered for various services and applications including instant financial information transmission between financial markets \cite{wang2005novel,GLaughlin14}, wireless sensor and actuator networks for industrial automation \cite{MBrachmann13} and cyber-physical systems for controlling power grids \cite{JAllen12}. As various 5G applications such as tactile internet, remote control of industrial systems, autonomous vehicle  are emerged in which both latency and reliability are critical requirement  \cite{promwongsa2020comprehensive}, ultra low latency and ultra low reliability were gradually started to be considered jointly, resulting in emergence of a new service class. This new service class was defined as URLLC \cite{Petar1}. In 3GPP release 15 to 18, a comprehensive set of requirements for URLLC has been specified.

In release 15, some application scenarios of URLLC- such as edge computing, network slicing, and service hosting of services, are identified. Next, few methods for supporting URLLC services to the 5G networks are proposed in release 16. The methods include redundant transmission, QoS monitoring, dynamic division of the packet delay budget, etc. In addition, some enhancements for the physical layer are proposed to support URLLC services. Few of them include sub-slot-based HARQ-ACK feedback, PUSCH enhancement, compact downlink control information format, etc. The following enhancements in RAN is proposed in release 17 to support URLLC services: physical layer feedback enhancement for HARQ-ACK and CSI reporting, prioritisation of traffic, intra-UE multiplexing, time synchronisation, etc.

%which was later coupled together that resulted in the concept of URLLC

%\begin{itemize}
%    \item Where does URLLC come from
%    \item 300-400 words
%    \item {\color{red} NHM}
%\end{itemize}

%=============================%
%         SUB-SECTION         %
%=============================%

\section{URLLC in 5G Verticals}\label{sec:5G_verticals}

The wireless technologies and systems before 5G acted mostly as connectivity solutions with limited capability to differentiate among different types of traffic with varied KPI requirements. However, 5G has taken a different approach and instead of acting as mere connectivity solution, it aims to create an ecosystem that takes into account the whole service requirements for various industry verticals such as automobile and IoT  with each having varied requirement in terms of  data rate, quality of service, reliability and latency \cite{MImran2019WB}. %\cite{XLi2017IC} 

\begin{figure*}[!t]
 \centering
 \includegraphics[width=0.8\textwidth]{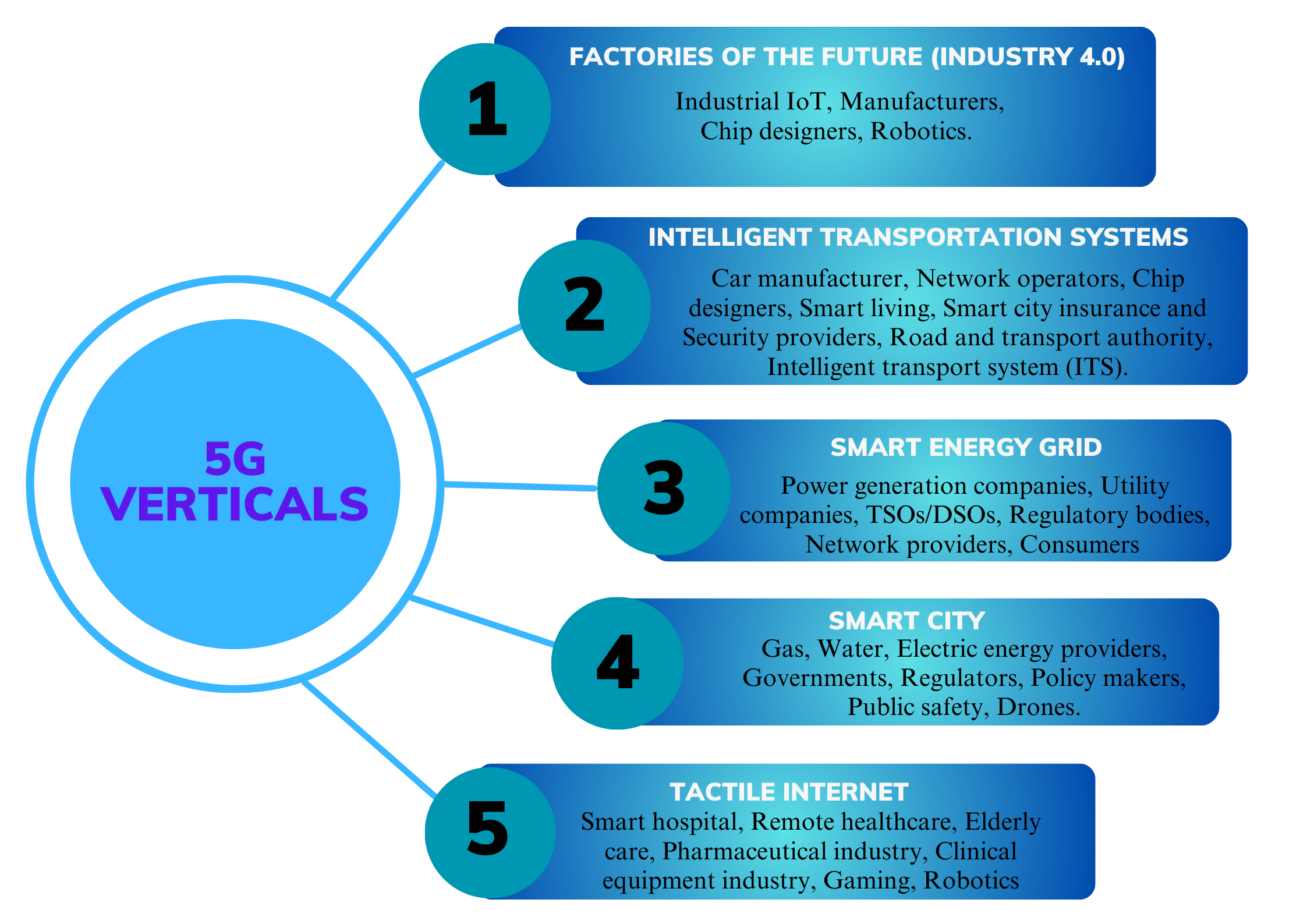}
 \caption{5G Verticals and their stakeholders.} 
 \label{fig_2}
 \end{figure*}

The standardisation initiative and designing solution for the eMBB and mMTC has progressed well as they do not have much difference from previous generations except for the scale of devices/services. URLLC solutions, however, still require significant exploration both from business and technical point of view in order to become a reality since the relevant use cases have significantly different combinations of requirements compared to use cases of the previous generation wireless systems. Therefore, a comprehensive discussion on latency and reliability requirements of various relevant use cases are required in order to develop a fully complying 5G ecosystem. The figure \ref{fig_2} demonstrates various 5G verticals and associated issues in relation to URLLC. Rest of the section provides an overview on various use cases in different 5G verticals from URLLC viewpoint.

\label{sub:useCases}
%applications, requirements, market share/forecast (including state of the art solutions - see example in bracket)

\subsection{Industry 4.0 and Factories of the Future}

Rapid proliferation of Artificial Intelligence (AI)/Machine Learning (ML), IoT and other advanced technologies are significantly enhancing the flexibility, agility and productivity \cite{AMahmood}. 
This emerging new paradigm, popularly known as industry 4.0  %\cite{mourtzis2021smart} 
or factories of the future (FoF) in more general sense, are adopting these technologies to transform the legacy factories into highly effective, flexible, smart and ubiquitously connected industries \cite{sanusi24}. %\cite{EFFRA19}. 
Connectivity is one of the key components of the  FoF and URLLC enabled wireless connectivity is expected to play a crucial role in realising FoF or industry 4.0 \cite{nakimuli2021deployment}. 

There are a number of potential factory functions which are highly reliable on URLLC such as remote monitoring, process control, motion control of the moving parts of the manufacturing equipment, mobile robots \cite{ho2024energyEfficiency} and plant asset management etc.%\cite{SGangakhedkar2018ICC}, 
The user plain reliability and E2E latency requirements for motion control are $1-10^{-5}$ and 1 ms, respectively. In motion control sensor and actuator, data transmit through uplink and downlink, respectively, with a E2E delay of 1$\mu$s and both use isochronous transmission. The factory automation is specified in \cite{ts3Gpp22661}. The reliability requirement  is $1-10^{-4}$ and the user plain E2E latency requirement is 10 ms with the jitter of 10 1$\mu$s. 
%\cite{AAIjaz2019PIEEE}
%The traffic stemmed from FoF is usually intermittent in nature and typically contains small payloads. This may lead to occasional severe congestion in the random access (RA) phase due to the procedure of the existing \emph{Random Access CHannel} (RACH) based network access system. This will certainly trigger several re-transmission requests and will result in failure to meet the latency requirement. One way of ensuring the latency with high degree of confidence is to employ a prioritised reservation scheme in random access for URLLC traffic %\cite{chen2017prioritized}, \cite{althumali2022priority}. 
A two stage RA with early data transmission (EDT) is proposed in \cite{MRaftopoulou2019EuCNC} where some are being transmitted while the link establishment process continues. It also proposes to map critical preambles to a pre-configured set of physical resource blocks (PRB) which saves significant amount of time. 
The work in \cite{johansson1} proposes to use Convolutional Code (CC) as Forward Error Correction Code (FEC) instead of Turbo Codes (TC) which will significantly reduce processing delay while maintaining the same level of error correction performance of TC for short packet lengths which  %. The results confirms that the non-backward compatible 5G air interface can 
consistently maintain stringent latency requirements in various factory scenarios. %\cite{IAshraf2016ETFA}. 

Provision for alarm signals is critical for safety of the future factories as any fault must be immediately forwarded to the control center. Pilots can be reserved dynamically to ensure guaranteed access for the alarm signals %\cite{EFitzgerald2019ArXiv}, 
\cite{karaca2020scheduling}. Reliability is another challenge in the factory and industrial setting where the RF environment is highly dynamic due to rapid movement of various components (such as robotic arms, rotors etc.) as well as humans. Cooperative Multi-Point (CoMP) transmission, a technology used to improve network performance for cell-edge users through the coordination among multiple base stations by implementing joint transmission, base station selection, and interference management, can be an effective solution in these situations as appropriate design can eradicate the need for re-transmission of a packet. This will ensure that the potential of incurring delay beyond threshold due to repeated transmission is avoided  \cite{MKhoshnevisan2019JSAC}.
%, \cite{dong2022comp}. 
However, CoMP is inherently resource inefficient. Therefore careful design and optimisation of the schemes are of paramount importance to provide a complete wireless solution for factory automation.%\cite{WLiang2019IEEEProc}
%\cite{liang2021experimental}.

\subsection{Intelligent Transport System (ITS)}

ITS is an innovative service that aims to improve traffic efficiency of various transport modes and allows users to get safer, smarter and efficient travel experience underpinned by massive data availability and wireless connectivity. In addition to the dominant theme of Connected and Autonomous Vehicles (CAV), ITS can also involve a variety of activities such as traffic regulation, road access control, emergency response to accidents, pollution control etc. 5G infrastructure Public Private Partnership (5G-PPP), which is an initiative by the European Commission and European ICT industries, laid out their vision for the future of automotive vehicles in a technical document. They have identified some key socio-economic drivers such as autonomous driving, enhanced road safety and traffic efficiency, intelligent navigation and digitalisation of transport and logistics and some emerging business models such as pay per drive or mobility as a service  \cite{5gppp2015}. Successful operation of these services require seamless, real-time and reliable information exchange among the vehicles, roadside infrastructure and other entities. Autonomous and self-driving cars are also becoming an integral part of ITS. Reliable communication between various entities is at the heart of ITS in the rapidly changing wireless propagation environment (as vehicles are moving at high speed). Thus reliability and latency is of paramount importance for ITS \cite{guo2022age}.

Both IEEE and 3GPP developed a number of standards for this purpose such as Dynamic Short Range Communication (DSRC) and Vehicle-to-everything (V2X). V2X involves various aspects including Vehicle to Vehicle (V2V), Vehicle to Infrastructure (V2I), Vehicle to Network (V2N), and Vehicle to Pedestrian (V2P) \cite{abdel20205g}. In 3GPP release 16,  technical report 3GPP TR 27.386 specified the  architectural enhancement needed for the 5G system to support the V2X services  \cite{3GPPTR23786}. Both 3GPP and the 5G Automotive Association (5GAA) standardised V2X services by the year 2020 \cite{SHussain2018CSCN}.

In vehicular communication, reliability requirement has the higher priority and latency is somewhat less challenging. 5G quality of service (QoS) Identifier (5QI) is proposed by 3GPP to ensure that quality requirements including reliability and latency are appropriately addressed. It also recommended to exploit advanced network architectures such as mobile edge computing (MEC) and network slicing (NS) to ensure extreme reliability and latency requirements \cite{SHUssain2019CSM}. In MEC, instead of storing and processing of user data at the central cloud server, these are carried out at edge nodes such as base station and access points. This reduces latency and bandwidth. On the other hand, a physical network is divided into a multiple virtual network with each virtual network optimised for a specific service and application. In vehicular networks, these services include vehicle-to-vehicle communications (V2V) and vehicle-to-infrastructure such as roadside unit (V2I). It has been shown that autonomous driving is particularly tricky as emergency scenarios may erupt which require sudden lane change or breaking to avoid collision. 
For autonomous driving, any disruption in communication may lead to fatal injury and therefore ultra reliability at the scale of 99.999\% must be ensured \cite{SHUssain2019CSM}. These scenarios may also require real time remote monitoring of the services. Apart from the URLLC, some aspects of intelligent vehicular communication may be data intensive (video monitoring) and therefore a mechanism needs to be in place to combine URLLC with eMBB provisions \cite{XSong2019Access}.

The level of autonomy in autonomous driving is divided into 6 steps with step 0 being no autonomy at all and level 5 being fully autonomous where no human intervention is anticipated \cite{HMa2016SAE}. With increasing levels of autonomy, the requirement of reliability and latency also becomes more stringent.
%The level of accuracy of positioning has to be very precise as any error in this will lead to wrong decision making regarding maneuvering, lane changing, braking etc. To enable this precision, an adequately large set of Positioning Reference Signals (PRS) should be available for DL, UL and sidelink \cite{CCasetti2019MST}. 
Rapid channel variation is another bottleneck in maintaining URLLC requirements as the channel state information becomes obsolete very quickly. Various solution has been proposed to solve this problem including URLLC design with network and resource slicing \cite{XGe2019TVT}, \cite{skondras2021network}, proactive learning, and Age of Information (AoI) awareness \cite{li2022age}, \cite{MAziz2019WCL}, where AoI represents the freshness of information at the receiver. Though a significant number of solutions have been proposed to make sure URLLC particularly suitable for vehicular communication purposes, fully autonomous vehicles and intelligent transport system design requires wide ranging considerations which needs further attention in URLLC design.

%{\color{red} NHM}

% \begin{figure*}[htp!]
%  \centering
%  \includegraphics[width=1\textwidth]{haptic.jpg}
%  \caption{Haptic Communication}
%  \label{fig_haptic}
%  \end{figure*}

\subsection{Smart Energy Grid}

Smart grid aims to revolutionise how power is generated, distributed and consumed by exploiting the recent rapid development in information, communication and sensing technologies \cite{HFarhangi2010PEM}. Innovation in 5G and URLLC also accelerated the pace of smart grid proliferation \cite{HHui2019EAE}. It is anticipated that contributions from various renewable energy sources including solar micro-grids and  wind turbines  will gradually increase. Managing the massive number of diverse sources along  with a massive number of households while ensuring the appropriate demand response modeling requires huge amounts of data collection on a regular interval. It also requires a very reliable channel to transmit command and control signals to intervene during emergency situations. Therefore, to ensure distributed control, a powerful platform for large scale real-time information acquisition, communication, storing and processing is required. The platform is vital for fault protection, control, monitoring and diagnosis of the grid  \cite{AMahmood}. Compared to other verticals, the smart grid has relatively less stringent requirements, with latency ranging from 3 to 20 ms and acceptable packet loss rate (reliability) of $1-10^{-6}$ \cite{PSchulz17ComMag}.

\subsection{Smart City}

Smart city is defined as an embedded technology based innovative city with sustainable environmental, economic and cultural infrastructure. The aim of the smart city is to enrich the living standard with improved city infrastructure, traffic management, governance, power management, water and waste management, health system, automated driving \cite{rao2018impact}. All the latest services of the 5G network should ideally  be available in a smart city. Data transmission range of the city is required up to several Gbps. Every point in the city is connected with the network, hence, a massive amount of devices are connected in the city. Many applications in the smart city are time critical. While plenty of works are found on the 5G network, a limited numbers of them address the URLLC issues for enabling smart cities.

A theoretical framework for development, deployment and in-life management of a 5G cloud native application is proposed in \cite{rusti2018smart}. % using MATILDA architecture \cite{gouvas2017design}. A generalised architecture, SmaLi-5G of a 5G network is proposed in \cite{oproiu20185g}. The architecture is designed from the mobile operator perspective for the application of smart lighting for Alba Iulia of Italy.
The distributed edge-based design is preferred over centralised cloud-based management to ensure low latency operations \cite{HHabibzadeh}, \cite{fang2021modeling}. For road traffic management, various elements of the traffic system such as street lights and road signs may be connected to a central database which can dynamically update based on real-time data analysis of the overall traffic scenario of the city. Some health and safety related emerging applications such as early earthquake warning systems and health monitoring require special attention to sensors and hardware issues to meet the latency and reliability requirements. Some works advocate using grant free transmissions and URLLC-eMBB multiplexing to ensure guaranteed and prioritised delivery of packets \cite{LDerrico}.

\subsection{Tactile Internet}

Tactile Internet is an emerging paradigm that enables real time control along with  touch and sense transmission in addition to traditional communication signals. In the tactile Internet, humans typically interact with the system such as teleoperation and immersive virtual reality. This requires control or haptic signal and sensing feedback to be transferred along with audio and video/image signals between the two ends in near real time %\cite{GFettweis}, 
\cite{shi2022versatile}. %The Figure \ref{fig_haptic} shows a schematic diagram of a haptic communication system that operates using tactile internet. 
Since the control and sense needs to be communicated in near real time, the end to end latency requires to be extremely low, typically in the range of $<10$ ms and the reliability needs to be extremely high, particularly where health and safety is at stake. The tactile internet brings a new set of possible applications in health care, industrial automation, robotics, education, gaming, autonomous driving, and smart utility management \cite{MsimsekJSAC}. The framework for tactile internet and various applications scenarios are being defined by IEEE standard association under \emph{IEEE P1918.1 Working Group on Tactile Internet}. 

 \begin{figure*}[htp!]
 \centering
 \includegraphics[width=0.7\textwidth]{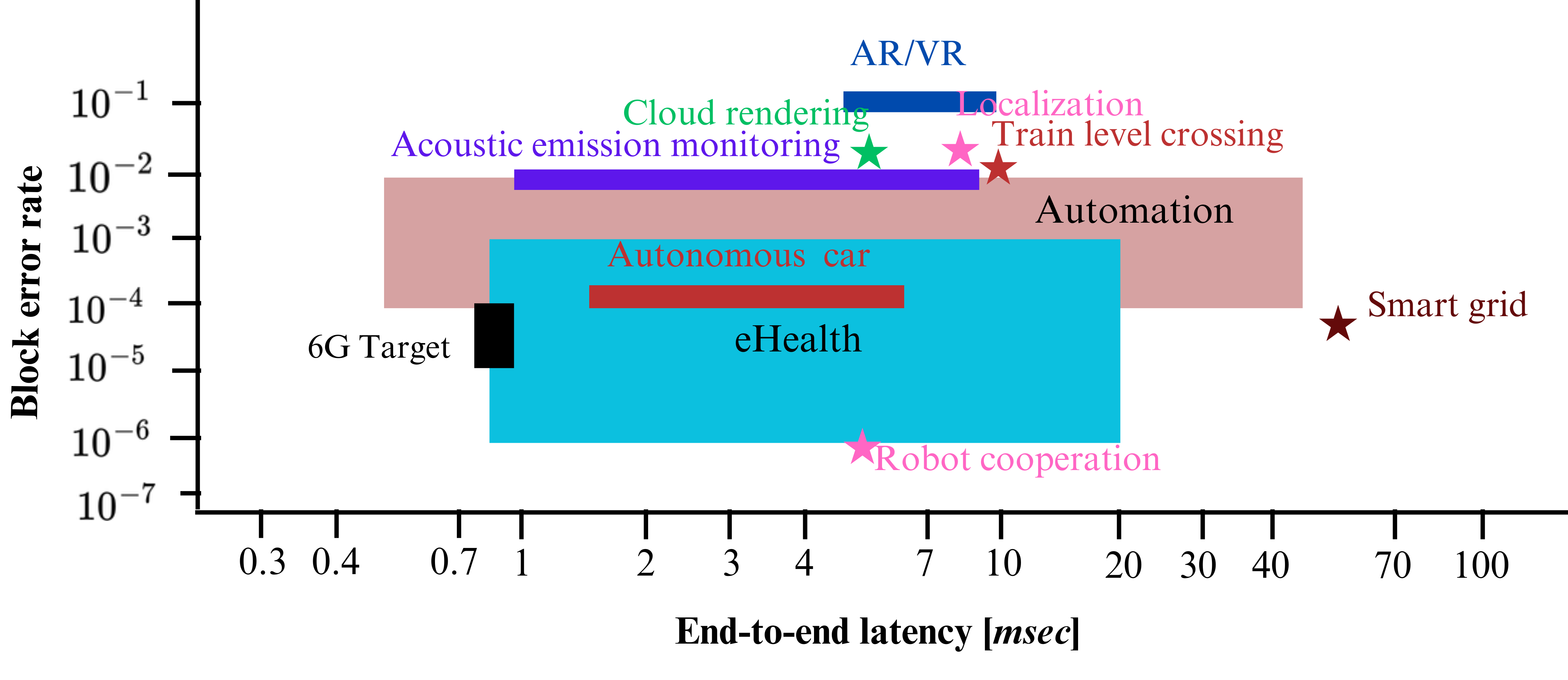}
 \caption{Latency and reliability requirements of various URLLC services \cite{MShirvanimoghaddam2018short}.}
 \label{fig_usecases}
 \end{figure*}

\begin{table*}[!htp]
\caption{Latency and reliability requirements of different use cases for automation} % title of Table
\centering % used for centering table
\begin{tabular}{|c|c|c|c|} % centered columns (4 columns) % centered columns (4 columns)
\hline\hline %inserts double horizontal lines
Scenario & End-to-end latency & Reliability & data rate \\
 & & &  \\ [0.5ex] % inserts table
%heading
\hline % inserts single horizontal line
Discrete automation (User Experience)  & 1 ms & 99.9999 & 1 Mbps to 10 Mbps \\ % inserting body of the
\hline
Discrete automation (motion control) & 10 ms & 99.99 & 10 Mbps \\ 
\hline
Process automation (remote control) & 50 ms & 99.9 & 1 Mbps to 100 Mbps \\ 
\hline
Process automation (monitoring) & 50 ms & 99.9 & 1 Mbps \\
\hline
Electricity distribution (medium voltage)& 25 ms & 99.9 & 10 Mbps \\
 \hline
Electricity distribution (high voltage)  & 5 ms & 99.9999 & 10 Mbps \\
\hline
Intelligent transport systems (infrastructure backhaul) & 10 ms & 99.9999 & 10 Mbps \\
\hline
Tactile interaction & 0.5 ms & [99.999] & [Low] \\
\hline
Remote control & [5 ms] & [99.999] & [up to 10 Mbps] \\
\hline %inserts single line
\end{tabular}
\label{table:usecases} % is used to refer this table in the text
\end{table*}

The latency and reliability requirement for tactile internet will vary depending on application scenarios. For remote surgery, extremely reliable (packet error rate of $10^-9$) and high quality audiovisual communication which requires data transmission rate in the range of several 100 Mbps. However, the latency requirement varies depending on the component of the communication. For audio and video, the remote surgery requires a number of other information to be transferred such as  vibration, touch, pressure, velocity and torque for synchronisation of master-slave robotic surgical equipment. The torque requires latency of approximately 1ms \cite{KSKim19}. The synchronisation of all these components makes challenges more difficult to meet. A system design approach considering PHY/MAC and cross-layer perspective for URLLC application has been advocated in \cite{CLi19a}.

%Traditional QoS metrics are not sufficient for performance evaluation and bench-marking of tactile internet services. To date, there is no universally agreed or defined benchmark of QoS metric for haptic communication over tactile internet.  In \cite{JSachs19}, major challenges including the lack of objective quality metrics for evaluating the performance of various haptic interaction has been discussed.

\section{Challenges in URLLC Design}\label{sec:challenges_urllcDesign}

While URLLC has offered a promising solution for a number of 5G verticals, achieving stringent latency and reliability requirements remains a major challenge in realising full potential of 5G and beyond. In a well managed environment, managing those requirements are somewhat possible in the link layer. However, in a network spread over a wide geographic area and in multihop scenarios, it is extremely difficult to meet URLLC targets in end to end fashion in network layer as well as in upper layers. In this section, the challenges from various technical perspectives are discussed in detail.

3GPP has specified requirements for the 5G system for different use cases and applications. The complete service requirements and KPIs for private networks are specified in Release 16 which includes various applications such as  industrial automation, the tactile internet, augmented reality (AR), and virtual reality (VR), where AR and VR are immersive technologies that blend physical and digital worlds (AR) or create fully virtual environments (VR). These requirements are specified by 3GPP Radio Access Network (RAN) and Services and Systems Aspects groups. The high reliability and low latency are the most important KPIs of URLLC. These requirements depend on applications. The general version of requirements are as follows.

 \begin{figure*}[!htp]
 \centering
 \includegraphics[width=0.8\textwidth]{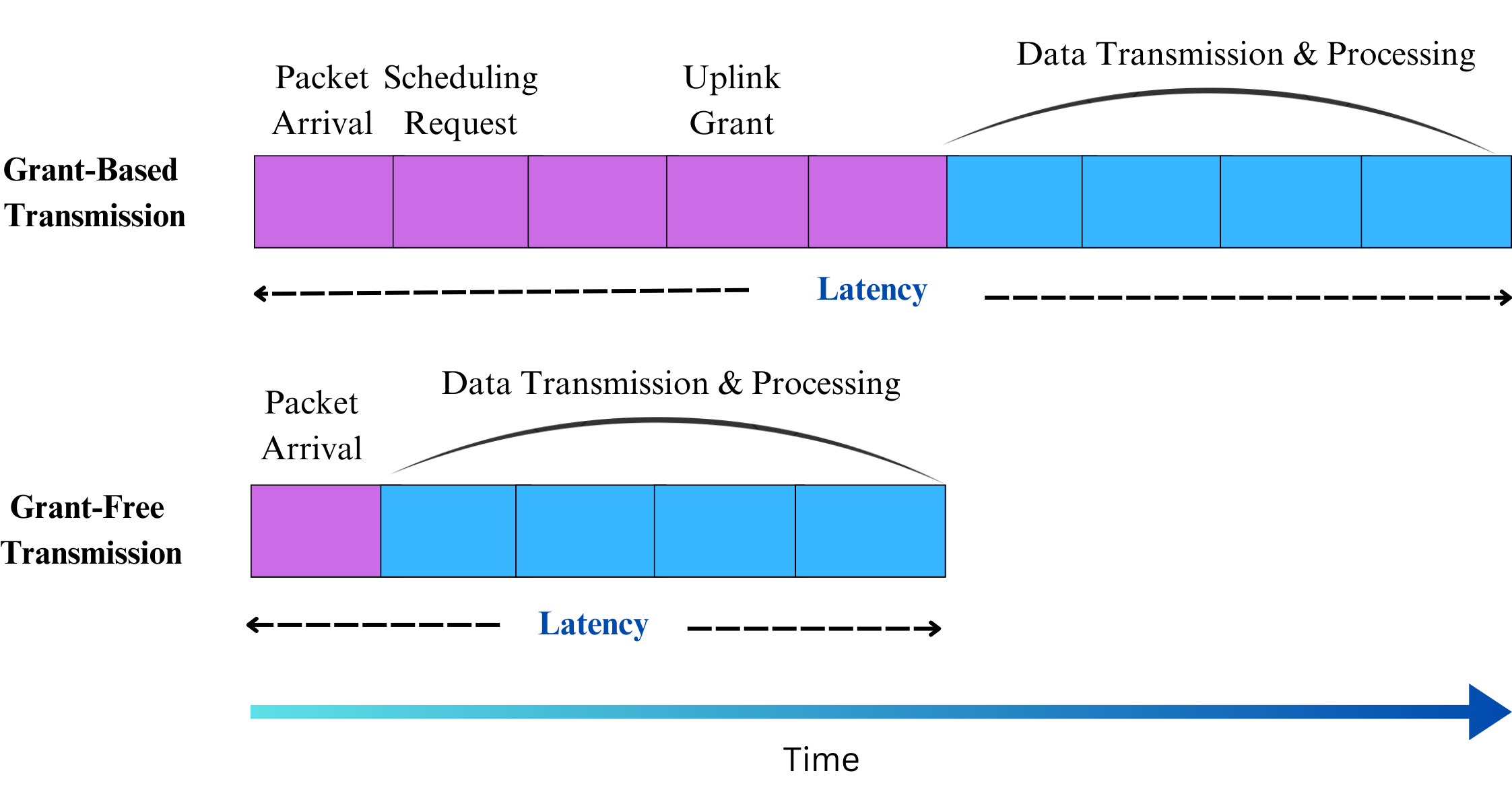}
 \caption{Time window of grant-based and grant-free scheme}
 \label{fig_3_gb_gf}
 \end{figure*}

\subsubsection{Ultra High Reliability}
The general version of requirements are specified in \cite{tr3Gpp38913}. One of the most important requirements is high reliability which is specified in the release as follows. The user-plane reliability requirement is specified as $1-10^{-5}$ (i.e. 99.999\%) for short message transmission. This requirement is application specific. For example, according to 3GPP \cite{ts3Gpp22661} reliability requirement of ITS is $1-10^{-4}$. The extreme reliability requirement of $1-10^{-9}$ is also specified for some use cases \cite{osseiran2015manufacturing}, %\cite{seimensag2016}, 
\cite{chen2018ultra}.
%\cite{holfeld2016wireless}

However, to increase the reliability, generally data packets with large metadata (including preamble, end-device ID, frame check sequence authentication etc) is required. This contradicts the short packet size requirement of URLLC. Thus achieving high reliability with short packets is challenging in 5G.

\subsubsection{Low Latency}
As discussed above, with the emergence of new applications requiring real time interventions and interactions, latency requirements are becoming increasingly stringent. In 3GPP Release 15, the user-plane and link level (both DL and UL) latency requirements are specified as 1 ms and 0.5 ms respectively.

The specified general reliability and latency requirements are, however, inadequate for a variety of use cases as shown in Figure \ref{fig_usecases}. Automotive use cases are an important consideration of 5G URLLC. The assisted, co-operative and tele-operated driving are the main areas of the applications. Several things are incorporated in these systems including a video feed, a GPS-positioning on a map and current weather conditions. In this case 99.999\% user plane reliability is required. On the other hand the latency requirements are 10 ms and 20 ms respectively, for assisted co-operative and tele-operated driving.

The latency and reliability requirements of factory automation vary depending on the specific applications. For some applications the requirement is extremely high. For example, reliability requirement is specified as $1-10^{-9}$ or more along with user-plane latency as 1 ms for local monitoring and control environment and 5 ms for remote monitoring and control environment %\cite{osseiran2015manufacturing}, 
\cite{chen2018ultra}, \cite{holfeld2016wireless}. The process automation is aimed at reducing human intervention and time to deliver products. The user-plain reliability, E2E latency and jitter  requirements for process automation are specified as $1-10^{-6}$, 50 ms and 20 ms respectively. The detailed requirement about different use cases is presented in the Table \ref{table:usecases}.

\subsection{Overhead for control signalling}

Control signal overhead has a strong influence on latency performance of URLLC packet transmission. The control signal is required to achieve high reliability. However, this additional bit increases the packet length and hence increases the latency of the transmission. Therefore, finding a trade-off between appropriate length of control signal to meet the reliability requirement while ensuring the latency constraints remain a major challenge in URLLC design  \cite{soret1}. To reduce the control signal overhead different techniques are used including semi-persistent scheduling and grant-free scheduling.

In standard scheduling a UE sends a scheduling request (through control signal) to the base station. The base station allocates a RB to the UE based on a predefined algorithm. To reduce the signal overhead in every RB semi-persistent scheduling is an alternative way %\cite{jiang2007principle}, 
\cite{song2022semi} in which a control signal overhead is used to pre-allocate several predefined RBs to a specific UE over a certain time sequence. A UE under this policy knows in advance the TTI/RB couples where it should decode data from physical downlink shared channel (PDSCH), which is responsible for carrying user data and higher-layer signaling information from the base station to UEs (or encode information to physical uplink shared channel (PUSCH) that is used for carrying user data from UEs to base station. This eliminates the requirement of additional overhead physical downlink control channel (PDCCH), a channel used for carrying downlink control information. This technique is applicable for the periodic data transmission of UEs. Since the periodic reservation is not applicable for sporadic data transmission, hence the control overhead reduction is challenging. 

Multiple access techniques have a great influence on the fulfillment of URLLC requirements due to the spectral efficiency, which affects reliability; computational complexity, which affects latency, and the choice between grant-free and grant-based access mechanisms. The grant-free transmission is suitable for aperiodic uplink transmission in which a UE sends message without any scheduling request and grant issuing and thus reduces the control signal overhead \cite{PSchulz17ComMag}. A comparison between the grant-based and grant-free transmissions is shown in Fig.~\ref{fig_3_gb_gf}. It is evident that the grant-free transmission is much faster compared to the grant-based transmission due to the reduced control overhead. For both UL and DL transmissions, grant-free non-orthogonal multiple access (NOMA) has been proposed for URLLC in various literature \cite{mahmood2019uplink} \cite{dougan2019noma}. The grant-free NOMA removes the grant-request and the scheduling process, thereby reducing the control signal overhead and the latency without substantially decreasing the reliability. Although the grant-free NOMA has great potential to reduce the control signal overhead, it requires complex computation as well as a large number of iterations which affects latency performance. Thus further investigation is required for successful implementation of grant-free NOMA schemes in URLLC based applications. While NOMA reduces latency due to the grant-free access, it is not theoretically suitable for multi-antenna systems \cite{katwe2024rsma}. This limitation can be overcome using space division multiple access (SDMA) which allows simultaneous use of time and frequency by beamforming. Furthermore, it is less complex compared to NOMA as it does not exploit successive interference cancellation (SIC) method. However, this method requires sufficient isolation between the receiver to limit interference. Recently proposed rate splitting multiple access (RSMA) harvests the benefits of both NOMA and SDMA. It divides the user message into private and common parts. RSMA can be considered as a generalisation of NOMA and SDMA when the private message and common message vanish, respectively. It has much lower computational overhead as it requires only one SIC while NOMA requires multiple SICs. Multiple access techniques should support short packet size, grant-free access, high user mobility, high spectral efficiency, etc. Furthermore, intelligent reflecting surface-aided multiple access is expected to play a vital role in URLLC, especially for cell-edge users, by ensuring improved reliability.

\subsection{Co-existence with other systems}
One of the key features of the 5G wireless system is that different services such as URLLC, eMBB and mMTC must be able to co-exist. However, this poses a significant challenge as each of them has different requirements which often have opposite characteristics. As discussed widely in this article, URLLC services have very stringent latency and reliability requirements compared to eMBB and mMTC. On the other hand, eMBB has much higher data rate requirements compared to URLLC and mMTC. eMBB can tolerate much higher latency and low reliability compared to other two. For mMTC, the minimum connection density requirement is 1000000 devices per km$^2$ which is significantly higher than eMBB resulting in very different connectivity and control signal requirements.

Therefore coexistence of URLLC with these heterogeneous systems is a major challenge owing to diverse system requirements. To satisfy those requirements, 3GPP standardisation body has proposed an innovative superposition/puncturing framework that uses the multiplexing traffic concept in 5G cellular systems.

In 5G, the URLLC services are given the highest priority. The BS should transmit the URLLC packet immediately regardless of whether the URLLC service request arises in a predefined scheduled period or in the middle of eMBB or mMTC transmission \cite{anand2018joint}. This means, to support the URLLC services, ongoing eMBB and mMTC packets should be stopped immediately without any kind of notification. Since this interruption in eMBB and mMTC services are not communicated to the devices in use, reception quality of the eMBB and mMTC services will potentially degrade severely. This problem is dubbed as a coexistence problem in the 3GPP 5G-NR discussion. This is a significant concern for non-URLLC traffic in 5G-NR. To solve the problem several mechanisms to protect the ongoing services are discussed in literature but requires thorough further investigation before implementations.

Since mMTC services require connectivity for short duration with moderate latency, managing them is relatively easier. Thus, the literature on coexistence mechanisms are mostly focused on coexistence challenges between eMBB and URLLC 
%\cite{gidlund2017will}, 
\cite{shi2022risk}. Due to growing demand for URLLC and eMBB services, 3GPP RAN WG1 is aiming to standardise the slot structure. The group proposes a channel to support superposition and puncturing of the slot  \cite{chairman3Gppwg1}. In addition to standardisation activities for existence, recent works focus on system design from different perspectives including overheads, packet size and control channel structure %\cite{pedersen2016flexible}, 
\cite{rinaldi20215g}, 
%\cite{durisi1}
. Joint optimisation for radio resource allocation for eMBB and URLLC is proposed in \cite{you2018resource}. Based on system level simulation and queue modeling, \cite{CPLi2017EuCNC} partitions the bandwidth eMBB and URLLC efficiently. The work \cite{popovski20185g}, %\cite{kassab2018coexistence}
use information theoretic approach to obtain expression for the possible average eMBB rates with puncturing the eMBB traffic by URLLC users. 

Coexistence of eMBB and URLLC traffic using puncturing/superposition based mechanisms with joint scheduling is discussed in \cite{anand2018joint}. The imbalance in user load distribution of eMBB and URLLC among next-generation node Bs (gNBs) may result in the fluctuation of the service satisfaction of individual users of eMBB. To overcome this problem, UAV-assisted relay-based system is proposed in \cite{tian2024intelligentCoexistence} to connect multiple gNBs and macro base stations for multiplexing of eMBB and URLLC. The optimal strategy for load distribution is carried out using deep Reinforcement Machine Learning (RML). This method increases the data rate and user fairness of eMBB while reducing user satisfaction  significantly. A less complex eMBB/URLLC multiplexing scheme is proposed in \cite{shi2024puncturingMaching} dividing the problem into two probelms: (a) alloction of bandwidth to eMBBs, solved by unsupervised deep learning, and (b) allocation of bandwidth and power between eMBBs and URLLC, solved by two-sided matching game. 
However, several challenges need to be addressed before realising the full potential of URLLC and its successful coexistence with other services.

\subsection{Resource Optimisation for URLLC}
Optimising resources for URLLC is very challenging, particularly in the presence of other types of traffic, such as eMBB. Since the characteristics of these two systems vary widely and so is the requirement, meeting the combined need of both systems while ensuring efficient resource optimisation is a very challenging task. Also, data rate requirements and how frequent the data needs to be transmitted vary widely depending on specific URLLC applications. It makes the optimisation of resource allocation further challenging. Due to this complex nature of the traffic, fixed or predefined spectrum/RB allocation for URLLC traffic leads to inefficient utilisation of the spectrum pool \cite{CPLi2017EuCNC}.

Using system level simulation and queuing analysis, the authors in \cite{CPLi2017EuCNC} demonstrated that wideband allocation leads to better utilisation for hard latency requirements. It is also shown that the queuing has a significant impact on the spectrum utilisation as a longer queue will result in dropped packets due to hard latency constraints. Power optimisation is also another aspect which may lead to improved efficiency. In \cite{CXiao_JSAC19}, transmit power optimisation conditions are identified which were implemented to minimise the transmit power of the MIMO-NOMA system while ensuring latency and reliability requirements. 
%Frame_Structure_Combined.png
 \begin{figure*}[!t]
 \centering
 \includegraphics[width=1\textwidth]{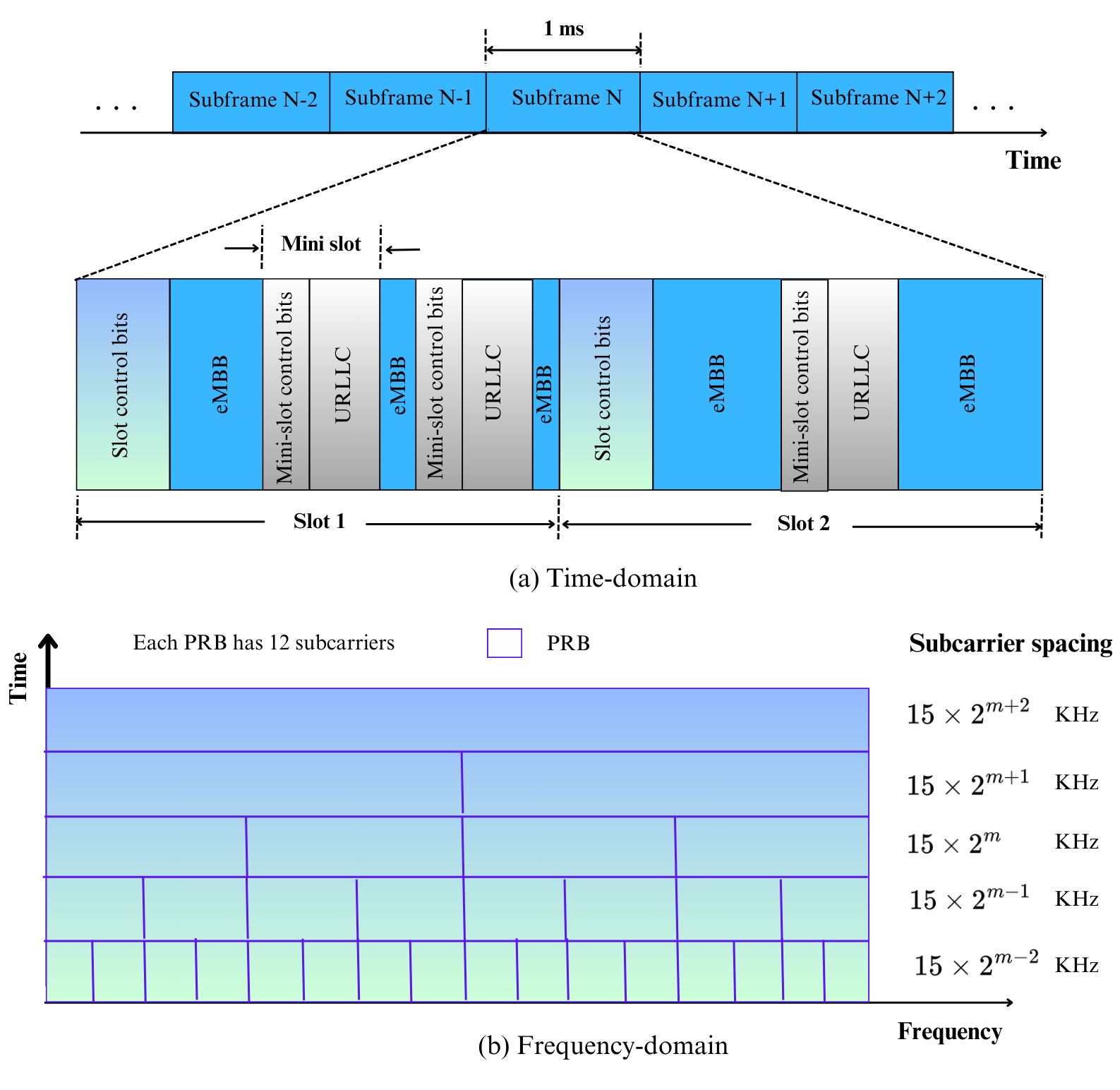}
 \caption{Time and frequency domain view of 5G NR frame structure \cite{lien20175g}}
 \label{fig_frame_structure}
 \end{figure*}

\begin{table*}[htp!]
\caption{Summary of different frame structure} % title of Table
\centering % used for centering table
\begin{tabular}{|c|c|c|c|c|c|c|c|} % centered columns (4 columns) % centered columns (4 columns)
\hline\hline %inserts double horizontal lines
Reference & Subframe & No. of slots & Slot duration & Minislot & subcarrier & No. of OFDM  & OFDM symbol \\
& duration & per subframe &  & duration & spacing & symbols per slot & duration \\
% & & & & & & & & \\ [0.5ex] % inserts table
%heading
\hline % inserts single horizontal line
\cite{lien20175g} & 1ms & Integer no. of slots & Variable & Variable & 15 kHz & 14 & Variable \\ 
\hline
\cite{iwabuchi20175g} & N/A & N/A & 0.25 ms & N/A & 60 kHz  & 6 & 16.67 $\mu$s \\ [0.5ex] 
\hline 
\cite{kela2015novel} & 0.144 ms & N/A & N/A & N/A & 312.5 kHz & N/A & 3.2 $\mu$s \\ [0.5ex] 
\hline 
\cite{pedersen2015flexible} & N/A & N/A & Unit time $\Delta$t=0.2 or 0.25 & N/A & 32 kHz & Integer no. of slots & Variable \\ [0.5ex] % inserts table
%heading
\hline

\cite{you2018resource} & N/A & Not specified & Not specified & Not specified & 15, 30, 60 kHz & 14 & Not specified \\ [0.5ex] 
\hline

\cite{vihriala2016numerology} & Variable & N/A & N/A & N/A & 15, 30, 60 kHz & N/A & 66.7, 33.33, 16.67 $\mu$s \\ [0.5ex] 
\hline

\cite{pedersen2016flexible} & Variable & N/A & N/A & N/A & 32, 16 kHz & N/A & 33.33, 66.66 $\mu$s \\ [0.5ex] 
\hline

\end{tabular}
\label{table:frame_structure_summary} % is used to refer this table in the text
\end{table*}

Securing the communication in URLLC is particularly challenging due to short packet size and hence physical layer security (PLS) becomes an important technique for reducing the vulnerability. However, the implication on resource allocations need to be explored thoroughly. For example, in mission-critical IoT applications, PLS technique may improve resource utilisation as it will either reduce or eliminate the need for separate channel usage for cryptographic security key exchange \cite{HRenTCom2020}. However, optimisation of power allocation and beamforming vector design is challenging due to non-convex nature leading to increase in computational complexity and latency compromise.

Extremely high reliability requirement is met by employing various re-transmission mechanism to ensure guaranteed packet delivery even if a packet lost in the transmission process. However, this re-transmission leads to resource inefficiency as in case of good channel condition redundant packet transmission occurs. Finding the optimal number of retransmissions is very challenging as the wireless channel quality varies widely, particularly when the receiver or transmitter is not stationary. Another challenge is that due to latency requirement, traditional ACK/NACK based re-transmission may not work. Therefore some intelligent solutions such as blank re-transmission in reserved radio resource \cite{elayoubi2019radio}, adaptive partial  re-transmission with hybrid automatic repeat request (ARQ) \cite{JYeo}, re-transmission in a different hopped frequency \cite{CSun} can be designed.

Moreover, the combined effect of queuing delay, end to end delay, transmission error probability makes the task of resource allocation while meeting the hard latency and reliability threshold extremely challenging. Thus careful study of these effects are required. Most of the works are evaluated under certain fixed scenarios and therefore further thorough investigation as well as rigorous analytical modeling are required to find optimal resource allocation strategy for URLLC in 5G and beyond.

\section{Physical Layer Design for URLLC}\label{sec:physicalLayer}

%{\color{red} Editor: ME \newline length: 3 pages}

There are a number of target KPIs for 5G which include 20 Gb/s peak data rate, 100 Mb/s user experienced data rate, 10 Mb/s/m$^2$ area traffic capacity, $10^6$ devices/km$^2$ connection density, 1 ms latency, mobility up to 500 km/h, backward compatibility to LTE/LTE-Advanced (LTE-A), and forward compatibility to potential future evolution \cite{lien20175g}. To integrate these requirements in 5G the fundamental changes in the physical layer is inevitable especially to meet stringent URLLC requirements. This section addresses the physical layer design issues for URLLC.

\subsection{Frame Structure}
\label{sub:PHY}
The user plane latency is the sum of frame alignment and queuing delay, transmission delay and the receiver processing delay. The frames were designed in LTE-Advanced with the aim of meeting user plane latency of 4 ms \cite{iwabuchi20175g}. However, the same frame structure will not be suitable for 5G as the latency requirement is much lower. Thus a new frame structure is required to reduce the transmission delay.

Considering the requirements of various services of 5G NR, 3GPP has proposed a frame structure in Release 15. In the time domain, all the downlink and uplink transmissions are structured into frames with 10 ms duration. Each frame is divided into subframes of 1 ms duration and subframes are further divided into slots. Each slot consists of 14 orthogonal frequency division multiplexing (OFDM) symbols with 15 kHz subcarrier spacing and normal cyclic prefix (CP). The OFDM symbols in a slot can be used for either downlink or  uplink. This frame structure is more flexible compared to the structure of LTE and is illustrated in Figure \ref{fig_frame_structure}(a). Each slot has a control signal at the beginning and/or at the end of an OFDM symbol as shown in the figure. A unique feature of the structure is that as soon as the resource request arrives, the system allocates the resource without requiring to wait to fill up the full slot. This enables the system to meet the resource request for the URLLC service within the latency boundary. To support the small size packet transmission especially for URLLC packets, the slot is divided into mini slots. Table \ref{table:frame_structure_summary} shows the summary of frame structures.

In the frequency domain, the unit physical resource block (PRB) consists of 12 subcarrier \cite{lien20175g}. Subcarriers are equally spaced and have equal CP overhead. The 5G NR supports multiple subcarrier spacing and similarly supports PRBs with flexible bandwidth. When the PRBs of the different bandwidth are multiplexed in the time domain the boundaries of the PRBs must be aligned as shown in Figure \ref{fig_frame_structure}(b). The figure illustrates the PRB grid structure with 15 kHz$\times 2^m$ spacing, where $m$ is a positive integer. 

To reduce the latency, several approaches have been taken that exploit frame structure to minimise packet transmission time. A mini slot based LTE-Advanced frame structure is designed in \cite{3GPP2016comparison}, \cite{3GPP2016on} and wider subcarrier based frame structure is designed in \cite{kela2015novel}, \cite{3GPP2016onthe} for URLLC. User plane latency can be reduced by using a mini slot based frame with one OFDM symbol per slot \cite{3GPP2016comparison}. The authors in \cite{kela2015novel} proposed a novel frame structure with symbol length of 3.2 $\mu$s and TTI of 167.3 $\mu$s. It was able to achieve 5x latency reduction compared to LTE and maintain air interface latency of less than 1 ms in ultra dense small cell scenarios.

Retransmission mechanism is another important issue that plays a crucial role in latency and reliability constrained packet transmission. In the conventional acknowledgement (ACK) and negative acknowledgement(NACK) based packet retransmission scheme, a packet is retransmitted only when the packet is not received by the receiver which is sent as feedback (NACK). However, this multiple RTT in the process takes much longer than the delay tolerated by URLLC applications. %\, \cite{3GPP2016overview} 
Considering the problem, in \cite{pedersen2015flexible} ACK/NACK-less retransmission policy is proposed. The method sends the same packet multiple times hence the reliability of the transmission is increased significantly at the added cost of resource wastage due to retransmission.  
Similar approach is taken in \cite{iwabuchi20175g} where wider subcarrier spacing of 60 kHz was used with frame length is 10 ms. Frames were divided into 40 slots of 0.25 ms duration. In the proposed structure, first two slots of each frame are merged to incorporate additional control signals e.g. a synchronisation signal. Their field trial with ACK/NACk less approach achieved reliability of 99.999\% and less than 1 ms of latency in both the DL and UL. Another approach is to use variable or scalable frame length in time domain to optimise resource allocation while achieving latency and reliability target \cite{you2018resource} \cite{vihriala2016numerology}.
%In \cite{vihriala2016numerology} a OFDM numerology and frame structure is proposed for 5G network. The proposed numerology and frame structure is scalable and can be used for all the proposed carrier frequencies. 

\subsection{Optimum Packet Structure}
The existing long packet structure cannot guarantee the latency requirement of URLLC. Moreover, the time critical information is generally short in size. Thus a short packet structure is more suitable for latency critical applications. However, compatibility of short packet structure with the current standard remains a major challenge as most 5G standards are based on long packet structure.

For low latency communications, a one-shot random access based system with short data packet is proposed in \cite{lee2017packet} where both pilot and data packet are transmitted simultaneously. The virtual piloting system is used in the work, where the most reliable data packet is used to estimate the channel. The data packet is selected as the virtual pilot based on mean square error estimation. However, the paper shows that the reception quality of the receiver is improved with the frame structure.

In the pilot based frame structure, pilot signal is required for controlling purpose. The target of pilot based frame structure is to find intelligent and efficient ways to minimise the pilot overhead in short packet transmission. % \cite{mousaei2017optimizing}. %\cite{ji2018sparse}
In \cite{mousaei2017optimizing} pilot overhead optimisation technique is proposed which takes packet length and packet error rate into consideration. In \cite{ji2018sparse} a short packet transmission is proposed based on sparse vector coding (SVC). In the SVC, the information is mapped into the position of a sparse vector. The mapped coded vector is then transmitted. The simulation result demonstrated that the designed system meets the performance threshold of URLLC. An unified pilot optimisation is proposed in \cite{jindal2010unified}. The paper shows that the optimisation for a multi-antenna system is effectively the same as a single antenna system.

A joint pilot power, pilot length and block length optimisation technique is proposed in \cite{lin2021pilot}. The low complexity based joint algorithm outperforms the existing algorithms under continuous fading. Based on the analysis, a low-complexity joint pilot power, pilot length and block length optimisation (JPLLO) algorithm is proposed, which achieves a near-optimal performance and a dramatic complexity reduction compared to exhaustive search, converging within only 1–3 iterations. The JPLLO algorithm also significantly outperforms the previous joint optimisation algorithm under continuous fading.

A more radical approach to reducing overhead is pilot-less packet transmission. Authors in \cite{ji2019pilot} proposed a coding scheme known as pilot-less sparse vector coding (PL-SVC). The PL-SVC performs sparse vector conversion and uses random spreading for encoding. The unique feature of PL-SVC over SVC transmission is that it maps the input as a composite of sparse vector and fading channels. This is done without the channel state information and with low computational complexity. The proposed PL-SVC transmission can also be applied to SIMO and MIMO transmission systems with a simple extension of the scheme.

% \begin{table*}[htp!]
% \caption{Summary of the major numerology} % title of Table
% \centering % used for centering table
% \begin{tabular}{|c|c|c|c|c|} % centered columns (4 columns) % centered columns (4 columns)
% \hline\hline %inserts double horizontal lines
% Reference & Sub carrier & cyclic prefix & OFDM symbol & No. of OFDM \\
% & spacing & duration ($\mu$s)& duration ($\mu$s) & symbols per slot \\
% % & & & & & & & & \\ [0.5ex] % inserts table
% %heading
% \hline
% \cite{vihriala2016numerology} & 15, 30, 60 kHz & 4.69, 33.33, 16.67, and $1.17/2^L$ & 66.7, 33.33, 16.67 $16.67/2^L$ & N/A\\ [0.5ex] 

% \hline % inserts single horizontal line
% \cite{zaidi2016waveform} & 15, 30, 60 kHz and above & 4.69  & 66.67  & 7\\ 
% \hline
% \cite{Josue2018waves} & 15, 30 and 60 kHz & 4.69, 2.34, 1.17 & 66.67, 33.33 and 16.67 & 14 \\ [0.5ex] 
% \hline 
% \cite{yazar2018flexibility} & Variable starting from 15 kHz & Variable starting from 16.67  & Variable starting from 66.67  & 7 and 14 \\ [0.5ex] 
% \hline 
% \cite{ZLi2018ISWCS} & 15, 30, 60, 120 and 240 kHz & 4.7, 2.3, 1.2, 0.59, and 0.29  & 66.7 33.3 16.7 8.33 4.17 & variable, up to 14 \\ [0.5ex] 
% \hline 

% \end{tabular}
% \label{table:numerology_summary} % is used to refer this table in the text
% \end{table*}

In pilot based packet transmission, optimum power of pilot transmission and pilot length are the main concerns. The optimal design of the packet is application specific. Thus the application specific constraints need careful consideration in the design. On the other hand, in pilot-less packet transmission, low complexity coding is an important concern as the stringent time requirement of the short packet transmission. However, for both cases the size of the packet should be optimised to meet the latency and reliability constraints.

%\textcolor{red}{ Some concluding remark for this subsection is required.}.

\subsection{Waveform and Numerology}

 %\subsubsection*{Waveform}
 
% In OFDM based LTE systems, uniform time-frequency (also known as RB) allocation is used which is not very flexible. In LTE around 10\% spectrum is used as a guard band which leads to further inefficiency. Finally, in LTE stringent synchronisation is required to mitigate the inter-carrier interference. To maintain this requirement, the BS sends timing advance signals to UEs. On the other hand data intensive traffic of 5G such as eMBB will require much higher spectrum efficiency compared to LTE. Thus 
 
 6G networks are expected to support a wide variety of services and scenarios. In addition, the targeted mobility for 6G speed will be 1000 km/h \cite{jiang2021road}. These factors affect the performance of URLLC. With the traditional waveforms, it is not possible to satisfy these requirements. New waveforms have to be designed. A waveform affects both latency and reliability. Contributory factors to latency include symbol duration, overhead, processing time, and so on. The newly designed waveform should use short symbol duration, fewer overhead bits, and reduced computational complexity. On the other hand, the reliability is impacted by the waveforms' robustness against channel impairments, error correction capability, and spectral efficiency. The high-speed mobility requirement of 6G makes the waveform design more challenging. To address this, delay-Doppler domain waveforms, such as OTFS and orthogonal delay-Doppler alignment multiplexing (ODDM), have recently drawn the researchers' attention. In addition, the peak-to-average (PAPR) and out-of-band emission (OOBE) properties of waveforms also indirectly impact the reliability. The performance of OFDM, GFDM, and OTFS is compared in \cite{sheikh2024comparative} against the impulsive noise channel, where OTFS is found to outperform others in terms of PAPR and BER. In \cite{tarboush2022single}, performance of OFDM, DFT-spread-OFDM, and OTFS are compared over a wideband THz channel. DFTS-spread-OFDM outperforms others in terms of BER with an increased computational complexity. However, OFTS provides the lowest latency among them. In \cite{xiao2024rethinking}, delay-Doppler alignment modulation (DDAM) is reported to achieve the lowest PAPR and BER with the least computational complexity compared to OTFS and OFDM.
 
 The inflexible time-frequency allocation of LTE is not suitable for 5G as a massive number of devices are connected under a single BS which will incur significant resources to fulfill this requirement. % Therefore the rigid and uniform structure of RBs in LTE is not suitable for  5G in general and URLLC in particular. 
 5G also requires flexibility in resource allocation in order to meet the variable demand from heterogeneous traffic types including eMBB, mMTC and URLLC. The most logical way forward is to enable coexistence of waveforms  in a mixed numerology setup \cite{SEldessoki17}, %\cite{guan2016ultra}. 
 which will help to achieve higher spectrum efficiency and will also enable asynchronous transmission.

For 5G, various alternatives to conventional OFDM have been considered. It includes quadrature amplitude modulation filter bank multi-carrier (QAM-FBMC), %\cite{kim2015qam}, 
universal filtered multi-carrier (UFMC) 
%\cite{vakilian2013universal} 
and generalised frequency division multiplexing (GFDM). %\cite{michailow2014generalized}. 
These techniques offer better bandwidth efficiency, relaxed synchronisation requirements, and reduced inter-user interference. However, these techniques suffer from increased receiver complexity and difficulty in integrating MIMO \cite{lien20175g}.%, and specification impacts . %In \cite{kim2015qam}, QAM supported waveform is designed that provided the same symbol rate as CP-less OFDM. The designed waveform exploits the redundancies of CP and guard-band to improve overall efficiency. In \cite{vakilian2013universal} multi-carrier transmission is proposed that overcomes the inter-carrier interference (ICI) problem of OFDM. In the technique multiple BSs receive the waveform of UEs signals and send the receive signal to a central unit. The central unit then detects and processes the signals. A unified physical layer waveform is designed in \cite{michailow2014generalized} to enhance the data rate, reduce power consumption in battery driven communication devices and reduce response time for control system applications. Although uniform waveform improves latency performance which is crucial for URLLC, it is not suitable for heterogeneous services of 5G. 

Other efficient variants of OFDM based waveform includes filtered OFDM (f-OFDM), %\cite{zhang2015filtered}, 
windowing-OFDM (W-OFDM),
%\cite{qualcom2016owaveform}, 
orthogonal generalised frequency division multiplexing (OGFDM). %\cite{kadhum2019new}. 
In f-OFDM \cite{zhang2015filtered}, % is a flexible and spectrally efficient waveform. It overcomes the limitations of the conventional OFDM designed for 4G. The assigned bandwidth in the waveform is split in several subbands. 
the different subbands can be allocated for different service classes of 5G including eMBB, mMTC and URLLC. In contrast to unified numerology of OFDM, where the term numerology is understood as the parameterisation of waveforms including CP, subcarrier spacing, OFDM symbol duration and number of subcarriers per OFDM symbols, different numerology and waveform could be used in the different subbands for more efficient spectrum utilisation. W-OFDM utilises the spectrum in the same way but uses the windowing concept to reduce out of band emission and thereby limits the need of guard bands. OGFDM improves the channel capacity, reduces the BER and increases the number of connected users. This mixed numerology and flexible waveform design enables 5G to accommodate heterogeneous service classes including URLLC \cite{zhang2018mixed}.

\subsection{Channel Coding and Modulation}

In wireless communication various channel coding and modulation schemes are dynamically chosen together to mitigate the effect of fading and other channel impairments. When the channel condition is bad, higher coding length and lower order modulation schemes are usually chosen. However, many traditional channel coding schemes are not suitable for URLLC in their current form due to introduction of short packet length to meet strict latency requirements. Also, many of them will not be compatible with very strict reliability requirements.

\begin{table}[!htp]
\caption{Complexity analysis of coding schemes [adapted from \cite{sybis2016channel}]} 
\centering % used for centering table
\begin{tabular}{|c|c|c|c|} 
\hline\hline %inserts double horizontal lines
Reference & Coding scheme & Block length & Complexity (\%) \\

\hline %inserts single line
\cite{berrou1993near} & Turbo & 40 bits & 100\% \\
\hline %inserts single line
\cite{sybis2016channel} & LDPC MSA & 40 bits & 100.7\% \\
\hline
 \cite{sybis2016channel} & LDPC SPA & 40 bits & 98.0\% \\
\hline
\cite{arikan2009channel} & Polar & 40 bits & 1.5\% \\
\hline
\cite{tal2015list} & Polar SCL & 40 bits & 100.5\% \\
\hline
\cite{maiya2012low} & Convolution & 40 bits & 66.7\% \\
\hline
\cite{berrou1993near} & Turbo & 200 bits & 100\% \\
\hline
\cite{sybis2016channel} & LDPC MSA & 200 bits & 100.7\% \\
\hline
\cite{sybis2016channel} & LDPC SPA & 200 bits & 98.0\% \\
\hline
\cite{arikan2009channel} & Polar & 200 bits & 1.6\% \\
\hline
\cite{tal2015list} & Polar SCL & 200 bits & 100.7\% \\
\hline
\cite{maiya2012low} & Convolution & 200 bits & 66.7\% \\
\hline
\end{tabular}
\label{table:coding} % is used to refer this table in the text
\end{table}

Turbo \cite{berrou1993near}, low-density parity-check (LDPC) \cite{gallager1962low}, polar \cite{arikan2009channel} and convolution \cite{maiya2012low} codes have been adopted as the potential candidates channel coding scheme in 5G services. Turbo code is a primary candidate of 5G as it has been used in its predecessor LTE. However, the complexity of implementation and typical block error rate (BLER) performance (in the range of $10^{-3}$ and $10^{-4}$), makes it unsuitable for URLLC \cite{sybis2016channel}. Therefore, for selecting appropriate channel coding mechanism, careful attention should be given to a number of factors including complexity of the algorithm, flexibility in code rate and length, time consumed in coding and decoding, and error performance \cite{sybis2016channel}.

Different types of channel coding are proposed including variants of LDPC, polar codes, tail-biting convolutional codes (TB-CC) and turbo codes\cite{MShirvanimoghaddam2018short}. Table\ref{table:coding} compares various encoding schemes with their relative complexity. In the aforementioned work of \cite{sybis2016channel}, two different encoding schemes known as suboptimal mini-sum algorithm (MSA) and optimal sum-product algorithm (SPA) are proposed where the latter has slightly lower complexity compared to the former algorithm. In \cite{wu2018low} a raptor-like protograph is defined to design the LDPC for URLLC short packet communication. The protograph-based raptor-like LDPC (PBRL-LDPC) is a generalisation of Richardson-Urbanke (RU) protograph in \cite{divsalar2005low} for low-rate protographs which supports rate-compatible construction and lead to better performance for short length code.

One challenge with LDPC is that it has an error floor for short packets of 40 byte size. In this range, polar codes outperforms LDPC and doesn't exhibit any error floor \cite{MShirvanimoghaddam2018short}. For polar codes, successive cancellation (SC) may incur large latency which can be mitigated by the use of successive cancellation list (SCL) decoding, which keeps a list of most 
likely  decoding  paths  at  all  times. Further improvement can be achieved by (CRC)-aided SCL (CA-SCL) \cite{KNiu2012CL}. Another advantage of polar coding is that it can achieve 1-bit granularity. One challenge however is the increased complexity with cancellation list size. Thus a careful balance between them is usually recommended. Also, in order to cope with the channel variation, coding schemes need to be flexible to incorporate various adaptive modulation and coding (AMC) schemes. However, there is a tremendous scope of work to find low complexity coding and decoding schemes that meet both latency and reliability requirements.

\subsection{MIMO and Massive MIMO Technologies}

With the emergence of mission critical applications researchers integrate different technologies to attain the URLLC in 5G. Multiple-input and multiple-output (MIMO) is such a technique in which multi-connectivity is used to achieve the services of 5G. For example, in \cite{CXiao_JSAC19} NOMA with MIMO and 4 antennas are used to enhance connectivity, reliability and latency. In the paper the authors used stochastic network calculus to find the critical condition for optimal power allocation to the system to achieve the reliability and latency. In \cite{tang2016improving} the authors propose a spatial modulation (SM) based detection scheme to reduce the latency and increase the reliability. The SM-MIMO is a low complexity algorithm that reduces the latency and increases the reliability in the software defined radio simulation platform. In \cite{tarneberg2017utilizing} the authors use massive MIMO for the tactile Internet and show that the URLLC can be achieved with the MIMO technology. The paper concludes that the results can generally be applied to any URLLC applications including robotics, control and IoT. In \cite{casciano2019enabling} a realistic indoor factory scenario is studied to enable reliable communication to the actuators in the factory. The authors simulate the environment with 16 antennas with a new path-loss model to reflect the real scenario of the factory. The study shows that the requirement of URLLC cannot be satisfied if the modeling of the environment is not accurate. The summary of the technologies with number of antennas used and applicability for the URLLC is presented in Table \ref{table:mimo_technology}.

\begin{table}[!htp]
\caption{Overview of MIMO technology for URLLC} 
\centering % used for centering table
\begin{tabular}{ |c|c|c|c| } 
\hline\hline %inserts double horizontal lines
References & No. of antenna used & Consider URLLC? & Achieved URLLC? \\
\hline %inserts single line

\cite{CXiao_JSAC19} & 4 & Yes & Yes\\
\hline %inserts single line
 \cite{tang2016improving} & 36 & Latency only & Latency achieved\\
\hline %inserts single line
 \cite{tarneberg2017utilizing} & 18 (minimum) & Yes & Yes \\
\hline %inserts single line
\cite{casciano2019enabling} & 16 & Yes & No\\
\hline %inserts single line

\cite{kashima2016large} & 64 & No & No\\
\hline %inserts single line
\cite{xu20163d} & 6 & Yes & No\\
\hline %inserts single line
\cite{busari2019terahertz} & 64 & Yes & Yes\\
\hline %inserts single line
 
\cite{vu2017ultra} & 32 & Yes & Yes\\
\hline %inserts single line
\end{tabular}
\label{table:mimo_technology} % is used to refer this table in the text
\end{table}

%\subsection{Massive MIMO aspects}\label{sub:mMIMO}

Massive MIMO uses a large number of antenna arrays and is capable of supporting multiple users to communicate over the same time-frequency resource block. This simultaneous use of the same resource block by several users offers the opportunity of providing similar latency and reliability for all of them along with multi-fold increase in data rate compared to conventional MIMO systems. The ultra high reliability can be offered by an interesting concept called \emph{channel hardening} effect \cite{LLu14}, in which case, the channel becomes stable and the channel coefficients are deterministic due to the deployment of a large number of transmitting antennas.  

The work in \cite{WTarneberg17} evaluated the performance bounds of massive MIMO and recommended the system configuration to ensure achieving the latency requirement. %It also discussed the advantages and disadvantages of massive MIMO for use in tactile internet.
The work in \cite{TVu17} proposed a  utility-delay control approach using Lyapunov technique for  intelligently adapting to queue length and channel dynamics. The technique achieves 99.00\% reliability while significantly reducing the latency compared to some other available techniques.

URLLC design of cell-free massive MIMO system for smart city is investigated in \cite{peng2022resourceUplink, peng2023resourceDownlink}. The cell-free massive MIMO system satisfying URLLC requirements for uplink is proposed in \cite{peng2022resourceUplink}. The performance of the full-pilot zero forcing and maximal ratio combining are investigated under the imperfect channel state information(CSI) and finite blocklength condition, where the CSI is understood as the information that a UE and/or base station has about the communication channel between them. The investigation reveals that the increase in the number of user devices improves the weighted sum-rate in the cell-free massive MIMO system, while that has no impact in the centralised massive MIMO system. The impact of the cell-free massive MIMO system for the downlink under the same scenario is investigated in \cite{peng2023resourceDownlink}. In \cite{zhang2024performanceRician}, the performance of the cell-free massive MIMO-enabled URLLC is investigated over the Rician fading channel with phase shifts. The requirements of URLLC are fulfilled via optimised power allocation. The minimum mean squared
error estimation (MMSE) is found to outperform the linear MMSE concerning the number of supported user devices. In \cite{huang2024performanceSE_EE}, the impact of the short blocklength transmission on the energy and spectral efficiency is investigated for a cell-free massive MIMO-enabled URLLC system. It is found that the cell-free architecture with URLLC requirements achieves higher spectral efficiency sacrificing energy efficiency. In addition, a tighter latency requirement leads to a decrease in spectral efficiency. 

%ISAC MIMO
Recently, researchers have started using the massive MIMO in conjunction with the integrated sensing and communication (ISAC) concept to improve the spatial diversity, beamforming gains, and spectral efficiency \cite{ding2022joint, behdad2024joint}. However, this integration burdened the base station to perform sensing data processing and analysis. To overcome this, an ISAC system is proposed for multi-antenna systems in \cite{ding2022joint}, where the data processing is done in an edge server. The ISAC base station is equipped with a MIMO system. A precoder is designed by leveraging quadratic transform-based fractional programming to minimise the transmit power while satisfying the requirements of both URLLC and ISAC.  In \cite{behdad2024joint}, the authors propose a method for joint optimisation of power allocation and blocklength for transmission and end-to-end energy consumption minimisation by formulating two non-convex optimisation problems of a downlink cell-free massive MIMO system with URLLC and multi-static sensing. The study shows that the integration of sensing reduces blocklength, which is further reduced by the end-to-end energy minimisation. Furthermore, the integration of ISAC degrades energy efficiency due to the energy consumption of receive access points. A bi-static MIMO ISAC system is proposed in \cite{nikbakht2025mimo} to enable both eMBB and URLLC transmissions. The URLLC transmission is only triggered based on the sensing data. Dirty paper coding is used for interference cancellation caused by the sensing signal, sensing and eMBB transmission during regular eMBB transmission and URLLC transmission, respectively. This method outperforms power-sharing methods. An ISAC-assisted beamforming method is proposed in \cite{yuan2023orthogonal} for vehicular networks. In this method, the locations and directions of vehicles are extracted from the received OTFS signal. Analyzing such information extracted in last few reception, the locations of the vehicles are predicted and beamforming is done toward the predicted locations.  

% \begin{figure*}[!t]
% \centering
% \includegraphics[width=1\textwidth]{Diversity.png}
% \caption{Conceptual illustration of three major diversity techniques}
% \label{stf_diversity}
% \end{figure*}

\subsection{mmWave Communications}

To mitigate the ever-increasing high data rate wireless communication system, more spectra and new technologies are required. Millimeter-wave (mmWave) communication is considered as a promising candidate for the 5G cellular network. A vast amount of spectrum is available above 10 GHz, which has the potential to deliver gigabits of data rate to individual users. Therefore, to meet the growing requirements, it is expected to embrace the 60 GHz mmWave ISM band as an enabling technology with URLLC support, especially for industrial wireless communication. However, mmWave communication faces several challenging issues, including high path loss and penetration losses, which lead to severe degradation of reliability over relatively larger distances.

% \begin{figure*}[!t]
% \centering
% \includegraphics[width=1\textwidth]{mmWave_network.png}
% \caption{A mmWave network}
% \label{mmWave_netowrk}
% \end{figure*}

In \cite{vu2017ultra} the authors explore the issues of URLLC in mmWave enabled massive MIMO networks. The network utility maximisation problem is formulated to address the issues and solved using convex-concave procedure \cite{lipp2016variations}. In \cite{vu2018ultra} the authors investigated the blockage problem in mmWave massive MIMO enabled gigabyte wireless network with URLLC services. A distributed reinforcement learning based approach is adopted in the paper to overcome the problems which would ensure reliability requirements.  Apart from physical layer issues, higher layer issues such as congestion control, core network architecture and MAC layer design also affect the latency and reliability requirement for systems operating on mmWave frequencies.  In order to provide more scalability and flexibility,  core network architecture needs to be modified to bring services and data
physically closer to the end user \cite{ford2017achieving}. A deep learning based low overhead beam selection technique is proposed in \cite{echigo2021deep}. The deep learning model is used to measure the quality of beam by partial beam measurement. The method reduces the overhead in beam training while maintaining the QoS. In \cite{ding2021context} a context-aware beam tracking method is used to help the base station to determine when to trigger beam sweeping in 5G mmWave vehicular communication. The paper uses the noisy and quantised beam-specific reference signal received power as feedback from the vehicle to make the decision.

Since NLoS  is a major challenge in mmWave regime, some works have been proposed to differentiate light of sight (LoS) and non-light of sight (NLoS) in the 5G NR frame structure with latter having double frame duration (0.1ms) compared to the former (0.05ms) to ensure latency requirements (0.1ms RTT in physical layer) \cite{levanen2014radio}. The authors in \cite{dutta2017frame} on the other hand, proposes an mmWave MAC layer flexible frame structure that supports different features including highly granular transmission times, dynamic location of control signals in the frame, extended messaging and efficient directional control signal multiplexing capability.

mmWave frequencies are also explored for various applications including vehicular and high speed railway communications with certain latency and reliability constraints. For example, \cite{yoshioka2016field} explores the beamforming and latency performance of 5G mmWave radio access in an outdoor environment. The experiment is performed up to 20 km/h vehicular speed condition with continuous LOS alignment. The paper also redesign the frame structure to enable URLLC services in the outdoor vehicular environment. \cite{noh2017mmwave} presents a real world implementation based mmWave mobile backhaul transceiver for high speed train communication systems. A hierarchical network architecture is used in the experiment in which a user accesses the backhaul using an onboard relay installed in the train. The system was implemented in the Seoul subway line 8 and found 1.25 Gbps download speed in most of the railway paths. High speed communication suffers from narrow beamforming and path loss. The problem is worse for high speed vehicles. To solve the problem optimal non-uniform mmWave beamforming is proposed in \cite{cui2018optimal}. The paper designs a URLLC system for signaling in safety-critical high speed railway signaling. In \cite{zhang2021design} a joint sensing and communication integrated system is proposed to support the dynamic frame structure configuration for both sensing and communication functions for 5G. The method can fulfill the low latency and high data rate requirements in raw sensing and data sharing among CAVs. All these works  demonstrated promising performance with significant boost in data rate achievement though much more works are needed to realise the full potential of mmWave for URLLC applications.

\subsection{Diversity Techniques}

Diversity techniques are widely considered for improving reliability of transmission as it combats frequency selective fading, co-channel and multi-user interference by exploiting different aspects of the transmission (e.g., time, frequency, space). Diversity techniques can exploit space, frequency, time, polarisation, user distribution and cooperation. %The Figure \ref{stf_diversity} demonstrates the concept of three major diversity techniques: frequency, time and space diversity.

%In the techniques space, frequency and time diversity are mostly investigated. The technique fundamentally uses redundant communication to increase the reliability. The recent techniques invented by researchers can be used to mitigate the reliability and latency requirements through the low complexity algorithms.

% \begin{figure*}[!t]
% \centering
% \includegraphics[scale=0.8]{Space_time_frequency_diversity.png}
% \caption{Space, time and frequency diversity for wireless network}
% \label{stf_diversity}
% \end{figure*}

\subsubsection{Space Diversity}

Space diversity exploits the fact that if transmitter or receiver antennas are separated in space then the signals will experience different fading/interference due to the fact that each propagation path is independent of the others.  To increase the reliability using the spatial domain, the authors in \cite{wu2017signal} proposed dynamic multiplexing of downlink channels to transmit eMBB and URLLC data on the same time/frequency resource. The paper uses a puncturing scheme with space diversity in which part of eMBB data is scheduled for URLLC data transmission. %To prevent the performance degradation due to the puncturing technique special diversity is used. 
Similar puncturing techniques are proposed for industrial wireless network systems in \cite{gashema2018spatial} which uses unlicensed bands for capacity enhancement and thereby ensures latency constraints. The work in \cite{gao2016improving} proposes a transmission scheme with cooperative automatic repeat request to support the URLLC which demonstrates significant improvement in link quality and hence the reliability of the transmission. Spatial diversity with a MIMO system can be a proment way to further enhance the reliability. In \cite{ozyurt2018performance} the authors incorporate signal space diversity (SSD) into the MIMO system for space diversity and analysed the performance with phase-shift keying under slow and flat Rayleigh fading channel with correlated receiver antenna. In \cite{panigrahi2017feasibility} the authors investigate the feasibility of using a large antenna array at the receiver to achieve the URLLC for up-link communication. Both coherent and non-coherent MIMO receivers are investigated in the paper. The non-coherent MIMO receiver is found to be a promising technique with reasonable SNR to achieve the URLLC. A probabilistic shaping scheme for MIMO systems with SSD is proposed in \cite{kang2022probabilistic}. The technique improves the shaping and diversity gain using optimal rotation angles of QAM signals.

 %the signal propagates through differIt is shown that using several links instead of a single links increase the reliability of the communication \cite{ohmann2014achieving}

\subsubsection{Time Diversity}
In time diversity technique multiple copies of the same packet are transmitted at different times to increase the reliability. The time difference between packet retransmission is greater than the channel coherence time. In \cite{peon2017applying} the authors propose space diversity based techniques to handle different time and safety critical heterogeneous traffic with transmission and retransmission time limit to increase latency and reliability. The method increases the reliability of the traffic under different interference scenarios. In \cite{kojima2015improved}, the authors propose time diversity based technique for helicopter-satellite communication where signal blockage at the rotor blade is a major problem. Later the authors enhance the performance of the communication system in \cite{sato2018accurate} with a novel channel estimation using first Fourier transform (FFT). Several papers, for example \cite{rao2018packet,centenaro2019system,michalopoulos2019data} use packet or data duplication to increase the reliability of the network. Different application scenarios are considered including stationary, and mobility (pedestrians, vehicular) in the studies.

%In the communication the blockage of received signal for rotor blades is an important problem. The method used the time diversity to overcome the periodic blockage of the received signal and achieved a remarkable improvement over conventional technique. 

\subsubsection{Frequency Diversity}

The frequency diversity exploits the fact that different frequencies will experience different and independent fading. Therefore,  transmission of same data packet at different frequencies will increase reliability as one may experience deep fading while the other may experience better channel condition leading to increased reliability. To achieve the frequency diversity gain, different strategies are investigated in literature. For example, in \cite{boyd2019non} two different models with coded multichannel random access schemes are proposed for URLLC with non-orthogonal contention-based access with frequency diversity. Contention-based random access with two different collision avoidance policies are investigated in the paper. The authors in \cite{senthil2019improved} presents a modified symmetric symbol repetition (MSSR) to reduce the inter-carrier interference (ICI). The proposed MSSR scheme uses frequency diversity to improve the carrier-to-interference ratio (CIR) by reducing the ICI. The authors in \cite{DBLP:journals/corr/SheYQ16a} investigate the effect of frequency diversity on reliability and the required bandwidth. A two-state transmission model is used in the study with finite blocklength channel codes. Joint time and frequency diversity is another way to improve the performance. For example, A joint time-frequency diversity technique is used in \cite{zhao2019joint} for grant-free uplink transmission. In the transmission scheme the packets are transmitted multiple times both on time and frequency domain to increase the reliability.
In \cite{wu2019urllc} both time and frequency diversities are combined to increase the performance of URLLC. The paper uses a stochastic theory based approach with two retransmission policies to improve the performance. In the literature some other diversity techniques could be found including interface diversity \cite{JNielsen2}, polarisation diversity \cite{desai2020wideband}, cooperative diversity \cite{kurma2022urllc}, and so on. However, the summary of different diversity techniques are presented in Table \ref{table:diversity_technique}.

\begin{table*}[!htp]
\caption{Summary of various diversity techniques} 
\centering % used for centering table
\begin{tabular}{|L{11mm}|L{10mm}|L{101mm}|} 
\hline %inserts double horizontal lines
Diversity technique & Reference & Key features \\
\hline %inserts single line

\multirow{6}{*}{Space}& \cite{wu2017signal} &

\begin{itemize} 
 \item Dynamically multiplex the eMBB and URLLC data on the same time\/frequency resource.
 \item A part of the radio resource of eMBB is scheduled dynamically for URLLC data.
 \end{itemize} \\
\cline{2-2} \cline{3-3}

 & \cite{gashema2018spatial} & 
 \begin{itemize} 
 \item Combine the unlicensed spectrum with license spectrum to increase the capacity of the network.
 \item Special diversity is used to achieve the reliability requirement of URLLC.
 \end{itemize}  \\
\cline{2-2} \cline{3-3}

 & \cite{gao2016improving} & 
 \begin{itemize} 
 \item Data transmission scheme based on special diversity with cooperated ARQ is used to achieve the reliability requirement of URLLC.
\end{itemize} \\
\cline{2-2} \cline{3-3}
 
 & \cite{ozyurt2018performance} & 
 \begin{itemize} 
 \item SSD is incorporated into the MIMO system.
 \item A realistic exponential correlated model is adopted. 
 \end{itemize} \\
\cline{2-2} \cline{3-3}

 & \cite{panigrahi2017feasibility} & 
 \begin{itemize} 
 \item Use of large antenna arrays at the receiver is investigated.
 \item Non-coherent MIMO receiver is found as a promising option to achieve the URLLC for uplink under acceptable SNR. 
 \end{itemize} \\

\hline

\multirow{6}{*} {Time} & \cite{peon2017applying} & 
 \begin{itemize} 
 \item Time diversity is used in the MAC protocol to increase the reliability.
 \item The protocol uses the time diversity to handle heterogeneous time critical traffic. 
 \end{itemize} \\
\cline{2-2} \cline{3-3}

& \cite{kojima2015improved} & 
\begin{itemize} 
 \item The time diversity is used to remove the periodic blockage problem in helicopter satellite communication systems due to rotor blades.
 \end{itemize} \\
\cline{2-2} \cline{3-3}

 & \cite{sato2018accurate} & 
 \begin{itemize} 
 \item A novel channel blockage to predict the blockage information with FFT and inverse FFT.
 \end{itemize} \\
\cline{2-2} \cline{3-3}
 
 & \cite{rao2018packet,centenaro2019system,michalopoulos2019data} & 
 \begin{itemize} 
 \item The packet is duplicated over time.
 \end{itemize} \\

\hline

\multirow{6}{*} {Frequency} & \cite{boyd2019non} & 
\begin{itemize} 
 \item Coded multichannel random access scheme is used for uplink URLLC data transmission.
 \item Frequency diversity is used in the random access for packet repetition.
 \end{itemize} \\
\cline{2-2} \cline{3-3}

%\hline %inserts single line
& \cite{senthil2019improved} & 
\begin{itemize} 
\item MSSR exploits the Frequency diversity to achieve a better bit error rate.
\item MSSR scheme improves CIR by using ICI.
\end{itemize} \\
\cline{2-2} \cline{3-3}

& \cite{DBLP:journals/corr/SheYQ16a}  & 
\begin{itemize} 
\item Frequency diversity is used to achieve the reliability in uplink transmission with massive type devices in tactile internet.
\item Two-state transmission model with finite blocklength channel code is used in the system.
\end{itemize} \\
\cline{2-2} \cline{3-3}

& \cite{zhao2019joint} & 
\begin{itemize} 
\item Joint time-frequency based grant-free uplink transmission scheme is designed.
\item Packets are duplicated multiple times both on time and frequency domain to increase the reliability.
\end{itemize} \\
\hline

\end{tabular}
\label{table:diversity_technique} % is used to refer this table in the text
\end{table*}

%=============
%Authors in \cite{CXiao_JSAC19} developed an analytical model for ensuring latency target while also maintaining reliability requirement.

%=============================%
%         SUB-SECTION         %
%=============================%

\section{MAC Layer Design for URLLC}\label{sec:MACLayerDesign}

\subsection{Scheduling and Resource Allocation}
MAC layer design for URLLC predominantly focuses on the scheduling techniques of resources among different URLLC users. The scheduler assigns time-frequency resources to user equipment and is responsible for deciding how the uplink and downlink channels are used by the base station and UEs of a cell. The scheduler is responsible to assure the QoS of different services of the 5G network. However, this subsection investigates different scheduling and resource allocation schemes of 5G URLLC packets.

In \cite{ADestounis} the authors proposed a scheduling technique to achieve latency with reliability guarantee. Multiple unreliable transmissions are combined to achieve the reliability. The scheduling problem is converted to the URLLC service level agreement (SLA) satisfaction (USS) problem. For a small number of active users the problem is solved using the dynamic programming technique and for a large number of active users the problem is NP-hard. A knapsack-inspired technique is adopted to solve the NP-hard problem.

A joint scheduling of URLLC and eMBB traffic is proposed in \cite{anand2020jointJournal}. In the technique time is divided into slots with 1 ms duration. Each slot is further divided into 8 mini slots. The eMBB traffic is scheduled at slot boundaries and due to time latency requirement, the URLLC traffic is scheduled at minislot boundaries. The joint scheduling problem is formulated as a joint optimisation problem. Maximum utilisation of resources is required for eMBB with minimum impedance by URLLC while immediately satisfying the URLLC demand. With these constraints the problem is solved with three different models of superposition/puncturing the eMBB traffic.

\begin{table*}[!htp]
\caption{Summary of Scheduling Algorithms}

\centering % used for centering table
\begin{tabular}{|L{0.8cm}|L{11cm}|c|c|c|c|} 
\hline %inserts double horizontal lines
Ref. No. & Features & uplink & downlink & eMBB & URLLC \\
\hline %inserts single line
\cite{ADestounis} 
& 
%\begin{minipage} [t] {0.4\textwidth}
\begin{itemize} 
 \item Multiple retransmission policy is adopted to achieve reliability.
 \item The scheduling problem is converted to a USS problem.
 \item	Dynamic programming techniques are adopted to solve the problem for small numbers of active users and for large numbers a heuristic technique is applied.
\end{itemize} & & \checkmark & &  \checkmark \\
%\end{minipage}
\hline %inserts single line

\cite{anand2020jointJournal}
&
%\begin{minipage} [t] {0.4\textwidth}
\begin{itemize} 
 \item Joint scheduling of URLLC and eMBB traffic is proposed.
 \item The eMBB traffic is scheduled at slot boundaries with 1 ms duration and URLLC traffic is scheduled immediately at minislot boundaries with 0.125 ms duration.
 \item The joint scheduling problem is solved with three different modules: linear, convex and threshold models.
\end{itemize} & \checkmark & \checkmark & \checkmark  & \checkmark \\
%\end{minipage} \\
\hline %inserts single line

\cite{Abreu1}
&
%\begin{minipage} [t] {0.4\textwidth}
\begin{itemize} 
 \item A group of users pre-schedule resources for retransmission.
 \item An alternative retransmission scheme to provide the HARQ retransmission capability for URLLC.
 \item Reduces the control signal transmission and latency, and increases the reliability.
\end{itemize} & \checkmark & \checkmark & &  \checkmark \\
%\end{minipage} \\
\hline %inserts single line
 
\cite{GPocovi} 
&
%\begin{minipage} [t] {0.4\textwidth}
\begin{itemize} 
 \item Allocates resources for URLLC and eMBB traffic on a shared channel.
 \item Dynamically adjusts the block error probability of URLLC traffic.
 \item Reduces the inter-cell interference in the scheduling process. 
\end{itemize} & & \checkmark & \checkmark &  \checkmark \\
%\end{minipage} \\
\hline %inserts single line

\cite{esswie2018opportunistic} 
&
%\begin{minipage} [t] {0.4\textwidth}
\begin{itemize} 
 \item The joint scheduling algorithm that provides quality URLLC and eMBB services.
 \item Maximum possible eMBB ergodic capacity is achieved with the guaranteed URLLC service.
 \item In a particular time if the existing radio resources are not sufficient for URLLC then the NSBPS scheduler forcibly schedules the traffic in the next eMBB transmission. 
\end{itemize} & & \checkmark & \checkmark &  \checkmark \\
%\end{minipage} \\
\hline %inserts single line

\cite{she2017radio} 
&
%\begin{minipage} [t] {0.4\textwidth}
\begin{itemize} 
 \item Identifies the appropriate tools to allocate the radio resource under the delay, reliability and availability constraints.
 \item Optimised resource allocation tool is proposed to allocate the radio resource under the constraints.
 \item	Open problems for radio resource management in URLLC are discussed.
\end{itemize} & \checkmark & \checkmark & &  \checkmark \\
%\end{minipage} \\
\hline %inserts single line

\cite{shariatmadari2016optimized} 
&
%\begin{minipage} [t] {0.4\textwidth}
\begin{itemize} 
 \item Fixed and adaptive resource allocation strategies are proposed for URLLC considering the number of transmission attempts.
 \item Optimisation problem is formalised and sub-optimal allocation is proposed for adaptive allocation.
 \item The sub-optimal allocation reduces the computational complexity with negligible difference from the optimal solution.
\end{itemize} & \checkmark & \checkmark & &  \checkmark \\ 
%\end{minipage} \\

\hline %inserts single line

\cite{sun2017energy} 
&
%\begin{minipage} [t] {0.4\textwidth}
\begin{itemize} 
 \item An energy efficient resource allocation technique is proposed.
 \item The transmit power, bandwidth and number of active antennas are jointly optimised to maximise the system energy efficiency.
 \item The optimal problem is non-convex hence an approximation is used to achieve the global solution. 
\end{itemize} & \checkmark & \checkmark & &  \checkmark \\
%\end{minipage} \\
\hline %inserts single line

\cite{hytonen2017coordinated} 
&
%\begin{minipage} [t] {0.4\textwidth}
\begin{itemize} 
 \item Coordination based multi-cell resource allocation methods for URLLC.
 \item The technique shows the superiority in three different scenarios including single-frequency network, narrowband muting and macro-diversity with soft combining. 
 \item A powerful technique to increase the reliability of URLLC packet transmission without violating the latency constraint. 
\end{itemize} & \checkmark & & &  \checkmark \\
%\end{minipage} \\
\hline %inserts single line

\cite{anand2018resource} 
&
%\begin{minipage} [t] {0.4\textwidth}
\begin{itemize} 
 \item Resource allocation and HARQ optimisation technique is proposed URLLC traffic.
 \item Queuing network model is used to characterise the design choice on the maximum URLLC traffic support of the network. 
 \item Based on the analysis the required level of system parameters is computed.   
 \end{itemize} & \checkmark & \checkmark & & \checkmark \\
%\end{minipage} \\
\hline %inserts single line

\cite{AAzari19} 
&
%\begin{minipage} [t] {0.4\textwidth}
\begin{itemize} 
 \item Resource allocation technique for non-scheduled URLLC packet along with the scheduled URLLC transmission.
 \item Distributed hierarchical ML concept is used to solve the problem.
 \item A hybrid OMA/NOMA is proposed to allocate the resource.
\end{itemize} & \checkmark & & & \checkmark \\
%\end{minipage} \\

\hline %inserts single line
\end{tabular}
\label{table:scheduling} % is used to refer this table in the text
\end{table*}

In \cite{Abreu1} a pre-scheduling mechanism is proposed for retransmission of packets to reduce the control signaling overhead. In the technique a group of users shares pre-schedule resources to reduce retransmission. This reduces the control signal transmission and latency, and increases the reliability of the network. An important feature of the technique is that it does not lose excessive capacity if the retransmission fails.

In \cite{GPocovi} a joint link adaptation and scheduling scheme is proposed for both URLLC and eMBB traffic on a shared channel. The policy dynamically adjusts the block error probability of URLLC traffic according to the instantaneous experienced load per cell. The dynamic resource allocation technique is simple but effective as it provides the efficient distribution of resources for the two services.

In \cite{esswie2018opportunistic} a null-space-based preemptive schedule (NSBPS) algorithm is proposed for eMBB and URLLC traffic in densely populated 5G networks. The scheduler dynamically cross optimised a joint optimisation problem where the QoS for URLLC traffic is guaranteed. If the current resource for URLLC traffic does not satisfy the requirements of URLLC the NSBPS starts working. The NSBPS scheduler forcibly reschedules the URLLC traffic in the ongoing eMBB transmission.

In \cite{she2017radio} a radio resource management technique is proposed considering the queuing delay violation probability for URLLC service in the 5G network. The packet delay and loss components in URLLC and network availability are analysed to ensure the quality of service to users. An optimised resource allocation scheme is proposed to allocate the radio resource under the delay, reliability and availability constraints. The paper concludes with major challenges in radio resource management and open problems in radio resource management in URLLC.

A resource allocation policy with optimum number of transmission is proposed for URLLC in \cite{shariatmadari2016optimized}. Two allocation schemes are introduced in the paper including fixed and adaptive allocation. In fixed allocation the maximum number allowed transmission attempts is set before the actual transmission. In adaptive allocation the number of transmission attempts are variable and set during the data transmission. To find the number of transmission attempts an optimisation problem is formulated and a sub-optimal solution is provided to decrease the computational complexity with a negligible performance degradation compared to the optimal solution.

An energy efficient resource allocation technique is proposed in \cite{sun2017energy} without compromising the resource usage efficiency in terms of delay and overall packet loss. The delay components include queuing delay and transmission delay. The packet loss includes the queuing delay validation, transmission error for finite block length channel coding and proactive packet dropping in deep fading channel condition. Under these objectives an optimisation problem is formalised to jointly optimises transmit power, bandwidth and number of active antennas. A global optimal solution is approximated for the problem.

A coordination based multi-cell resource allocation method for URLLC is presented in \cite{hytonen2017coordinated} where the neighboring cells coordinate to increase the reliability. Three coordinated allocation methods namely single-frequency network, narrowband muting and macro-diversity with soft combining are presented for indoor environments. The analysis shows that the inter-cell coordination is a powerful technique to increase the reliability of URLLC packet transmission with the latency constraint.

A resource allocation and hybrid automatic repeat request (HARQ) optimisation technique is proposed in \cite{anand2018resource}. Based on the queuing system model the paper investigates the impacts of several design choices. The system can support the maximum URLLC load including system parameter, resource allocation, and HARQ schemes. Using the analysis the minimum required system bandwidth to support a given URLLC load is computed, optimal OFDM allocation strategy determined and different parameters of the HARQ scheme are optimised.

A resource allocation technique for non-scheduled URLLC traffic with the co-existence of scheduled traffic is proposed in \cite{AAzari19}. Machine learning (ML) is used to exploit the special/temporal correlation of the user behavior and use of radio resources, and allocate the resource for the non-scheduled traffic. A distributed risk-aware ML solution is proposed to allocate the non-scheduled URLLC traffic. 
 
A robust URLLC packet scheduling for OFDM system is proposed in \cite{cheng2020robust}. An optimisation technique is used to minimise the PRB assignment and power allocation under the required delay and reliability constraints. A low complexity successive convex approximation based method is used to solve the problem that yields the sub-optimal solution. Although the technique shows efficient performance in a simplified environment, a realistic scenario with a multi-cell environment is required to evaluate the performance study.

An uplink resource scheduling technique for Smart Grid (SG) Neighborhood Area Network (SGNAN) is proposed in \cite{zhu2019priority}. SGNAN consists of intelligent household electrical appliances and the home gateway that collects data and sends it to the data concentrator unit of NAN. Different delay, bandwidth and reliability are required for different home area networks including residential, business and industry. To meet the different quality demands the paper proposes the priority based URLLC scheduling algorithm for the SG. A similar work of resource scheduling in heterogeneous cellular networks for SG is found in \cite{wu2020uplink}. The optimisation technique is used to maximise the system throughput and first-order Taylor expansion is used to approximate the solution. However, different latencies  are required for different networks of the SG. Hence the results with variable latency are required in the study.

A joint time-slot scheduling, sub-band scheduling and power allocation (JTSP) scheme for 6G terahertz mesh network is proposed in \cite{yu2020joint}. A mixed integer programming problem is formulated in the paper. A sub-optimal Greedy Shrinking Algorithms (GSA) is proposed in the paper. The GSA reduces the computational complexity of the optimisation problem.

\subsection{Multi-Connectivity}
\label{sub:MC}

%{\color{red} FT}
Employing more than one connection to improve reliability has been considered as the promising technology. For example, in 3GPP, multi-connectivity (MC) is proposed mainly to improve communication reliability. As the name indicates, more than one communication link is created to convey the same message reliably which exploits diversity of the channel. Multiple links ensures that if one link fails to meet the QoS requirement due to adverse conditions in the link path, one of the other links may still experience good channel condition and therefore will be able to meet the QoS requirement. There are several ways of realising MC including frequency and spatial diversity, career aggregation (CA) and dual connectivity (DC) \cite{Awolf}.

Multi-connectivity enhances the performance of 5G networks for URLLC service in terms of outage probability as stated in \cite{mahmood01}. Significant improvement can be achieved in MC compared to the single-connectivity network with the cost of almost doubling the used radio resources. The paper shows that with a given block error rate, the multi-connectivity network improves more than an order of magnitude of outage probability over the single-connectivity network.

In \cite{THobler} the authors introduce the concept of mission reliability and utilise the mean time to first failure (MTTFF) metric in wireless networks. The mission reliability is defined as the probability that a device can successfully transmit a message within a very small amount of time. The MTTFF metric is the expected mission duration without any failure. The matrices are utilised in a multi-connectivity system with Rayleigh fading channel and demonstrated the trade-off among mission duration, mission reliability and the number of links used. The paper also discusses how the investigations can help to design 5G and future networks with URLLC service.

In \cite{nhmahmood1} the authors present a conceptual multi-connectivity model and then a novel admission mechanism is proposed to control the number of multi-connectivity users at a time. The technique is evaluated with a system level simulation and 23\% latency reduction and 75\% reliability improvement are recorded over a single connectivity network.

In \cite{CShe} the authors define the network availability and then establish a framework to maximise the available range. The network availability is defined as the probability that the reliability and the latency requirement that can be satisfied in a wireless network. The available range is the maximal communication distance that satisfies the network availability requirement. The framework achieves the availability requirement by exploiting the multi-connectivity. The paper uses the device-to-device (D2D) and cellular links to transmit packets.

In \cite{THolber2} the authors investigate two performance metrics namely availability and reliability metrics. Time based reliability matrices such as mean up time, mean down time or mean time to failure are the most common metrics in reliability theory that are used in the paper to jointly analyse the availability and reliability of URLLC. A multi-connectivity system with Rayleigh fading channel is used in the investigation. 

A downlink transmission power optimisation method with multiple transmissions of the same reduced-sised packet is proposed in \cite{xue2024cooperativeDeep} in multi-connectivity scenario for vehicular networks. Replicas of each packet are transmitted over multiple links. The reliability of the transmission is ensured as long as the transmission over at least one link is successful. Taking into account the CSI, the transmit power is allocated cooperatively among multiple links using deep reinforcement learning.

\subsection{Scheduling Approach for URLLC}

%Enhancing the LTE, especially LTE MAC protocol is the first consideration of researchers to achieve the 5G requirements. 
In this section we discuss the MAC scheduling approaches that are adopted in various literature.  Different scheduling algorithms emphasise different parameters including maximising throughput, increasing fairness among users to satisfy the QoS requirements. %\cite{capozzi2013downlink} investigates the downlink related works and \cite{abu2013uplink} investigates the uplink related works in the literature. 
Moreover, the schedulers can be categorised into channel-aware, channel-unaware, QoS-aware, QoS-unaware and QoS-aware/channel-aware scheduler. For URLLC QoS and channel aware scheduler are required. The QoS-aware ensures the latency requirement and the channel-aware scheduler increases the reliability of the protocol.

%Blind equal throughput (BET), maximum throughput (MT) and Proportional fair (PF) are the fundamental scheduling algorithms in LTE that are QoS-unaware. BET ensures equal exponential weighted moving average throughput and MT ensures the maximum throughput. PF scheduler combines BET and MT that maximises total throughput providing the fairness among the UE. These schedulers ignore the deadline of the message and hence are not applicable for URLLC. The earliest deadline first (EDF) and largest weighted delay first (LWDF) are the schedulers that consider the delay of messages, which are QoS-unaware. EDF considers the deadline of the packet in which the packet with the earliest deadline is scheduled at first. The LWDS on the other hand considers the weighted average delay in the queue and schedules the packets at first with the largest weighted average delay.

There are several schedulers that maximise the channel utilisation with guaranteed delay constraints. The schedulers that are proposed in LTE also suitable for URLLC with this characteristics includes modified LWDF (M-LWDF) \cite{andrews2001providing}, exponential/proportional fair (EXP/PF) \cite{basukala2009performance} and the frame level scheduler (FLS) \cite{piro2011two} which are QoS-aware/channel-aware schedulers. M-LWDF combines LWDF and PF to improve the priority of packets that approach close to the deadline. But the modified protocol does not reflect the packet that approaches the deadline limit exceed since the modification increases the priority linearly over time.  The limitation of M-LWDF is eliminated in EXP/PF by introducing the exponential priority function. Even when the exponential function is used the scheduler does not completely reflect the priority for the packets of imminent deadline that is required for URLLC.  The FLS at first determines the resource required within a 10 ms frame to fulfill the delay requirement of packets. Then the resources are allocated according to the PF algorithm within the frame time. The additional complexity of resource estimation and 10 ms frame time consideration are not completely suitable for the delay requirement of URLLC. The work in \cite{MIHossain19} proposes an efficient algorithm that exploits channel condition and QoS requirement. The work demonstrates superior performance compared to both the M-LWDF and EXP/PF schemes.

The control signal is a source of error which is reduced with semi-persistent scheduling (SPS). To support periodic traffic with tight delay requirements the SPS was introduced in LTE for VoIP services. To avoid excessive control overhead caused by multiple assignment/grant messages the SPS pre-schedule the traffic with certain periodicity. Several research works could be found on SPS, for example in \cite{amjad2018latency} a latency reduction technique is proposed to achieve the latency requirement of URLLC for uplink communication of LTE. The paper proposes the parameter adjustment technique to reduce the latency for narrowband LTE. Another work on narrowband LTE scheduling for latency could be found in \cite{arnjad2018latency}. The paper studies different scheduling techniques including dynamic, semi-persistent scheduling for uplink communication. In \cite{IAshraf2016ETFA} a realistic factory environment is simulated to analyse the performance of different radio-interface design for both LTE and 5G. The factory environment is defined as the realistic scenario with a dedicated wireless environment. The paper shows that for the fixed factory environment the LTE system can fulfill the requirements of URLLC. A comprehensive coverage of URLLC scheduling algorithm could be found in \cite{haque2023survey}.

A scheduling method for heterogeneous traffic is proposed in \cite{zhao2024jointBeam}, where ISAC is employed for sensing an event, e.g., a vehicle, and aperiodic traffic is triggered in addition to the regular traffic to a user. A joint beamforming and scheduling is designed by exploiting Markov decision process to maximise the successful transmission of both types of traffic satisfying the requirements of URLLC.

\section{Cross-Layer Approaches for URLLC}\label{sec:crossLayer}

Cross-layer design approaches for URLLC are investigated in this section.  In this approach, data from two or more layers are accessed to design the protocol. In most of the literature, PHY and MAC layer data are used but link-layer and network-layer data are also used in some papers. Table \ref{table:cross_layer_summary} summarise the cross-layer techniques used to achieve the URLLC requirements.

\subsection{ARQ/HARQ}

ARQ and HARQ are the important mechanisms that balances the spectrum efficiency and reliability in wireless networks. ARQ uses the cyclic redundancy check (CRC) to detect the error and initiates retransmission if the error occurs. HARQ uses forward error correction coding to resolve an error and initiates retransmission if the error cannot be corrected. The HARQ increases the reliability but retransmission within 1 ms is not possible for time varying channels. With these limitations, the ARQ/ARQ cannot support the URLLC requirements. Cooperative version of ARQ increases the reliability under severe fading conditions. The performance of the cooperative ARQ with short retransmission delay in terms of reliability is analysed in \cite{serror2015channel}. The study found the sharp reduction of block error rate (BLER). The use of CSI in HARQ improves the performance of the protocol. \cite{shariatmadari2015analysis} presents the performance of HARQ with the 
impact of CSI feedback accuracy. The paper concludes that only one retransmission is possible to ensure the low latency of URLLC and precise CSI feedback can improve the reliability of the network.

The idea of cooperative ARQ comes from the overhearing feature due to the broadcast nature of wireless transmission. A number of secondary nodes receive the transmit message. In cooperative ARQ, secondary receivers retransmit the message in a cooperative manner if the primary receiver fails to receive the message. Several works could be found that incorporate the cooperative technique. For example, in \cite{hwang2017adaptive} a cross-layer design of joint adaptive modulation and coding (AMC) and cooperative ARQ is proposed in cognitive radio systems to improve the performance. The scheme maximises the spectral efficiency (SE) while maintaining the average packet loss with a predefined level. An optimisation problem is formed to achieve the SE and a technique is proposed to obtain the suboptimal solution. The minimum packet error rate (PER) for the secondary systems is obtained and the corresponding suboptimal AMC policy is derived to achieve the PER. A cross-layer framework is proposed in \cite{chiu2018cross} to design and performance evaluation of cooperative ARQ in mobile networks with opportunistic multi-point relaying (OMPR). The paper studied three different OMPR models and their combinations are studied. The study shows that using two active relay stations for opportunistic distributed space-time coding is a robust and effective choice to provide near optimal system throughput for vehicular systems that maintain typical speed with less energy consumption and lower protocol time complexity.

In \cite{serror2016performance} the authors propose cooperative ARQ to achieve reliability and latency constraints for mission-critical communication. The cooperative ARQ is analytically studied for multi-user transmission patterns with time division multiple access (TDMA). The technique is applied in two different scenarios including centralised and distributed relays and shows that both can achieve the target criteria of mission-critical application. A node-cooperative ARQ is proposed in \cite{dianati2006node} for wireless ad hoc networks. The paper shows that the cooperation among the small number of nodes increases the performance significantly in terms of throughput, average delay and delay jitter. Both the analytical model and simulation results are used to show the effectiveness of the scheme. A cooperative ARQ protocol with a source, a destination and a relay is analysed in \cite{yu2006cooperative}. The paper proposes three different ARQ schemes and shows the effectiveness of the protocol compared to the existing cooperative ARQ protocols.

Several works could be found on cross-layer design with HARQ. For example, in \cite{selmi2016efficient} a cross-layer design is proposed to optimise the throughput for MIMO wireless systems. The technique combines power control, AMC and MIMO technology at the physical layer with ARQ at the data-link layer. The paper implements the system for both ARQ and HARQ and found the higher throughput with the HARQ. In \cite{chen2018cross} the authors propose a cross-layer protocol that combined HARQ and the turbo code to improve peak-signal-to-noise ratio. The study shows that the proposed technique outperforms the existing with maximum two retransmission.

The above modifications on HARQ do not satisfy the requirements of URLLC. Hence a number of modifications are proposed for URLLC services. For example, in \cite{avranas2019throughput} the authors propose incremental redundancy HARQ (IR-HARQ) for short packet communication. An optimisation problem is formalised to tune the IR-HARQ mechanism for throughput maximisation. A dynamic programming algorithm is proposed to solve the problem and shows the optimality with numerical results. In \cite{deghel2018joint} a joint link adaptation and HARQ scheme is proposed for URLLC. An optimisation technique is used to find the maximal number of HARQ transmission and compute optimal modulation and coding scheme (MCS) levels to maximise the spectral efficiency. The channel diversity and flexibility is improved in the scheme. In \cite{imamura2017low} a HARQ technique is proposed that exploits the channel decoding that is mainly responsible for high latency. The technique uses the CSI in the decoding process. The method uses the early retransmission policy before the channel decoding process is completed.

\begin{table*}[!htp]
\caption{Summary of Cross-layer Technique for URLLC} 
\centering % used for centering table
\begin{tabular}{|L{1.9cm}|L{1.5cm}|L{1.0cm}|L{11.4cm}|} 
\hline\hline %inserts double horizontal lines
Topics & Basic & References & Key features \\
& technology & & \\
\hline %inserts single line

\multirow{6}{*}{ARQ/HARQ} & Cooperative ARQ & \cite{hwang2017adaptive,chiu2018cross,serror2016performance,dianati2006node,yu2006cooperative} &

\begin{itemize} 
 \item Cooperation is initiated with a relay node.
 \item Relay node retransmit the message.
 \item Relay can be centralised or distributed.
 \item Improve reliability and latency to achieve URLLC.
 \end{itemize} \\
\cline{2-2} \cline{3-3} \cline{4-4}

 & Mostly optimisation of parameters & \cite{selmi2016efficient,chen2018cross,avranas2019throughput,deghel2018joint,imamura2017low} &
 \begin{itemize} 
 \item Optimised transmit power level and appropriate MCS are jointly selected.
 \item Turbo code is combined with HARQ and the maximum number of allowed retransmissions is selected.
 \item Optimised the parameters of IR-HARQ including no. of information bits, no. of transmission rounds and blocklength-power allocation.
 \item CSI is used in the channel decoding process.
 \item Reduced latency and increased reliability. 
\end{itemize}  \\

\hline

\multirow{6}{*} {and retransmission} Transmission & Mostly parameter optimised transmission & \cite{she2016cross,she1,DBLP:journals/corr/SheYQ16a,devassy2019reliable} &
 \begin{itemize} 
 \item Packet transmission is adopted with a packet dropping policy.
 \item Packet dropping is based on optimisation of different quantities including packet error probability, queue delay violation probability and packet dropping probability.
 \item Each device transmits packets through multiple subchannels with a small error probability if the channel gain exceeds a threshold.
 \item The no. of subchannels, bandwidth of each subchannels and the device threshold is optimised.
 \end{itemize} \\
\cline{2-2} \cline{3-3} \cline{4-4}

& Mostly optimisation to find no. of retransmission & \cite{CSun,avranas2018energy,gao2016improving,serror2015channel} &
\begin{itemize} 
 \item Frequency hopping retransmission to achieve stringent QoS requirements of URLLC.
 \item Optimised retransmission with resource allocation.
 \item Dynamic programming based energy minimisation with optimised retransmission, blocklength and power per round.
 \item Cooperative retransmission is used to maintain the strict latency requirement.
 \end{itemize} \\

\hline
Resource allocation & Scheduling with resource allocation with optimisation &  \cite{elayoubi2019radio,vora2018effective,zhang2019statistical,femenias2017downlink} & 
\begin{itemize}
\item Resource allocation policy is proposed for periodic and sporadic traffic. 
\item Downlink resource allocation is proposed that incorporates eMBB, URLLC and eMTC traffic.
\item Energy efficient optimised resource allocation for URLLC.
\item The resource allocation policy incorporates the data-link control layer and physical layer information.
\end{itemize}
\\
\hline
\end{tabular}
\label{table:cross_layer_summary} % is used to refer this table in the text
\end{table*}

\subsection{Transmission and Retransmission Policy}

Transmission and retransmission policies affect the performance of URLLC significantly. If a packet transmission fails, the retransmission is the only option to achieve the ultra-high reliability. However the retransmission must be completed within the required ultra-low (e.g., 1 ms) latency constraints. Hence, a transmission and retransmission policy is required to mitigate the stringent QoS requirements.

\subsubsection{Transmission Policy}

A cross-layer optimised transmission policy is proposed in \cite{she2016cross} to ensure ultra-low E2E delay and ultra-high reliability. The paper proves that required transmit power is unbounded to achieve the requirements of URLLC when the allowed maximal queue delay is shorter than the channel coherence time. To achieve the requirement with finite power a proactive packet dropping policy is adopted in the paper. The transmission policy is proposed with optimised packet error probability, queuing delay violation probability and packet dropping probability. \cite{she1} proposes transmission policy with optimised packet dropping, power allocation and bandwidth allocation policy to minimise the transmit power with stringent reliability and delay constraints. The solution depends on both CSI and queue state information.

%In \cite{DBLP:journals/corr/SheYQ16a} an uplink transmission policy is proposed with massive machine type devices in tactile internet. Based on channel conditions a two state transmission policy is adopted in the paper. If the channel gain exceeds a predefined threshold then the packet will be transmitted with a small error probability; otherwise, there will be a packet loss. To achieve reliability multiple subchannels are assigned to each active device.  The paper proposes an optimisation technique to find the optimal number of subchannels, bandwidth of each subchannel and transmission threshold for each device to ensure the reliability. 

A reliable transmission of a short packet for URLLC is proposed in \cite{devassy2019reliable}. The paper evaluates the delay violation and peak age violation probabilities. The peak age of information is a performance metric which indicates the maximum elapsed time since the last received update at the destination. The paper accounts the undetected errors for different computation that are harmful for mission critical communications.

\subsubsection{Retransmission Policy}

Retransmission is the most common way of increasing and maintaining the ultra-high reliability requirement despite the fact that it reduces spectrum efficiency. A retransmission policy is proposed in \cite{CSun} that uses the frequency hopping technique for retransmission to improve the reliability in deep fading condition. The paper investigates how to satisfy the QoS requirement with E2E delay components including transmission delay, queuing delay, and backhaul latency. Additionally the packet loss component including transmission error probability and E2E delay violation probability are also investigated. All these parameters are used to find the optimised resource allocation for retransmission. In \cite{avranas2018energy}, a tradeoff between latency and consumed energy is studied for finite blocklength constraints and feedback delay.

A transmission and retransmission policy is proposed in the factory environment for M2M communication in \cite{gao2016improving} that efficiently increases the reliability to support the URLLC. Spatial diversity is adopted in the paper and uses cooperative ARQ to fulfill the reliability requirement. In the paper, the base station only initiates the retransmission that is mounted in the factory environment. A centralised TDMA system with time diversity and cooperative ARQ protocol is proposed in \cite{serror2015channel} to achieve the reliability and latency of URLLC. In the protocol if the transmission is failed then the cooperative retransmission is started by a node that receives the packet and successfully decodes the packet.

\subsection{Cross Layer Resource Allocation}

In \cite{elayoubi2019radio}, a cross-layer resource allocation scheme is proposed for two different traffic patterns namely periodic and sporadic. For periodic traffic, sufficient radio resources are reserved for packet transmission and some resources are reserved for packet retransmission of the packets that are lost due to bad channel condition. The resources for retransmission are reserved considering the fact that some packets may be lost due to bad radio conditions. For sporadic traffic a scheme is proposed that combines grant-free contention-based scheme with packet repetition. In order to increase the reliability the resources are allocated to each replica based on the contention.

In \cite{vora2018effective} a cross-layer downlink scheduling and resource allocation scheme is proposed for 5G network. The paper adopts a dynamic programming technique based on optimum allocation at the base station to support diverse 5G use-cases. The work then extended for the joint allocation of MTC, URLLC and eMBB traffic in the network.

In \cite{zhang2019statistical} a cross-layer energy efficient optimised scheme is proposed for URLLC. The finite block-length coding (FBC) based technique guarantees the statistical delay bound QoS requirements of URLLC. The paper formulates an effective energy efficiency optimisation problem with the proposed FBC scheme. Then a solution is proposed to determine the optimal average power allocation for the finite block-length.

A downlink cross-layer resource allocation policy is presented in \cite{femenias2017downlink} that incorporates the data-link control layer and physical layer information. Based on the comprehensive comparison of OFDM and FBMC/OQAM modulation, the paper presents the framework for 5G MIMO-multicarrier that is able to embody the particularities of the two most promising wave modulation techniques for 5G networks.

\subsection{Security aspects of URLLC}
URLLC design is inherently vulnerable to security attacks. There are two aspects which are particularly challenging. Firstly, to increase reliability, packets are repeatedly transmitted so that if one packet is lost, one of the subsequent transmissions will reach the destination successfully. The number of repetitions depends on the channel condition. Now if unauthorised users have tried to receive the packet, repeated transmission increases its probability of packet capture and cryptographic key compromise.

Secondly, to improve latency, the packet sizes are kept deliberately short to minimise transmission time and queuing delay. The short packet size makes it impossible to include a stronger cryptographic key which increases the vulnerability of security attacks.

%Security is an important aspect of URLLC as it is vulnerable to eves-dropper due to short packet size which restricts use of redundant bits for enhancing security.  Subcarrier selection strategy for enhancing physical layer security for URLLC system \cite{JHamamreh}.

\subsubsection{Physical Layer Security}

Physical layer security is widely considered as an effective solution to eavesdropper attack on wireless transmission for example, CP-less OFDM for PLS \cite{JHamamreh1}, \cite{HRenTCom2020}. In \cite{pradhan2023blocklength}, to incorporate the physical layer security, the intercept probability of the eavesdroppers is amalgamated into the resource allocation problem to maximise the secrecy throughput and minimising the average power allocation per device. In \cite{xiang2020secure}, a QoS-sensitive user is paired with a security-required user. In this scenario, a HARQ scheme is designed for a cognitive NOMA system. An inverse relationship is observed between connection outage probability and secrecy outage probability. Additionally, the secrecy efficiency tradeoff improves with the increase of the maximum number of retransmissions. Security and reliability are ensured in \cite{dizdar2022rate} by enabling joint communication for secondary users and jamming of adversarial users for multi-antenna, multicarrier cognitive radio networks. RSMA in this scenario outperforms both NOMA and SDMA. However, the achieved secrecy comes at the expense of the overall sum-rate. Pilot contamination is a common threat to a grant-free transmission system, as the user identification and channel estimation are done from the position of the pilot subcarriers. In \cite{xu2021quantum}, a quantum learning based pilot econding scheme is proposed by a specially designed subcarrier activation pattern. This learning network enables early detection of unauthenticated codewords regarding the pilot subcarriers contaminated by the attacker. It ensures the fulfillment of  URLLC requirements even after pilot contamination.

\subsubsection{Network Slicing}
Network slicing segments the network resources among various services, thereby isolates each slice from another. This ensures better security as the users of one slice cannot access the resource of another network slice. Due to the shorter packet size constraint, higher security mechanisms cannot be applied in URLLC.  Thus, the network slicing technique can be a good solution for ensuring security in URLLC \cite{dos2021rate, liu2023networkslicing, yang2023networkslicing}. In \cite{dos2021rate}, network slicing between URLLC and eMBB users with rate splitting multiple access (RSMA) for URLLC transmission is proposed. RSMA is found to outperform orthogonal multiple access (OMA) when URLLC users' channel condition is better than that of the eMBB users. In a similar scenario, \cite{liu2018massive} investigates the situation where eMBB users experience better channel conditions compared to URLLC users. In this case, RSMA performs better than NOMA. Yang et al. \cite{yang2023networkslicing} investigate the the resource division for burst URLLC transmission and derive the minimum bandwidth requirement for ensuring low packet blocking. 

\subsubsection{Lightweight Encryption}
The short packet length in URLLC makes conventional strong cryptographic algorithms unsuitable for this application. For this reason, lightweight but effective cryptographic algorithms are needed. Limited efforts are devoted to develop suitable encryption mechanisms for URLLC. In \cite{seok2019secure}, combining authenticated encryption with associated data and elliptic curved cryptography, a lightweight security mechanism is developed for D2D communication in IoT. For V2X, a lightweight security mechanism is proposed in \cite{mustafa2020lightweight} exploiting parallel advanced encryption standard algorithm.

\subsubsection{Adversarial Attacks on Machine Learning}
The widespread applications of machine learning in URLLC systems (discussed in \ref{sec:ML}) leave URLLC vulnerable to various cyber attacks as the machine learning itself is vulnerable to various adversarial attacks. For example, deep reinforcement learning (DRL) is widely used in wireless communication for optimisation. \cite{qu2023adversarial} demonstrates that it can be fooled by deliberately supplying wrong information from colluding vehicles in a traffic control system. DRL can be made ineffective by compromising reward signals \cite{xu2023environment} and environment \cite{qiaoben2024understanding}. Federated learning is another widely used model for URLLC. It is attacked to either modify its behavior or compromise privacy by exposing data. These attacks can occur at different stages, either before (by injecting malicious data into the training set to corrupt the model) or after model training to modify the already trained model to mislead it in its predictions, causing it to make incorrect or biased decisions. Details of such attacks can be found in \cite{huang2024federated}, \cite{kumar2023impact}.

%{\color{red} Editor: NHM \newline length: 2 pages}

%%%%%%%%%%%%%%%%%%%%%%%%%%%%%%%
%%%%%%%%%%%%%%%%%%%%%%%%%%%%%%%
%%%                         %%%
%%%         SECTION         %%%
%%%                         %%%
%%%%%%%%%%%%%%%%%%%%%%%%%%%%%%%
%%%%%%%%%%%%%%%%%%%%%%%%%%%%%%%

%\section{Open research challenges and future directions towards 6G}

\bgroup
\def\arraystretch{.5}%
\begin{table*}[!htp]
\caption{Summary of ML Techniques for URLLC} 
\centering % used for centering table
\begin{tabular}{|L{1.8cm}|L{2.2cm}|L{1.2cm}|L{10.4cm}|} 
\hline %inserts double horizontal lines
ML Techniques & Learning Model & References & Key features \\
%& technology & & \\
\hline %inserts single line

\multirow{10}{*} {learning} Supervised & Autoencoder & \cite{strodthoff2019enhanced} &
%Supervised learning & Autoencoder & \cite{strodthoff2019enhanced} 

\begin{itemize} 
 \item Decode the message before the end of the transmission.
 \item An iterative decoder is combined with hard thresholding technique.
 \end{itemize} \\
\cline{2-2} \cline{3-3} \cline{4-4}

 & kNN and  & \cite{zhang2019machine} &
 \begin{itemize} 
 \item SAFE-TS for joint URLLC and eMBB scheduling.
 \item Flexible TTI scheduling.
\end{itemize}  \\
\cline{2-2} \cline{3-3} \cline{4-4}

 & Q-reinforcement & \cite{esswie2020online} &
 \begin{itemize} 
 \item Introduce a dual reinforcement machine learning model. 
 \item Optimise the DL and UL symbol structure to minimise the URLLC latency outage.
\end{itemize}  \\
\cline{2-2} \cline{3-3} \cline{4-4}

 & Random forest, kNN, SVC and others & \cite{weinand2019supervised} &
 \begin{itemize} 
 \item Supervised ML based message authentication for physical layer.
 \end{itemize}  \\
 \cline{2-2} \cline{3-3} \cline{4-4}

 & Decision tree & \cite{zhu2019supervised} &
 \begin{itemize} 
 \item Supervised learning based architecture for 5G networks.
 \item Supervised learning based resource allocation instead of traditional software based allocation.
 \end{itemize}  \\
 \cline{2-2} \cline{3-3} \cline{4-4}

  & DNN & \cite{abdelsadek2020resource} &
 \begin{itemize} 
 \item Supervised deep learning based resource allocation.
 \item Reduced computational complexity.
 \end{itemize}  \\
\hline

% UNSUPERVISED 
\multirow{10}{*} {learning} Unsupervised & DNN & \cite{zhao2023energy} &
 \begin{itemize} 
 \item Energy efficient power allocation for URLLC packets with unsupervised learning. 
  \end{itemize} \\
\cline{2-2} \cline{3-3} \cline{4-4}

& Unsupervised DNN  & \cite{sun2023unsupervised} &
\begin{itemize} 
 \item Unsupervised deep learning technique for resource allocation.
 \item DNN is used for static and instantaneous and static constraints. 
 \end{itemize} \\
 \cline{2-2} \cline{3-3} \cline{4-4}

& DNN & \cite{sun2019unsupervised} &
\begin{itemize} 
 \item Unsupervised deep learning for joint power and bandwidth allocation.
 \end{itemize} \\
 \cline{2-2} \cline{3-3} \cline{4-4}

& k-means & \cite{orim2019cluster} &
\begin{itemize} 
 \item Cluster-based random access scheme to eliminate network congestion and overload.
 \item Provide high priority to URLLC devices
 \end{itemize} \\
\hline

% Reinforcement Learning
Reinforcement learning & Q-reinforcement &  \cite{esswie2020online} & 
\begin{itemize}
 \item Introduce a dual reinforcement machine learning model. 
 \item Optimise the DL and UL symbol structure to minimise the URLLC latency outage.
\end{itemize} \\
\cline{2-2} \cline{3-3} \cline{4-4}

& DRL & \cite{li2020deep} &
\begin{itemize} 
 \item A joint scheduling that minimises the eMBB packet puncturing.
 \item Jointly optimises the URLLC and eMBB traffic with deep reinforcement learning technique.
 \end{itemize} \\
 \cline{2-2} \cline{3-3} \cline{4-4}

& CDRL and FDRL & \cite{liu2023machine} &
\begin{itemize} 
 \item ML algorithm for ubiquitous, deep connectivity and holographic connectivity. 
 \item Case study for downlink channel access with CDRL and FDRL.
 \end{itemize} \\
 \cline{2-2} \cline{3-3} \cline{4-4}

& DRL & \cite{huang2020deep} &
\begin{itemize} 
 \item Model-free DRL based optimisation for resource scheduling. 
 \item Fast and stable convergence of learning.
 \end{itemize} \\
 \cline{2-2} \cline{3-3} \cline{4-4}

& FRL & \cite{ali2021federated} &
\begin{itemize} 
 \item Handle densely deployed Wi-Fi and 5G/B5G coexistence with frequently changing situation.
 \item Shares learning estimates among local agents without sharing dataset for faster convergence.
 \end{itemize} \\
 \cline{2-2} \cline{3-3} \cline{4-4}

& FRL & \cite{pan2022asynchronous} &
\begin{itemize} 
 \item ASTEROID for task offloading and optimised RA under long-term URLLC constraints.
 \end{itemize} \\
 \cline{2-2} \cline{3-3} \cline{4-4}

& FRL & \cite{zhang2022federated} &
\begin{itemize} 
 \item Federated deep reinforcement learning based solution to coordinate two xAPPs.
 \end{itemize} \\

\hline
\end{tabular}
\label{table:ML_summary} % is used to refer this table in the text
\end{table*}
\egroup

\section{Machine Learning Based Approach for URLLC}\label{sec:ML}

Different machine learning techniques could be used to fulfill the requirements of URLLC. We will review the works from three aspects: supervised, unsupervised, and reinforcement learning approaches.

\subsection{Supervised Learning Approaches}

Supervised learning is a fundamental machine learning approach that involves training algorithms on labeled data to make predictions or classifications. %The algorithm learns from input-output pairs, mapping input features to their corresponding target labels. 
Several papers are available on supervised learning algorithms.
For example, in \cite{strodthoff2019enhanced}, the authors propose a machine learning-based Early HARQ (E-HARQ) feedback scheme to predict the outcome of the decoding process ahead of the end of the transmission. The proposed method reduces the packet retransmission time, making E-HARQ applicable for URLLC services, unlike traditional HARQ, which is not suitable for URLLC due to its long retransmission time. Instead of using only one feature, two features such as CSI and block error rate are employed in \cite{almarshed2023swift} for detecting-decodable packets using the support vector machine (SVM). Based on this information received from the receiver, the transmitter can decide early whether the packet should be retransmitted or not. This reduces the delay considerably.
 %Retransmission with E-HARQ under stringent latency requirements of URLLC increases reliability over regular HARQ.

A self-adaptive flexible TTI is proposed in \cite{zhang2019machine} for URLLC and eMBB coexistence scenarios. An ML-based flexible scheduling is applied in the method. Specifically, a random forest-based ensemble TTI decision algorithm is implemented to increase network performance.

A supervised learning-based physical layer message authentication technique is proposed in \cite{weinand2019supervised}. The work investigates the performance of the technique using different learning algorithms, including k-nearest neighbor, random forest, and support vector machine.

In \cite{zhu2019supervised}, the authors propose a supervised learning-based architecture for 5G to assure QoS. With the increasing complexity and dynamics of network resource requests, it is hard to allocate resources for different services, including URLLC, eMBB, and mMTC, using traditional software. The machine learning technique removes the traditional software-based resource allocation by employing the machine learning technique. The supervised learning-based technique also predicts future QoS-related anomalies in the network.

An optimised resource allocation with puncturing the eMBB is proposed in \cite{abdelsadek2020resource} to minimise eMBB loss rate. The paper optimises the resource allocation and scheduling process to minimise the eMBB puncturing by considering the QoS of eMBB and the transmission errors associated with the finite blocklength coding of the URLLC traffic. A low complexity deep supervised learning model is also proposed to predict the optimised resource allocation to be practically used in real-time operation.

In most scheduling schemes discussed in Section \ref{sec:MACLayerDesign} , URLLC traffic is algorithmically prioritised over eMBB services to ensure low latency. However, in \cite{hendaoui2024dynamic}, a proactive scheduling strategy is proposed using XGBoost classifier, where URLLC traffic delay is predicted a priori rather than waiting for URLLC traffic arrival to perform scheduling. This further reduces the latency incurred due to reactive scheduling.

\subsection{Unsupervised Learning Approaches}

Unsupervised learning is a type of machine learning where algorithms analyse unlabeled data to discover underlying patterns, relationships, and structures. Several works could be found in the literature on unsupervised learning algorithms for URLLC. For example, In \cite{zhao2023energy}, the power allocation problem of URLLC is solved with an unsupervised machine learning algorithm. First, an energy-efficient power allocation problem is formulated, and then the problem is solved using Deep Neural Network (DNN). The DNN is trained offline with a primal-dual unsupervised algorithm and can then be used in real-time power allocation for URLLC packets

 A unified framework with unsupervised deep learning is proposed in \cite{sun2023unsupervised} for the URLLC resource allocation problem in 5G networks. The framework utilises a technique to solve the problem with instantaneous and static constraints. The problem of constraints optimisation is converted into a functional optimisation problem with instantaneous constraints. Then the unsupervised DNN is employed to find the value of the Lagrange multiplier for the problem.% The earlier version of the work can be found in \cite{sun2019learning}.

A resource allocation technique is proposed in \cite{sun2019unsupervised} using an unsupervised deep learning algorithm. A joint power and bandwidth allocation problem is formulated to minimise bandwidth requirements while satisfying the QoS requirements of URLLC. Then deep learning is employed to find the optimal solution in a general scenario.

In \cite{orim2019cluster}, the authors 
uses the k-means clustering algorithm to control the network access, reducing congestion, and network overload. To manage the massive access by the mMTC devices, a priority-based spatial aggregation algorithm is employed, considering the QoS requirements of URLLC.

\subsection{Reinforcement Learning Approaches}

Reinforcement learning is a machine learning paradigm where an agent learns by trial and error to maximise cumulative rewards. Through interactions with an environment, the agent receives feedback in the form of rewards or penalties. This subsection presents the recent studies on reinforcement learning-based techniques used in URLLC.

\subsubsection{General Approaches}

A dual Reinforcement Machine Learning (RML) technique is proposed in \cite{esswie2020online} to optimise URLLC latency outage for 5G new radio TDD networks. The proposed solution uses real-time capacity and latency statistics of the network. To obtain the statistics, the network periodically monitors the quantities and provides them to the RML algorithm to estimate the upcoming DL and UL radio patterns. These quantities are used to optimise the URLLC outage latency performance.
In \cite{li2020deep}, the authors propose a scheduling technique that investigates the tradeoff between URLLC and eMBB transmission. To achieve the tradeoff, they jointly optimise the bandwidth allocation and overlapping position of URLLC users' traffic using %a reinforcement learning technique. The method use 
a deep deterministic policy gradient learning algorithm.

A reinforcement learning-based algorithm is proposed for the 6G URLLC vision in \cite{liu2023machine}. The vision is a broader extension of the three main 5G services, including enhanced ultrareliable and low latency communications (eURLLC), ultramassive machine-type communications (umMTC), and further enhanced mobile broadband (feMBB) \cite{ishteyaq2024unleashing}. The paper first categorises the three connectivity requirements: ubiquitous connectivity, deep connectivity, and holographic connectivity. ML algorithms are then proposed for three different scenarios: mobility URLLC, massive URLLC, and broadband URLLC. Additionally, the paper presents a case study that investigates downlink channel access problems, which are addressed using Centralised Deep Reinforcement Learning (CDRL) and Federated Deep Reinforcement Learning (FDRL).

In \cite{huang2020deep}, the authors propose dynamic multiplexing of eMBB and URLLC to minimise the adverse impact of preemptive puncturing on eMBB. To mitigate the impact, a model-free deep reinforcement learning-based optimisation is proposed. A DRL-based solution is proposed in \cite{robaglia2024deep} for NOMA uplink scheduling with URLLC requirements in IoT networks. Based on past observations and actions, this method tackles the problem of partial observable channels by exploiting the Markov decision process. Unlike the DRL model based on local data, a distributed DRL model is proposed in \cite{alsenwi2024distributed} for balancing the average and variance of eMBB data rate in multi-cell open radio access network. A central DRL model, which is actor-critic in principle and located in the cloud server, is trained based on the network state information sent by various edge agents distributed across the networks. The edge agents take resource allocation decision applying the centrally trained model. Though this method provides an impressive performance, the data privacy issue across multiple cell is not considered.

\subsubsection{Federated Reinforcement Learning Approaches}

Federated Reinforcement Learning (FRL) represents a groundbreaking approach tailored for the realm of URLLC. This innovative paradigm harmonises two pivotal concepts: Federated Learning and Reinforcement Learning, to address the unique challenges posed by URLLC applications. In FRL, multiple edge devices collaboratively train a reinforcement learning model while retaining their data locally, mitigating concerns related to data privacy and communication overhead. This decentralised training process not only ensures stringent latency requirements of URLLC but also adapts to dynamic and diverse environments \cite{hasan2024federated}. Several works address the FEL in 5G URLLC service. For example, in \cite{ali2021federated}, the authors propose an FRL-based channel resource allocation framework for 5G/B5G networks in the densely deployed Wi-Fi coexistence scenario, where the environment changes dynamically and frequently over time. Instead of an individual localised information-based RL algorithm, FRL-based learning is used, which exchanges the parameters among the agents without exchanging the dataset. 

In \cite{pan2022asynchronous}, the authors propose an Asynchronous federaTed deep Q-learning (DQN)-basEd and URLLC-aware cOmputatIon offloaDing algorithm (ASTEROID) for Space-Assisted Vehicular Networks (SAVN). ASTEROID is proposed to mitigate the limitations of deep reinforcement learning for highly dynamic SAVN. Traditional deep reinforcement learning is not suitable for the URLLC service due to its under-utilisation of environmental observations. ASTEROID maximises throughput while considering the long-term URLLC constraints.

In \cite{zhang2022federated}, the authors propose a coordination algorithm for multiple xAPPs in the open radio access network (O-RAN) for network slicing, where the O-RAN is a network architecture in which RAN is made more open, flexible, and interoperable by using open standards and interfaces. In O-RAN, different network functionalities are referred to as x-applications (xAPPs), which may be considered as independent agents; for example, power control and resource allocation \cite{bonati2021intelligence}. The proposed algorithm uses federated deep reinforcement learning for power control and slice-based resource allocation. Table \ref{table:ML_summary} shows the summary of supervised, unsupervised and reinforcement learning algorithms.

\section{URLLC in 6G: Challenges and Opportunities}\label{sec:6GChallenges_Opportunities}

Since the deployment of 5G started in 2019, researchers around the world started to discuss the vision for future wireless systems or so-called 6G. The emerging 6G era aims to revolutionise the communication paradigm by introducing some disruptive technologies such as programmable wireless environment, pervasive and distributed AI, high capacity low power consuming intelligent devices, terahertz spectrum, digital twin, integrated sensing and communication, and so on as shown in Fig. \ref{fig_6g_technologies}. % \cite{mohjazi2024journey, MGiordani6g, PYang6G}. . Table \ref{tab_ucases} summarises the key differences in terms of use cases between 5G and 6G. 
Researchers have also speculated some promising future applications such as remote surgery, connected autonomous vehicles, virtual and augmented reality, haptic communications which requires extremely low latency and/or high reliability as illustrated in Table \ref{tab_6gpapers}. Furthermore, researchers have envisioned that the future mission critical applications will require 0.1 ms and 1 ms latency for U-plane and C-plane respectively. This target requirement is extremely difficult to meet mainly due to technological limitations  as well as limitations due to reactive approach to latency and reliability problem \cite{JPark2020}. In the last couple of years, researchers have come up with many new technologies for enhancing communications. Of them, some technologies are envisioned to play a vital role in 6G. Ten enabling technologies for 6G are shown in Fig.~\ref{fig_6g_technologies} and shortly described below:

\begin{figure*}[!htp]
 \centering
 \includegraphics[width=0.7\textwidth]{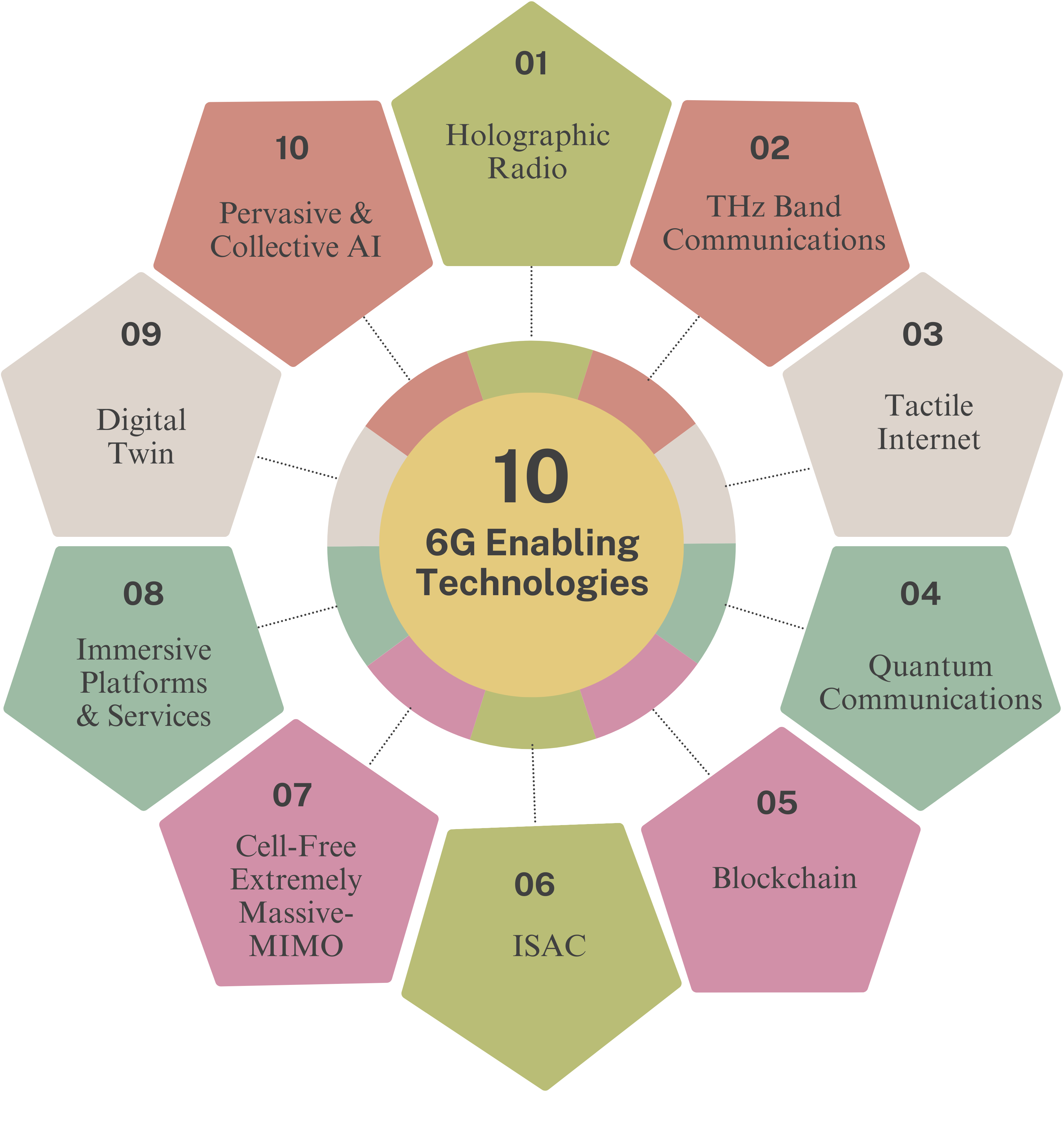}
 \caption{6G enabling technologies}
 \label{fig_6g_technologies}
 \end{figure*}

\textbf{Holographic Radio:} It is a method of designing antennas using metasurfaces to provide more efficient, flexible, and high-performance communications. This principle revolves around an array of antennas consisting of an enormous number of ultra-dense ultra-tiny antennas to produce high directive gain. It is implemented using a reconfigurable holographic surface which is a planner antenna capable of creating any radiation patterns. However, instead of phase-shifting, as is done in phased-array antennas, the amplitude of signals is changed to create the radiation patterns. This reduces the size and cost of the system \cite{zhang2023holographic}.   

\textbf{THz Band Communications:} Due to the unprecedented growth of wireless communication devices with data-hungry emerging applications, the THz band, spanning from 0.1 to 10 THz,  is seen as a highly promising technology for 6G. The use of THz band can alleviate spectrum scarcity. However, this frequency band suffers from high molecular absorption loss such as atmospheric loss, material loss, and so on. Furthermore, THz signals experience high attenuation due to scattering and reflection losses and cannot penetrate even small obstacles such as vehicles, furniture, foliage, and so on \cite{shafie2022terahertz}. For this reason, only line-of-sight communication is possible with this frequency band. Advanced channel estimation, antenna design, RIS, and signal processing systems are needed to properly exploit this abundant frequency band.

\textbf{Visible Light Communication (VLC):} Another avenue for reducing spectral scarcity is the use of visible light for communication. This is chiefly used for short-distance communication. Since this spectrum is unlicensed, any frequency within this range can be used. In addition, VLC is safer compared to radio frequency-based communication as the intruder cannot intercept the VLC signal because it cannot traverse objects such as walls. In 2011, VLC was successfully used to enable access to the Internet. Despite this success, a lot of challenges ahead to make it suitable for widespread applications. 
    
\textbf{Quantum Communications:} In contrast to the traditional communication that leverages the classical physics, the quantum communication exploits quantum properties such as superposition and entanglement. In this technology, information is encoded in quantum bits that can exists both 0 and 1 states simultaneously (superposition) and their states are correlated irrespective of distance. Quantum communication offers unparalleled security and is invulnerable to hacking without leaving a trace. Furthermore, due to the entanglement property, the maximum possible data rate can be doubled in the quantum paradigm. However, this technology is still in its early stages of development and requires a lot of further research to achieve its full potential and practical implementation \cite{chataut20246g}. 

\textbf{Blockchain:} The explosive growth of wireless devices leads to the deployment of highly dense network infrastructures that can be effectively managed using decentralised network architecture. Furthermore, 6G will introduce a wide variety of applications. Maintaining security and privacy in the decentralised collaborative environment will be highly challenging. Blockchain technology, a distributed ledger technology, is envisioned as a great solution to produce a transparent and trustless environment and networks \cite{khan2021blockchain}. It facilitates resource management, access control, privacy, availability, scalability, and accountability in 6G  \cite{hasan2024blockchain}.
    
\textbf{Integrated Sensing and Communication (ISAC):} Similar to THz communications and VLC, ISAC is another solution of spectrum scarcity. It is a unified system that shares the same spectrum, hardware, and signal processing to provide communication and sensing services. In addition, it reduces the cost, size, and energy consumption of communication and sensing equipment. ISAC is going to play a vital role in a wide variety of fields such as mobile communication, radar systems, agriculture, localisation, healthcare, smart cities, and so on.  
    
\textbf{Cell-Free Extremely Massive MIMO (CF-XM MIMO):} This is a paradigm shift technology for cellular networks. Instead of dividing a geographic area into multiple cells with a distinct base station, CF-XM MIMO randomly deploys a large number of access points throughout the area to reduce transceiver distance and enables multiple access points to support a user to increase diversity gain \cite{liu2024jointcoop}. Each AP in CF-XM MIMO is equipped with a massive antenna array, thereby increasing the received signal gain. A central processing unit performs user association, resource management, and all signal processing tasks. 
    
\textbf{Immersive Platforms \& Services:} Immersive platforms will revolutionise the way we interact with digital devices. They have applications in entertainment, healthcare, and education. These devices requires very large bandwidth, low latency, and high reliability. The bandwidth requirement ranges from 10 Mbps to 1 Tbps \cite{shen2023towardimmersive}. Meeting these demands is a great challenge for 6G networks. Combinations of multiple 6G technologies, such as THz bandwidth, CF-XM MIMO, digital twin, and holographic radio, need to be exploited to make such services viable.  
    
\textbf{Digital Twin:} It is a virtual representation of a physical object, environment, or system. It is created by collecting data from the real object or system, analysing the data, and simulating the object and systems. It is a great opportunity for engineers and network operators to exploit this technology to emulate/simulate the network system or specific scenarios such as traffic patterns, resource management, or user behaviors. By analysing the data obtained from the digital twin, network operators can proactively identify and troubleshoot issues. Digital twin technology is applicable in all phases of networks from design to deployment and expansion of the networks. In addition to the design and monitoring, an operational digital twin enables engineers to interact with the cyber-physical system and execute various actions, thereby enabling them to control the system in real-time \cite{khan2022digitaltwin}. 
    
\textbf{Pervasive \& Collective AI:} Wireless networks are becoming increasingly complex over time. Furthermore, diversity among various networks is also increasing. Thus, finding effective mathematical models or heuristic-based networking solutions is going to be increasingly more difficult. Instead, AI-drive network solutions can handle complicated networks more effectively \cite{baccour2023zero}. For this reason, AI has to be used in developing solutions for every aspect of communication systems. Furthermore, the vast capacity of 6G networks will enable data exchange among geographically dispersed devices with diverse AI algorithms. Algorithms have to be developed to implement collective AI to exploit such data for improving network performance. 

%\begin{table}[!htp] 
%\caption{Comparison between 5G and 6G use cases \cite{ftariq1}} \label{tab_ucases}
%    \centering
%\begin{tabular}{ |c|c|c|} 
% \hline
 
%{\bf Use case} & {\bf 5G} & {\bf 6G} \\ 
%\hline\hline

%Centre of gravity & user-centric & service-centric\\ 
%\hline

%Ultra-sensitive applications & not feasible & feasible\\ 
%\hline

%True AI & absent & present\\
%\hline

%Reliability & not extreme & extreme\\
%\hline

%VAR & partial & massive scale\\ 
% \hline

%Time buffer & not real-time & real-time\\ 
% \hline
 
%Capacity & 1-D (bps/Hz) or & 3-D (bps/Hz/m$^3$)\\
%& 2-D (bps/Hz/m$^2$) & \\
%\hline

%VLC & no & yes\\ 
% \hline

%Satellite integration & no & yes\\ 
% \hline

%WPT & no & yes\\ 
% \hline

%Smart city components & separate & integrated\\ 
% \hline

%Autonomous V2X & partially & fully\\ 
% \hline
 
%\end{tabular}

%\end{table}

\begin{table*}[!htp]
\caption{A comparative study of latency requirement, use cases and key enabling technology among various 6G approaches \cite{ftariq1} }
    \centering
    \setlength\tabcolsep{1.5pt}
\begin{tabular}{ |l||c|c|c|c|c|c|} 
 \hline
 
\backslashbox{\bf Attribute}{\bf Paper} & \makebox[2.8em]{\bf Yang et. al \cite{PYang6G}} & {\bf Zhang et. al \cite{ZZhang6g}} & {\bf Saad et al. \cite{WSaad6g}} & {\bf Giordani et al. \cite{MGiordani6g}} & {\bf Strinati et al. \cite{EStrinati6g}} & {\bf Tariq et al. \cite{ftariq1}}\\ 

%\bf{Attribute}&&&&&\\
\hline
\hline

\bf{Peak data rate} &1 Tbps & 1 Tbps& 1 Tbps& 5 Tbps (for VAR)& 1 Tbps& 1 Tbps\\ 
\hline

\hline
\bf{Latency} & $\leq$ 1ms& 0.01-0.1ms& $\leq$ 1ms & $\leq$1ms& 1ms& \vtop{\hbox{\strut Cplane: $\leq$ 1ms} 
\hbox{\strut Uplane: $\leq$ 0.1ms}}\\
\hline\hline
 &Fine Medicine,& Holographic Projection,& XR (VR/AR), & Teleportation,& High Precision  & Remote Surgery,\\ 

 &Intelligent &Tactile and Haptic,& Brain Communication & eHealth,&Manufacturing,&Haptic Comm,\\
 &Disaster Prediction, & Autonomous Driving,&Connected Robotics\& &VAR,& Smart Environment,&massive IoT enabled\\
\bf{Major Use Cases} & Surreal VR, &Internet of &Autonomous Systems&Industry 4.0,&Holographic &  Smart City, VAR,\\
 &3D Videos,&Nano-Bio Things, &&Robotics, &Communication& Autonomous Driving,\\
 &&Space travel &&Autonomous &&Automation and\\
 &&&&Transportation&&Manufacturing\\

\hline
\bf{Key Enabling } & Ultra Massive & THz Communication,& Tinycell,& THz Communication,& Novel Network  & THz Communication,\\ 
\bf{Technologies}&MIMO&Holographic & Ubiquitous Network&VLC,  ML& Architecture, VLC &Pervasive AI, EH\\
&OAM-MDM&Beamforming,  & energy transfer and , & 3D Networks,&AI at Network Edge& VLC, Blockchain\\
&Super Flexible&Quantum Comm,&harvesting, Transceiver& Cell-less architecture& Battery-less devices,& Cell Free Network\\
&Integrated Network,&AI/ML, LIS, VLC& with integrated&Energy harvesting& THz Communication,& Quantum Comm,\\
&Multi-Domain&Blockchain, & frequency band.& NFV, Backhaul.& distributed security& metasurface, OAM\\
&Index Modulation& &smart surface & & & WPT, Context. Comm.\\

\hline

\end{tabular}
\label{tab_6gpapers}
\end{table*}

%\subsection{URLLC and Machine Learning/Artificial Intelligence}
%\textcolor{red}{(about 0.5 page, FT)}

%Fortunately, widespread  proliferation of AI in every aspect of wireless system in general and machine learning in particular leads to some promising solutons for the problem discussed above. The AI/ML techniques will potentially change the approach of URLLC solutions from reactive regime to predictive modelling or proactive URLLC regime. The AI/ML techniques will enable to optimise the decision making within the time and reliability constraint. However, these require adequate high quality data availability as well as optimal design of the ML system itself. 
 
 %There has been several work on URLLC which attempted to exploit machine learning (ML) for achieving the stringent latency and reliability target. The works in \cite{AAzari19} used ML to make the resource allocation scheme risk aware.

%By employing a deep reinforcement learning method, namely deep Q-learning, we design an intelligent agent at the edge computing node to develop a real-time adaptive policy for computational resource allocation for offloaded tasks of multiple users in order to improve the average E2E reliability \cite{TYang}.

%The basic idea of DPre is to explore and exploit the correlation of data acquisition and access behavior between nodes through static and dynamic learning mechanisms in order to make judicious resource per-allocation decisions \cite{MLi}.

\section{6G URLLC Research Directions}\label{sec:6GResearchDirections}

\begin{figure*}[t]
 \centering
 \includegraphics[width=0.9\textwidth]{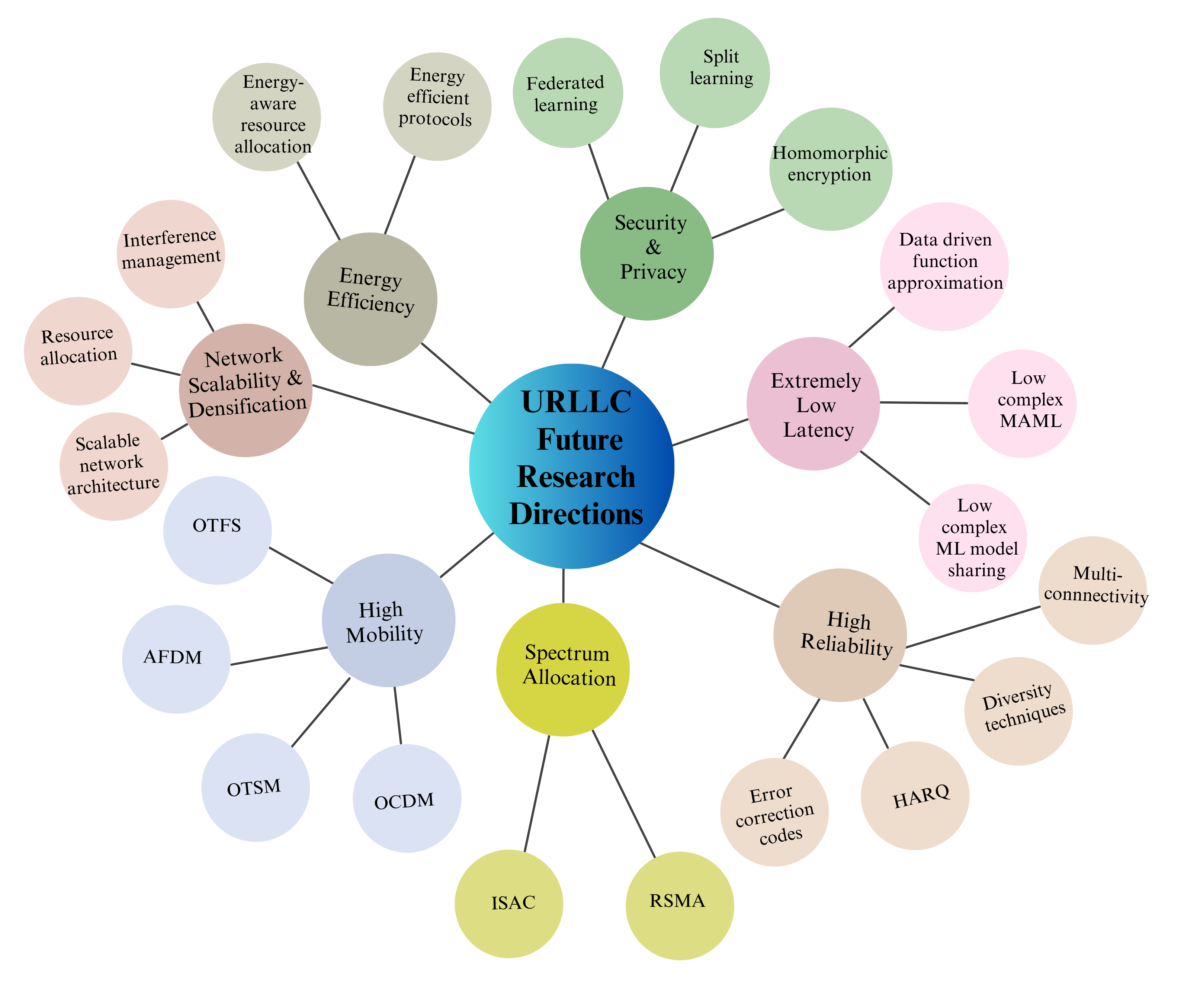}
 \captionsetup{justification=centering}
 \caption{URLLC Research Challenges.}
 \label{fig_research_directions}
 \end{figure*}

In 6G systems, Ultra-Reliable Low Latency Communications (URLLC) aims to provide extremely reliable and low-latency communication services, opening up possibilities for mission-critical applications, industrial automation, autonomous systems, and more. While 6G networks are still in the research phase, there are several challenges that researchers are likely to face in the development of URLLC capabilities \cite{tariq20236g}. Potential research directions related to URLLC in 6G are illustrated in Fig.~\ref{fig_research_directions} , covering various domains of communication systems. Here are some of the main challenges and research directions:

 \textbf{Extremely Low Latency:} URLLC applications demand ultra-low latency, typically in the sub-millisecond range. Achieving such low latency consistently across the network is a significant challenge. Researchers will need to design new protocols, optimise network architectures, and explore innovative transmission techniques to minimise latency. With the prolific use of data-driven solutions, it is possible to achieve a significant reduction of latency. For example, in \cite{xu2021faster}, the decoding problem of the resource assignment information from the physical downlink control channel at the user level is transformed into a signal classification problem. Such function approximation significantly reduces the latency introduced at the receiving end.  As the propagation time is not the primary contributor to latency \cite{she2021tutorial}, more research efforts should be devoted to approximate the functionalities of various blocks in the receiver exploiting data-driven methods. Another avenue of research should be focusing on making ML suitable for massive heterogeneous networks. Most of the existing work considers scenarios where heterogeneity among the devices and networks is limited. However, the next-generation mobile networks and devices are going to be highly heterogeneous in terms of bandwidth availability/requirement, network/device sizes, applications, latency requirements, and so on. In this scenario, not all networks and/or devices will be able to support ML model parameter exchange at the same speed. This will affect the latency. To overcome this problem, researchers need to come up with new model exchange algorithms for such heterogeneous networks/devices. The network topology is expected to change dynamically, which will require retraining machine learning model frequently. However, the current best-performing ML model is deep learning which requires a significant training time. For this reason, model agnostic meta-learning (MAML) can be a good candidate solution. However, most applications have a specific latency requirement. A thorough investigation should be carried out to know whether the existing MAML can meet the requirements. It might be required to explore the modification of the existing MAML to make it suitable to meet the network constraints.   

\textbf{Extremely High  Reliability:} URLLC in 6G systems is expected to provide even higher levels of reliability, with very low packet error rates and 99.99999\% or better reliability \cite{pourkabirian2024vision}, than in 5G. Maintaining reliable communication links, even in highly dynamic and challenging environments, will require advanced error correction codes, diversity techniques, and interference mitigation strategies. Further investigation is needed on the resource allocation in the network slicing and multi-connectivity scenarios. Improvement in the reliability in most cases adversely affects the latency \cite{pourkabirian2024vision}. A thorough analysis of the trade-offs among the latency and reliability is needed for various reliability and latency improvement methods.

 \textbf{Communication and Control Co-design:} One of the key feature of future communications system is to have the ability to physically interact with remote system in real time such as remote surgery, tactile internet and so on. While the traditional design approach is to have separated design where the device is designed for a specific purpose first and then communication system is designed separately  to provide connectivity as per requirement for the control system of the device. However, in order to optimise the performance, it is essential to design both together which gives rise the concept of 'communication control co-design' \cite{GZhao2018ComMag}. This codesign approach will be beneficial for a number of 5G verticals including vehicular platooning \cite{TZeng2018ICC}. There are many challenges that need to be considered as information and control theory is not well investigated in the URLLC context. While both the information theory domain and control theory domain are quite mature, not much have been done in terms of combining this theories together.

 \textbf{Application Specific Design:} Many emerging applications have very specific and stringent requirement in terms of latency and  reliability as well as other QoS metrics. For example, to ensure distributed intelligence in robotics and autonomous systems applications seamless wireless connectivity with extremely low latency is required. In particular, there is a very specific pattern of data communication in cloud robotics paradigm which uses wireless connections for offloading high complexity and data intensive computation tasks to the edge/cloud platform. It also has much higher upload data rate compared to downlaod data rate requirement which is in contrast with most common applications. Since, most of the existing design of URLLC protocols has been explored by the research community in an application-agnostic manner, they may not offer the best possible or globally optimum performance. Therefore, alternative approaches that are based on co-design of control systems, inference engines, and communication protocols must be explored in the context of specific applications.

\textbf{Mobility considerations}: 6G networks are expected to support highly mobile devices, including autonomous vehicles and drones. Ensuring seamless connectivity and uninterrupted URLLC services during high-speed mobility poses challenges in terms of handover mechanisms, beamforming, and efficient resource allocation. OFDM, the most popular waveform, suffers from the problem of large Doppler spread in the time-varying channels. Recently proposed orthogonal time frequency space (OTFS) modulation has drawn significant attention due to its robustness in this scenario \cite{aldababsa2024survey}. The research on OTFS is still in its nascent state and a lot of effort is required to make it suitable to meet the requirements of URLLC. Research should be devoted to designing a multi-objective low-complex precoder to support high mobility and reduced peak-to-average power ratio (PAPR). Three more latest developed waveforms for dealing with extremely highly mobile devices are orthogonal chirp-division multiplexing (OCDM) \cite{haif2024novelOCDM}, orthogonal time sequency multiplexing (OTSM) \cite{thaj2021orthogonaltime} and Affine Frequency Division Multiplexing
(AFDM) \cite{bemani2023affine}. Still, no research exists to evaluate the performance of these three waveforms in light of URLLC requirements. Furthermore, URLLC design of the communication systems exploiting these waveforms is a promising research field.  In addition, the high-speed vehicles lead to frequent handover which increases processing latency. This is still an open research problem in 6G.

 \textbf{Network scalability and densification:} 6G is expected to support a large number of connected devices, leading to an increase in network density. Researchers need to address the challenges of managing interference, optimising resource allocation, and designing scalable network architectures to accommodate the diverse requirements of URLLC applications.

 \textbf{Energy efficiency:} URLLC applications often operate on battery-powered devices or have stringent energy constraints. URLLC protocols are in general power hungry as they often initiates retransmission to ensure high reliability. Therefore, balancing the need for low latency and high reliability with energy efficiency poses a significant challenge. It needs attention from the researchers to develop power-saving mechanisms, efficient protocols, and energy-aware resource management techniques.

 \textbf{Security and privacy:} As the number of connected devices and critical applications that rely on URLLC grows, ensuring robust security and privacy becomes crucial. Researchers must address the challenges of secure transmission, authentication, encryption, and privacy preservation in 6G URLLC systems \cite{zaman2023comprehensive}. Federated learning is applied in various fields for its effectiveness in preserving privacy by eliminating the need for training data sharing. However, it requires a considerable bandwidth for sharing all the parameters of machine learning models. This consumption of bandwidth results in the degradation of latency. Further investigation is needed to modify the federated learning architecture to reduce the required communications between the servers and clients. Recently proposed split learning overcomes some of the problems associated with the federated learning. %Usually, it divides the machine learning model into two segments. The forward pass of the data through the first segment is carried out on the client side. The output of the last layer of the first segment is sent to the server to be passed through the second layer. The gradient from the first layer of the second segment is sent back to the last layer of the first segment on the client side. 
As it requires sending only the weights of one layer, the bandwidth consumption for model training is considerably less compared to that of federated learning. Limited work \cite{wu2023split, lin2024efficient, kim2023bargaining, lin2024split,khan2023resource, khan2023joint,xu2023accelerating} has been done to date on the applicability of split learning for URLLC. In \cite{wu2023split, lin2024efficient}, latency is reduced on the client side via parallel training. The segment size is optimised in \cite{kim2023bargaining, xu2023accelerating} to ensure tradeoffs among various performance criteria. The split learning is applied in 6G edge networks \cite{lin2024split} and IoT \cite{khan2023resource, khan2023joint}. However, no work has been found yet assessing split learning in the light of URLLC requirements. The reliability of wireless networks employing split learning is yet to be investigated. Alongside the split learning, homomorphic encryption and federated learning/split learning with differential privacy can also be used to ensure privacy. However, as these methods are computationally expensive, investigations can be carried out to find a trade-off among privacy, latency, and reliability.

 \textbf{Spectrum allocation and utilisation:} Efficient spectrum management is critical for URLLC services. Researchers will need to explore new spectrum sharing mechanisms, dynamic spectrum access techniques, innovative spectrum utilisation strategies, advanced interference mitigation, and new frequency bands to accommodate the increasing demand of bandwidth for stringent requirements of URLLC applications. ISAC has recently drawn significant attention due to its spectrum efficiency. Some of its potential applications include autonomous vehicles and smart factories where high reliability and low latency are highly essential. To satisfy the sensing and URLLC requirements, different techniques including transmit power minimisation \cite{ding2022joint}, joint beamforming and scheduling \cite{zhao2024jointBeam}, and energy efficiency maximisation via optimal power allocation \cite{behdad2024interplay, behdad2024joint} are proposed. An amalgamation of various spectrum-efficient technologies, including RSMA \cite{lyu2024rate-splitting}, OTFS/orthogonal delay-Doppler multiplexing (ODDM), and ISAC, in future 6G networks is worth investigating in light of URLLC requirements. Furthermore, URLLC design of RSMA-assisted ISAC is another interesting research topic. In the last decade, many waveforms, such as variants of OFDM, FBMC and its variants, generalised frequency modulation and its variants, UFMC and its variants, have been developed in a bid to improve spectrum efficiency. Although considerable effort is devoted to assessing their effectiveness in terms of spectrum efficiency, there is no comprehensive comparative study of the waveforms in the light of URLLC requirements.

%Addressing these challenges will require interdisciplinary research efforts involving communication theory, signal processing, network architecture, machine learning, and other fields. Furthermore, close collaboration with industry stakeholders and standardisation bodies will be essential to ensure the practical deployment of URLLC capabilities in 6G systems. It is important to note that 6G is still an area of ongoing research, and the challenges mentioned here are speculative based on the current understanding of URLLC and the anticipated requirements of 6G systems. As research progresses and more information becomes available, the specific challenges for URLLC in 6G may evolve or be refined.

\section{Conclusions} \label{sec:conclusions}

This survey paper has traced the evolution of URLLC over the last decade and how it shaped the 5G innovation leading to its anticipated role in future 6G communications. It provides a comprehensive coverage of enabling technologies, performance trade-offs and open challenges. The analysis highlights the fact that in order to meet stringent requirement of beyond 5G systems, researchers need to think beyond incremental improvements and focus on rethinking the wireless system design where reliability, latency, scalability, and efficiency are addressed in a holistic and integrated manner.\\
The paper identifies several important emerging research directions. Firstly, AI and ML will take the role of fundamental driving force for optimising the performance and achieving the latency and reliability requirement, particularly in dynamic and diverse application scenarios. This will allow to implement predictive decision making and resource allocation which is vital for managing latency constrained systems. Secondly, cross-layer design that considers PHY, MAC and other layers holistically will be indispensable part of the URLLC system to mitigate inherent trade-offs between latency, reliability and energy efficiency. Thus, multi-objective optimisation will also gain similar importance. Finally, emerging network architecture and enabling technologies such as IRS, ISAC, edge intelligence and non-terrestrial networks radically expands the application and design space of URLLC which offers innovative ways to push the boundary of the traditional cellular systems. URLLC will play crucial role in future mission critical applications that requires seamless mobility, context aware adaptation and much wider coverage compared to traditional systems.\\
In conclusion, URLLC in 6G represents both a formidable engineering challenge and a transformative opportunity. By pursuing the research directions discussed in this paper, a future network can be designed that not only connect people and machines but also underpin critical societal functions with unprecedented levels of reliability and responsiveness. Thus, URLLC will evolve from a specialised service in 5G into a defining pillar of 6G connectivity, shaping the future of communications for decades to come.

\section*{Acknowledgment}
The authors would like to thank Dr. Nurul Huda Mahmood of Oulu University, Finland for his valuable guidance in initial structuring of this article.

\bibliographystyle{IEEEtran}
\bibliography{IEEEabrv,list_papers}

\begin{thebibliography}{100}
\providecommand{\url}[1]{#1}
\csname url@samestyle\endcsname
\providecommand{\newblock}{\relax}
\providecommand{\bibinfo}[2]{#2}
\providecommand{\BIBentrySTDinterwordspacing}{\spaceskip=0pt\relax}
\providecommand{\BIBentryALTinterwordstretchfactor}{4}
\providecommand{\BIBentryALTinterwordspacing}{\spaceskip=\fontdimen2\font plus
\BIBentryALTinterwordstretchfactor\fontdimen3\font minus \fontdimen4\font\relax}
\providecommand{\BIBforeignlanguage}[2]{{%
\expandafter\ifx\csname l@#1\endcsname\relax
\typeout{** WARNING: IEEEtran.bst: No hyphenation pattern has been}%
\typeout{** loaded for the language `#1'. Using the pattern for}%
\typeout{** the default language instead.}%
\else
\language=\csname l@#1\endcsname
\fi
#2}}
\providecommand{\BIBdecl}{\relax}
\BIBdecl

\bibitem{ftariq1}
F.~Tariq, M.~R.~A. Khandaker, K.~Wong, M.~A. Imran, M.~Bennis, and M.~Debbah, ``A speculative study on {6G},'' \emph{IEEE Wireless Communication}, vol.~99, 2020.

\bibitem{JAndrews2014}
J.~G. Andrews, S.~Buzzi, W.~Choi, S.~V. Hanly, A.~Lozano, A.~C. Soong, and J.~C. Zhang, ``What will {5G} be?'' \emph{IEEE Journal on selected areas in communications}, vol.~32, no.~6, pp. 1065--1082, 2014.

\bibitem{islam2025artificial}
M.~M. Islam, M.~L. Baptista, and F.~Tariq, \emph{Artificial Intelligence for Smart Manufacturing and Industry X. 0}.\hskip 1em plus 0.5em minus 0.4em\relax Springer, 2025.

\bibitem{AGupta15}
A.~{Gupta} and R.~K. {Jha}, ``A survey of {5G} network: Architecture and emerging technologies,'' \emph{IEEE Access}, vol.~3, pp. 1206--1232, 2015.

\bibitem{wijethilaka2021survey}
S.~Wijethilaka and M.~Liyanage, ``Survey on network slicing for internet of things realization in 5g networks,'' \emph{IEEE Communications Surveys \& Tutorials}, vol.~23, no.~2, pp. 957--994, 2021.

\bibitem{fuad02}
N.~H. {Mahmood}, M.~{Lauridsen}, G.~{Berardinelli}, D.~{Catania}, and P.~{Mogensen}, ``Radio resource management techniques for embb and mmtc services in {5G} dense small cell scenarios,'' in \emph{2016 IEEE 84th Vehicular Technology Conference (VTC-Fall)}, Sep. 2016, pp. 1--5.

\bibitem{filali2022dynamic}
A.~Filali, Z.~Mlika, S.~Cherkaoui, and A.~Kobbane, ``Dynamic sdn-based radio access network slicing with deep reinforcement learning for urllc and embb services,'' \emph{IEEE Transactions on Network Science and Engineering}, 2022.

\bibitem{CBockelmann}
C.~{Bockelmann}, N.~K. {Pratas}, G.~{Wunder}, S.~{Saur}, M.~{Navarro}, D.~{Gregoratti}, G.~{Vivier}, E.~{De Carvalho}, Y.~{Ji}, C.~{Stefanovic}, P.~{Popovski}, Q.~{Wang}, M.~{Schellmann}, E.~{Kosmatos}, P.~{Demestichas}, M.~{Raceala-Motoc}, P.~{Jung}, S.~{Stanczak}, and A.~{Dekorsy}, ``Towards massive connectivity support for scalable mmtc communications in {5G} networks,'' \emph{IEEE Access}, vol.~6, pp. 28\,969--28\,992, 2018.

\bibitem{MShafi17}
M.~{Shafi}, A.~F. {Molisch}, P.~J. {Smith}, T.~{Haustein}, P.~{Zhu}, P.~{De Silva}, F.~{Tufvesson}, A.~{Benjebbour}, and G.~{Wunder}, ``5g: A tutorial overview of standards, trials, challenges, deployment, and practice,'' \emph{IEEE Journal on Selected Areas in Communications}, vol.~35, no.~6, pp. 1201--1221, June 2017.

\bibitem{popovski2}
P.~Popovski, G.~Mange, A.~Roos, T.~Rosowski, G.~Zimmermann, J.~S{\"o}der \emph{et~al.}, ``Deliverable d6. 3 intermediate system evaluation results,'' \emph{ICT-317669-METIS/D6. 3}, 2014.

\bibitem{MsimsekJSAC}
M.~{Simsek}, A.~{Aijaz}, M.~{Dohler}, J.~{Sachs}, and G.~{Fettweis}, ``5g-enabled tactile internet,'' \emph{IEEE Journal on Selected Areas in Communications}, vol.~34, no.~3, pp. 460--473, March 2016.

\bibitem{alliance2019verticals}
N.~Alliance, ``Verticals urllc use cases and requirements,'' \emph{NGMN Alliance}, 2019.

\bibitem{5gppp1}
G.~Association \emph{et~al.}, ``‘{5G} and e-health,'' \emph{5GPPP, White Paper, Oct}, 2015.

\bibitem{challacombe1}
B.~Challacombe, L.~Kavoussi, and P.~Dasgupta, ``Trans-oceanic telerobotic surgery,'' \emph{BJU international}, vol.~92, no.~7, pp. 678--680, 2003.

\bibitem{she1}
C.~She, C.~Yang, and T.~Q. Quek, ``Cross-layer optimization for ultra-reliable and low-latency radio access networks,'' \emph{IEEE Transactions on Wireless Communications}, vol.~17, no.~1, pp. 127--141, 2018.

\bibitem{soret1}
B.~Soret, P.~Mogensen, K.~I. Pedersen, and M.~C. Aguayo-Torres, ``Fundamental tradeoffs among reliability, latency and throughput in cellular networks,'' in \emph{Globecom Workshops (GC Wkshps), 2014}.\hskip 1em plus 0.5em minus 0.4em\relax IEEE, 2014, pp. 1391--1396.

\bibitem{Abreu1}
R.~Abreu, P.~Mogensen, and K.~I. Pedersen, ``Pre-scheduled resources for retransmissions in ultra-reliable and low latency communications,'' in \emph{2017 IEEE Wireless Communications and Networking Conference (WCNC)}, March 2017, pp. 1--5.

\bibitem{ji1}
H.~Ji, S.~Park, J.~Yeo, Y.~Kim, J.~Lee, and B.~Shim, ``Ultra-reliable and low-latency communications in {5G} downlink: Physical layer aspects,'' \emph{IEEE Wireless Communications}, vol.~25, no.~3, pp. 124--130, 2018.

\bibitem{johansson1}
N.~A. Johansson, Y.-P.~E. Wang, E.~Eriksson, and M.~Hessler, ``Radio access for ultra-reliable and low-latency {5G} communications,'' in \emph{Communication Workshop (ICCW), 2015 IEEE International Conference on}.\hskip 1em plus 0.5em minus 0.4em\relax IEEE, 2015, pp. 1184--1189.

\bibitem{durisi1}
G.~Durisi, T.~Koch, and P.~Popovski, ``Toward massive, ultrareliable, and low-latency wireless communication with short packets,'' \emph{Proceedings of the IEEE}, vol. 104, no.~9, pp. 1711--1726, 2016.

\bibitem{Survey2019PhyMac}
G.~J. {Sutton}, J.~{Zeng}, R.~P. {Liu}, W.~{Ni}, D.~N. {Nguyen}, B.~A. {Jayawickrama}, X.~{Huang}, M.~{Abolhasan}, Z.~{Zhang}, E.~{Dutkiewicz}, and T.~{Lv}, ``Enabling technologies for ultra-reliable and low latency communications: From phy and mac layer perspectives,'' \emph{IEEE Communications Surveys Tutorials}, vol.~21, no.~3, pp. 2488--2524, thirdquarter 2019.

\bibitem{siddiqi20195g}
M.~A. Siddiqi, H.~Yu, and J.~Joung, ``5g ultra-reliable low-latency communication implementation challenges and operational issues with iot devices,'' \emph{Electronics}, vol.~8, no.~9, p. 981, 2019.

\bibitem{shaik2024ai}
R.~B. Shaik, P.~Nagaradjane, I.~Ioannou, V.~Sittakul, V.~Vasiliou, and A.~Pitsillides, ``Ai/ml-aided capacity maximization strategies for urllc in 5g/6g wireless systems: A survey,'' \emph{Computer Networks}, p. 110506, 2024.

\bibitem{pradhan2024survey}
A.~Pradhan, S.~Das, M.~J. Piran, and Z.~Han, ``A survey on security of ultra/hyper reliable low latency communication: Recent advancements, challenges, and future directions,'' \emph{arXiv preprint arXiv:2404.08160}, 2024.

\bibitem{elgarhy2024energy}
O.~Elgarhy, L.~Reggiani, M.~M. Alam, A.~Zoha, R.~Ahmad, and A.~Kuusik, ``Energy efficiency and latency optimization for iot urllc and mmtc use cases,'' \emph{IEEE Access}, 2024.

\bibitem{khan2022urllc}
B.~S. Khan, S.~Jangsher, A.~Ahmed, and A.~Al-Dweik, ``Urllc and embb in 5g industrial iot: A survey,'' \emph{IEEE Open Journal of the Communications Society}, vol.~3, pp. 1134--1163, 2022.

\bibitem{ho2019next}
T.~M. Ho, T.~D. Tran, T.~T. Nguyen, S.~Kazmi, L.~B. Le, C.~S. Hong, and L.~Hanzo, ``Next-generation wireless solutions for the smart factory, smart vehicles, the smart grid and smart cities,'' \emph{arXiv preprint arXiv:1907.10102}, 2019.

\bibitem{siddiqui2023urllc}
M.~U.~A. Siddiqui, H.~Abumarshoud, L.~Bariah, S.~Muhaidat, M.~A. Imran, and L.~Mohjazi, ``Urllc in beyond 5g and 6g networks: An interference management perspective,'' \emph{IEEE Access}, 2023.

\bibitem{masaracchia2021uav}
A.~Masaracchia, Y.~Li, K.~K. Nguyen, C.~Yin, S.~R. Khosravirad, D.~B. Da~Costa, and T.~Q. Duong, ``Uav-enabled ultra-reliable low-latency communications for 6g: A comprehensive survey,'' \emph{IEEE Access}, 2021.

\bibitem{ali2021urllc}
R.~Ali, Y.~B. Zikria, A.~K. Bashir, S.~Garg, and H.~S. Kim, ``Urllc for 5g and beyond: Requirements, enabling incumbent technologies and network intelligence,'' \emph{IEEE Access}, vol.~9, pp. 67\,064--67\,095, 2021.

\bibitem{ts3Gpp36814}
``3\uppercase{GPP}, further advancements for e-utra physical layer aspects, technical report 36.814, release 9,'' Mar., 2010.

\bibitem{ts3Gpp22661}
3\uppercase{GPP TS} 22.261, ``Service requirements for the {5G} system; stage 1, release 15,'' 3GPP Technical Specification Group, Tech. Rep., 2017.

\bibitem{attaran2021impact}
M.~Attaran, ``The impact of 5g on the evolution of intelligent automation and industry digitization,'' \emph{Journal of Ambient Intelligence and Humanized Computing}, pp. 1--17, 2021.

\bibitem{ts3Gpp38913}
``3\uppercase{GPP}: Study on scenarios and requirements for next generation access technologies. technical report 38.913, release 14,'' Oct, 2016.

\bibitem{capozzi2013downlink}
F.~Capozzi, G.~Piro, L.~A. Grieco, G.~Boggia, and P.~Camarda, ``Downlink packet scheduling in lte cellular networks: Key design issues and a survey,'' \emph{IEEE Communications Surveys \& Tutorials}, vol.~15, no.~2, pp. 678--700, 2013.

\bibitem{hoffman1996hippi}
J.~Hoffman, ``Hippi-6400 technology dissemination,'' in \emph{Broadband Access Systems}, vol. 2917.\hskip 1em plus 0.5em minus 0.4em\relax SPIE, 1996, pp. 422--430.

\bibitem{wen2003ultra}
Y.~Wen and V.~W. Chan, ``Ultra-reliable communication over unreliable optical networks via lightpath diversity: system characterization and optimization,'' in \emph{GLOBECOM'03. IEEE Global Telecommunications Conference (IEEE Cat. No. 03CH37489)}, vol.~5.\hskip 1em plus 0.5em minus 0.4em\relax IEEE, 2003, pp. 2529--2535.

\bibitem{YWen05URC}
------, ``Ultra-reliable communication over vulnerable all-optical networks via lightpath diversity,'' \emph{IEEE Journal on Selected Areas in Communications}, vol.~23, no.~8, pp. 1572--1587, 2005.

\bibitem{CDombrowski11}
C.~Dombrowski and J.~Gross, ``Reducing outage probability over wireless channels under hard real-time constraints,'' in \emph{1st Workshop on Real-Time Wireless for Industrial Applications 2011 (RealWin 2011)}, 2011.

\bibitem{AOsseiran13}
A.~{Osseiran}, V.~{Braun}, T.~{Hidekazu}, P.~{Marsch}, H.~{Schotten}, H.~{Tullberg}, M.~A. {Uusitalo}, and M.~{Schellman}, ``The foundation of the mobile and wireless communications system for 2020 and beyond: Challenges, enablers and technology solutions,'' in \emph{2013 IEEE 77th Vehicular Technology Conference (VTC Spring)}, June 2013, pp. 1--5.

\bibitem{DSoldani1}
D.~Soldani, D.~Virette, G.~Cordara, and Q.~Zhang, ``An enriched multimedia experience for wireless networks in horizon 2020 and beyond,'' \emph{IEEE COMSOC MMTC E-Letter}, vol.~7, no.~8, 2012.

\bibitem{JSilvo13}
J.~{Silvo}, L.~M. {Eriksson}, M.~{Björkbom}, and S.~{Nethi}, ``Ultra-reliable and real-time communication in local wireless applications,'' in \emph{IECON 2013 - 39th Annual Conference of the IEEE Industrial Electronics Society}, Nov 2013, pp. 5611--5616.

\bibitem{RBaldemair13}
R.~{Baldemair}, E.~{Dahlman}, S.~{Parkvall}, Y.~{Selen}, K.~{Balachandran}, T.~{Irnich}, G.~{Fodor}, and H.~{Tullberg}, ``Future wireless communications,'' in \emph{2013 IEEE 77th Vehicular Technology Conference (VTC Spring)}, June 2013, pp. 1--5.

\bibitem{HSchotten14}
H.~D. Schotten, R.~Sattiraju, D.~G. Serrano, Z.~Ren, and P.~Fertl, ``Availability indication as key enabler for ultra-reliable communication in 5g,'' in \emph{2014 European Conference on Networks and Communications (EuCNC)}.\hskip 1em plus 0.5em minus 0.4em\relax IEEE, 2014, pp. 1--5.

\bibitem{popovski1}
P.~Popovski, ``Ultra-reliable communication in {5G} wireless systems,'' in \emph{5G for Ubiquitous Connectivity (5GU), 2014 1st International Conference on}.\hskip 1em plus 0.5em minus 0.4em\relax IEEE, 2014, pp. 146--151.

\bibitem{wang2005novel}
X.~WANG, ``A novel adaptive error correction control scheme for the ultra low-latency mobile network based on ofdm,'' \emph{Proc. IEICE Technical Report}, 2005.

\bibitem{GLaughlin14}
G.~Laughlin, A.~Aguirre, and J.~Grundfest, ``Information transmission between financial markets in chicago and new york,'' \emph{Financial Review}, vol.~49, no.~2, pp. 283--312, 2014.

\bibitem{MBrachmann13}
M.~Brachmann and S.~Santini, ``Towards the benchmarking of ultra-low latency communication protocols for wireless sensor and actuator networks,'' in \emph{Proceedings of the 11th ACM Conference on Embedded Networked Sensor Systems}.\hskip 1em plus 0.5em minus 0.4em\relax ACM, 2013, p.~55.

\bibitem{JAllen12}
J.~D. Allen, X.~Liu, I.~Lozano, and X.~Yuan, ``A cyber-physical approach to a wide-area actionable system for the power grid,'' in \emph{MILCOM 2012-2012 IEEE Military Communications Conference}.\hskip 1em plus 0.5em minus 0.4em\relax IEEE, 2012, pp. 1--6.

\bibitem{promwongsa2020comprehensive}
N.~Promwongsa, A.~Ebrahimzadeh, D.~Naboulsi, S.~Kianpisheh, F.~Belqasmi, R.~Glitho, N.~Crespi, and O.~Alfandi, ``A comprehensive survey of the tactile internet: State-of-the-art and research directions,'' \emph{IEEE Communications Surveys \& Tutorials}, vol.~23, no.~1, pp. 472--523, 2020.

\bibitem{Petar1}
P.~{Popovski}, J.~J. {Nielsen}, C.~{Stefanovic}, E.~d.~{Carvalho}, E.~{Strom}, K.~F. {Trillingsgaard}, A.~{Bana}, D.~M. {Kim}, R.~{Kotaba}, J.~{Park}, and R.~B. {Sorensen}, ``Wireless access for ultra-reliable low-latency communication: Principles and building blocks,'' \emph{IEEE Network}, vol.~32, no.~2, pp. 16--23, March 2018.

\bibitem{MImran2019WB}
M.~A. Imran., Y.~A. Sambo, and Q.~H. Abbasi, \emph{Enabling {5G} Communication Systems to Support Vertical Industries}.\hskip 1em plus 0.5em minus 0.4em\relax Wiley Online Library, 2019.

\bibitem{AMahmood}
A.~{Mahmood}, M.~I. {Ashraf}, M.~{Gidlund}, and J.~{Torsner}, ``Over-the-air time synchronization for urllc: Requirements, challenges and possible enablers,'' in \emph{2018 15th International Symposium on Wireless Communication Systems (ISWCS)}, Aug 2018, pp. 1--6.

\bibitem{sanusi24}
I.~O. Sanusi and K.~M. Nasr, ``A reinforcement learning approach for d2d spectrum sharing in wireless industrial urllc networks,'' \emph{IEEE Transactions on Network and Service Management}, pp. 1--1, 2024.

\bibitem{nakimuli2021deployment}
W.~Nakimuli, J.~Garcia-Reinoso, J.~E. Sierra-Garcia, P.~Serrano, and I.~Q. Fern{\'a}ndez, ``Deployment and evaluation of an industry 4.0 use case over 5g,'' \emph{IEEE Communications Magazine}, vol.~59, no.~7, pp. 14--20, 2021.

\bibitem{ho2024energyEfficiency}
T.~M. Ho, K.-K. Nguyen, and M.~Cheriet, ``Energy efficiency deep reinforcement learning for urllc in 5g mission-critical swarm robotics,'' \emph{IEEE Transactions on Network and Service Management}, 2024.

\bibitem{MRaftopoulou2019EuCNC}
M.~{Raftopoulou}, L.~{Jorguseski}, and R.~{Litjens}, ``Design and assessment of low-latency random access procedures in {5G} networks,'' in \emph{2019 European Conference on Networks and Communications (EuCNC)}, June 2019, pp. 406--411.

\bibitem{karaca2020scheduling}
M.~Karaca, ``Scheduling and dynamic pilot allocation for massive mimo with varying traffic,'' \emph{IEEE Wireless Communications Letters}, vol.~9, no.~12, pp. 2102--2106, 2020.

\bibitem{MKhoshnevisan2019JSAC}
M.~{Khoshnevisan}, V.~{Joseph}, P.~{Gupta}, F.~{Meshkati}, R.~{Prakash}, and P.~{Tinnakornsrisuphap}, ``{5G} industrial networks with comp for urllc and time sensitive network architecture,'' \emph{IEEE Journal on Selected Areas in Communications}, vol.~37, no.~4, pp. 947--959, April 2019.

\bibitem{5gppp2015}
G.~Association \emph{et~al.}, ``‘{5G} automotive vision,'' \emph{5GPPP, White Paper, Oct}, 2015.

\bibitem{guo2022age}
C.~Guo, X.~Wang, L.~Liang, and G.~Y. Li, ``Age of information, latency, and reliability in intelligent vehicular networks,'' \emph{IEEE Network}, 2022.

\bibitem{abdel20205g}
S.~A. Abdel~Hakeem, A.~A. Hady, and H.~Kim, ``5g-v2x: Standardization, architecture, use cases, network-slicing, and edge-computing,'' \emph{Wireless Networks}, vol.~26, no.~8, pp. 6015--6041, 2020.

\bibitem{3GPPTR23786}
``3gpp tr 23.786 study on architecture enhancements for eps and {5G} system to support advanced v2x services,'' June 2018.

\bibitem{SHussain2018CSCN}
S.~{Husain}, A.~{Kunz}, A.~{Prasad}, E.~{Pateromichelakis}, K.~{Samdanis}, and J.~{Song}, ``The road to {5G} {V2X: Ultra}-high reliable communications,'' in \emph{2018 IEEE Conference on Standards for Communications and Networking (CSCN)}, Oct 2018, pp. 1--6.

\bibitem{SHUssain2019CSM}
S.~S. {Husain}, A.~{Kunz}, A.~{Prasad}, E.~{Pateromichelakis}, and K.~{Samdanis}, ``Ultra-high reliable {5G} {V2X} communications,'' \emph{IEEE Communications Standards Magazine}, vol.~3, no.~2, pp. 46--52, June 2019.

\bibitem{XSong2019Access}
X.~{Song} and M.~{Yuan}, ``Performance analysis of one-way highway vehicular networks with dynamic multiplexing of embb and urllc traffics,'' \emph{IEEE Access}, vol.~7, pp. 118\,020--118\,029, 2019.

\bibitem{HMa2016SAE}
H.~S. Ma, E.~Zhang, S.~Li, Z.~Lv, and J.~Hu, ``A {V2X} design for {5G} network based on requirements of autonomous driving,'' SAE Technical Paper, Tech. Rep., 2016.

\bibitem{XGe2019TVT}
X.~{Ge}, ``Ultra-reliable low-latency communications in autonomous vehicular networks,'' \emph{IEEE Transactions on Vehicular Technology}, vol.~68, no.~5, pp. 5005--5016, May 2019.

\bibitem{skondras2021network}
E.~Skondras, E.~T. Michailidis, A.~Michalas, D.~J. Vergados, N.~I. Miridakis, and D.~D. Vergados, ``A network slicing framework for uav-aided vehicular networks,'' \emph{Drones}, vol.~5, no.~3, p.~70, 2021.

\bibitem{li2022age}
Z.~Li, L.~Xiang, and X.~Ge, ``Age of information modeling and optimization for fast information dissemination in vehicular social networks,'' \emph{IEEE Transactions on Vehicular Technology}, vol.~71, no.~5, pp. 5445--5459, 2022.

\bibitem{MAziz2019WCL}
M.~K. {Abdel-Aziz}, S.~{Samarakoon}, M.~{Bennis}, and W.~{Saad}, ``Ultra-reliable and low-latency vehicular communication: An active learning approach,'' \emph{IEEE Communications Letters}, pp. 1--1, 2019.

\bibitem{HFarhangi2010PEM}
H.~{Farhangi}, ``The path of the smart grid,'' \emph{IEEE Power and Energy Magazine}, vol.~8, no.~1, pp. 18--28, January 2010.

\bibitem{HHui2019EAE}
H.~Hui, Y.~Ding, Q.~Shi, F.~Li, Y.~Song, and J.~Yan, ``{5G} network-based internet of things for demand response in smart grid: A survey on application potential,'' \emph{Applied Energy}, vol. 257, p. 113972, 2020.

\bibitem{PSchulz17ComMag}
P.~{Schulz}, M.~{Matthe}, H.~{Klessig}, M.~{Simsek}, G.~{Fettweis}, J.~{Ansari}, S.~A. {Ashraf}, B.~{Almeroth}, J.~{Voigt}, I.~{Riedel}, A.~{Puschmann}, A.~{Mitschele-Thiel}, M.~{Muller}, T.~{Elste}, and M.~{Windisch}, ``Latency critical iot applications in 5g: Perspective on the design of radio interface and network architecture,'' \emph{IEEE Communications Magazine}, vol.~55, no.~2, pp. 70--78, February 2017.

\bibitem{rao2018impact}
S.~K. Rao and R.~Prasad, ``Impact of {5G} technologies on smart city implementation,'' \emph{Wireless Personal Communications}, vol. 100, no.~1, pp. 161--176, 2018.

\bibitem{rusti2018smart}
B.~Rusti, H.~Stefanescu, J.~Ghenta, and C.~Patachia, ``Smart city as a {5G} ready application,'' in \emph{2018 International Conference on Communications (COMM)}.\hskip 1em plus 0.5em minus 0.4em\relax IEEE, 2018, pp. 207--212.

\bibitem{HHabibzadeh}
H.~Habibzadeh, T.~Soyata, B.~Kantarci, A.~Boukerche, and C.~Kaptan, ``Sensing, communication and security planes: A new challenge for a smart city system design,'' \emph{Computer Networks}, vol. 144, pp. 163 -- 200, 2018.

\bibitem{fang2021modeling}
Y.~Fang, Z.~Shan, and W.~Wang, ``Modeling and key technologies of a data-driven smart city system,'' \emph{IEEE Access}, vol.~9, pp. 91\,244--91\,258, 2021.

\bibitem{LDerrico}
L.~{D’Errico}, F.~{Franchi}, F.~{Graziosi}, A.~{Marotta}, C.~{Rinaldi}, M.~{Boschi}, and A.~{Colarieti}, ``Structural health monitoring and earthquake early warning on {5G URLLC} network,'' in \emph{2019 IEEE 5th World Forum on Internet of Things (WF-IoT)}, 2019, pp. 783--786.

\bibitem{shi2022versatile}
X.~Shi, M.~Feng, G.~He, S.~Li, and T.~Jiang, ``A versatile experimental platform for tactile internet: Design guidelines and practical implementation,'' \emph{IEEE Network}, 2022.

\bibitem{MShirvanimoghaddam2018short}
M.~Shirvanimoghaddam, M.~S. Mohammadi, R.~Abbas, A.~Minja, C.~Yue, B.~Matuz, G.~Han, Z.~Lin, W.~Liu, Y.~Li \emph{et~al.}, ``Short block-length codes for ultra-reliable low latency communications,'' \emph{IEEE Communications Magazine}, vol.~57, no.~2, pp. 130--137, 2018.

\bibitem{KSKim19}
K.~S. {Kim}, D.~K. {Kim}, C.~{Chae}, S.~{Choi}, Y.~{Ko}, J.~{Kim}, Y.~{Lim}, M.~{Yang}, S.~{Kim}, B.~{Lim}, K.~{Lee}, and K.~L. {Ryu}, ``Ultrareliable and low-latency communication techniques for tactile internet services,'' \emph{Proceedings of the IEEE}, vol. 107, no.~2, pp. 376--393, Feb 2019.

\bibitem{CLi19a}
C.~{Li}, C.~{Li}, K.~{Hosseini}, S.~B. {Lee}, J.~{Jiang}, W.~{Chen}, G.~{Horn}, T.~{Ji}, J.~E. {Smee}, and J.~{Li}, ``5g-based systems design for tactile internet,'' \emph{Proceedings of the IEEE}, vol. 107, no.~2, pp. 307--324, Feb 2019.

\bibitem{tr3Gpp38913}
``3\uppercase{GPP}, “3gpp tr 38.913 v15.0.0: Study on scenarios and requirements for next generation access technologies; (release 15),'' Jan., 2018.

\bibitem{osseiran2015manufacturing}
A.~Osseiran, J.~Sachs, M.~Puleri \emph{et~al.}, ``Manufacturing reengineered: {Robots, 5G} and the industrial {IoT},'' \emph{Ericsson Business Review}, no.~4, 2015.

\bibitem{chen2018ultra}
H.~Chen, R.~Abbas, P.~Cheng, M.~Shirvanimoghaddam, W.~Hardjawana, W.~Bao, Y.~Li, and B.~Vucetic, ``Ultra-reliable low latency cellular networks: Use cases, challenges and approaches,'' \emph{IEEE Communications Magazine}, vol.~56, no.~12, pp. 119--125, 2018.

\bibitem{holfeld2016wireless}
B.~Holfeld, D.~Wieruch, T.~Wirth, L.~Thiele, S.~A. Ashraf, J.~Huschke, I.~Aktas, and J.~Ansari, ``Wireless communication for factory automation: An opportunity for lte and {5G} systems,'' \emph{IEEE Communications Magazine}, vol.~54, no.~6, pp. 36--43, 2016.

\bibitem{song2022semi}
H.~Song, K.~J. Kim, J.~Guo, P.~V. Orlik, and K.~Parsons, ``Semi-persistent scheduling scheme for low-latency and high-reliability transmissions in private 5g networks,'' in \emph{2022 IEEE International Conference on Communications Workshops (ICC Workshops)}.\hskip 1em plus 0.5em minus 0.4em\relax IEEE, 2022, pp. 651--656.

\bibitem{mahmood2019uplink}
N.~H. Mahmood, R.~Abreu, R.~B{\"o}hnke, M.~Schubert, G.~Berardinelli, and T.~H. Jacobsen, ``Uplink grant-free access solutions for urllc services in 5g new radio,'' in \emph{2019 16th International Symposium on Wireless Communication Systems (ISWCS)}.\hskip 1em plus 0.5em minus 0.4em\relax IEEE, 2019, pp. 607--612.

\bibitem{dougan2019noma}
S.~Do{\u{g}}an, A.~Tusha, and H.~Arslan, ``Noma with index modulation for uplink urllc through grant-free access,'' \emph{IEEE Journal of Selected Topics in Signal Processing}, vol.~13, no.~6, pp. 1249--1257, 2019.

\bibitem{katwe2024rsma}
M.~V. Katwe, R.~Deshpande, K.~Singh, M.-L. Ku, and B.~Clerckx, ``Rsma-enabled aerial ris-aided mu-mimo system for improved spectral-efficient urllc,'' \emph{IEEE Transactions on Vehicular Technology}, 2024.

\bibitem{anand2018joint}
A.~Anand, G.~De~Veciana, and S.~Shakkottai, ``Joint scheduling of urllc and embb traffic in {5G} wireless networks,'' in \emph{IEEE INFOCOM 2018-IEEE Conference on Computer Communications}.\hskip 1em plus 0.5em minus 0.4em\relax IEEE, 2018, pp. 1970--1978.

\bibitem{shi2022risk}
B.~Shi, F.-C. Zheng, C.~She, J.~Luo, and A.~G. Burr, ``Risk-resistant resource allocation for embb and urllc coexistence under m/g/1 queueing model,'' \emph{IEEE Transactions on Vehicular Technology}, vol.~71, no.~6, pp. 6279--6290, 2022.

\bibitem{chairman3Gppwg1}
``Chairman’s notes 3\uppercase{GPP}: 3\uppercase{GPP TSG RAN WG1} meeting 88bis \uppercase{A}vailable at $http://www.3gpp.org/ftp/tsg_ran /wg1_rl1/tsgr1_88b/report$,'' November, 2016.

\bibitem{rinaldi20215g}
F.~Rinaldi, A.~Raschella, and S.~Pizzi, ``5g nr system design: a concise survey of key features and capabilities,'' \emph{Wireless Networks}, vol.~27, no.~8, pp. 5173--5188, 2021.

\bibitem{you2018resource}
L.~You, Q.~Liao, N.~Pappas, and D.~Yuan, ``Resource optimization with flexible numerology and frame structure for heterogeneous services,'' \emph{IEEE Communications Letters}, vol.~22, no.~12, pp. 2579--2582, 2018.

\bibitem{CPLi2017EuCNC}
C.-P. Li, J.~Jiang, W.~Chen, T.~Ji, and J.~Smee, ``5g ultra-reliable and low-latency systems design,'' in \emph{2017 European Conference on Networks and Communications (EuCNC)}.\hskip 1em plus 0.5em minus 0.4em\relax IEEE, 2017, pp. 1--5.

\bibitem{popovski20185g}
P.~Popovski, K.~F. Trillingsgaard, O.~Simeone, and G.~Durisi, ``5g wireless network slicing for embb, urllc, and mmtc: A communication-theoretic view,'' \emph{Ieee Access}, vol.~6, pp. 55\,765--55\,779, 2018.

\bibitem{tian2024intelligentCoexistence}
M.~Tian, C.~Li, Y.~Hui, B.~Chen, W.~Yue, Y.~Fu, and Z.~Han, ``An intelligent coexistence strategy for embb/urllc traffic in multi-uav relay networks via deep reinforcement learning,'' \emph{IEEE Transactions on Wireless Communications}, 2024.

\bibitem{shi2024puncturingMaching}
B.~Shi, C.~She, F.-C. Zheng, L.~Gao, and G.~Li, ``Puncturing-based resource allocation for urllc and embb services via matching theory and unsupervised deep learning,'' \emph{IEEE Transactions on Vehicular Technology}, 2024.

\bibitem{CXiao_JSAC19}
C.~{Xiao}, J.~{Zeng}, W.~{Ni}, X.~{Su}, R.~P. {Liu}, T.~{Lv}, and J.~{Wang}, ``Downlink mimo-noma for ultra-reliable low-latency communications,'' \emph{IEEE Journal on Selected Areas in Communications}, vol.~37, no.~4, pp. 780--794, April 2019.

\bibitem{lien20175g}
S.-Y. Lien, S.-L. Shieh, Y.~Huang, B.~Su, Y.-L. Hsu, and H.-Y. Wei, ``{5G} new radio: Waveform, frame structure, multiple access, and initial access,'' \emph{IEEE communications magazine}, vol.~55, no.~6, pp. 64--71, 2017.

\bibitem{iwabuchi20175g}
M.~Iwabuchi, A.~Benjebbour, Y.~Kishiyama, G.~Ren, C.~Tang, T.~Tian, L.~Gu, T.~Takada, and T.~Kashima, ``{5G} field experimental trials on urllc using new frame structure,'' in \emph{2017 IEEE Globecom Workshops (GC Wkshps)}.\hskip 1em plus 0.5em minus 0.4em\relax IEEE, 2017, pp. 1--6.

\bibitem{kela2015novel}
P.~Kela, M.~Costa, J.~Salmi, K.~Leppanen, J.~Turkka, T.~Hiltunen, and M.~Hronec, ``A novel radio frame structure for {5G} dense outdoor radio access networks,'' in \emph{2015 IEEE 81st Vehicular Technology Conference (VTC Spring)}.\hskip 1em plus 0.5em minus 0.4em\relax IEEE, 2015, pp. 1--6.

\bibitem{pedersen2015flexible}
K.~Pedersen, F.~Frederiksen, G.~Berardinelli, and P.~Mogensen, ``A flexible frame structure for {5G} wide area,'' in \emph{2015 IEEE 82nd Vehicular Technology Conference (VTC2015-Fall)}.\hskip 1em plus 0.5em minus 0.4em\relax IEEE, 2015, pp. 1--5.

\bibitem{vihriala2016numerology}
J.~Vihri{\"a}l{\"a}, A.~A. Zaidi, V.~Venkatasubramanian, N.~He, E.~Tiirola, J.~Medbo, E.~L{\"a}hetkangas, K.~Werner, K.~Pajukoski, A.~Cedergren \emph{et~al.}, ``Numerology and frame structure for {5G} radio access,'' in \emph{2016 IEEE 27th Annual International Symposium on Personal, Indoor, and Mobile Radio Communications (PIMRC)}.\hskip 1em plus 0.5em minus 0.4em\relax IEEE, 2016, pp. 1--5.

\bibitem{pedersen2016flexible}
K.~I. Pedersen, G.~Berardinelli, F.~Frederiksen, P.~Mogensen, and A.~Szufarska, ``A flexible 5\uppercase{G} frame structure design for frequency-division duplex cases,'' \emph{IEEE Communications Magazine}, vol.~54, no.~3, pp. 53--59, 2016.

\bibitem{HRenTCom2020}
H.~Ren, C.~Pan, Y.~Deng, M.~Elkashlan, and A.~Nallanathan, ``Resource allocation for secure urllc in mission-critical iot scenarios,'' \emph{IEEE Transactions on Communications}, vol.~68, no.~9, pp. 5793--5807, 2020.

\bibitem{elayoubi2019radio}
S.~E. Elayoubi, P.~Brown, M.~Deghel, and A.~Galindo-Serrano, ``Radio resource allocation and retransmission schemes for urllc over {5G} networks,'' \emph{IEEE Journal on Selected Areas in Communications}, vol.~37, no.~4, pp. 896--904, 2019.

\bibitem{JYeo}
J.~Yeo, S.~Park, J.~Oh, Y.~Kim, and J.~Lee, ``Partial retransmission scheme for harq enhancement in {5G} wireless communications,'' in \emph{2017 IEEE Globecom Workshops (GC Wkshps)}, Dec 2017, pp. 1--5.

\bibitem{CSun}
C.~Sun, C.~She, and C.~Yang, ``Retransmission policy with frequency hopping for ultra-reliable and low-latency communications,'' in \emph{2018 IEEE International Conference on Communications (ICC)}, May 2018, pp. 1--6.

\bibitem{3GPP2016comparison}
``Comparison of slot and mini-slot based approaches for {URLLC, 3GPP R1-1609664, 3GPP Technical Specification Group},'' Oct, 2016.

\bibitem{3GPP2016on}
``{3GPP R1-1609663,} on the mini-slot structure, {3GPP Technical Specification Group},'' Oct, 2016.

\bibitem{3GPP2016onthe}
``On the \uppercase{URLLC} transmission formats for \uppercase{NR TDD}, {3GPP proposal R1-167269},'' Aug, 2016.

\bibitem{lee2017packet}
B.~Lee, S.~Park, D.~J. Love, H.~Ji, and B.~Shim, ``Packet structure and receiver design for low latency wireless communications with ultra-short packets,'' \emph{IEEE Transactions on Communications}, vol.~66, no.~2, pp. 796--807, 2017.

\bibitem{mousaei2017optimizing}
M.~Mousaei and B.~Smida, ``Optimizing pilot overhead for ultra-reliable short-packet transmission,'' in \emph{2017 IEEE International Conference on Communications (ICC)}.\hskip 1em plus 0.5em minus 0.4em\relax IEEE, 2017, pp. 1--5.

\bibitem{ji2018sparse}
H.~Ji, S.~Park, and B.~Shim, ``Sparse vector coding for ultra reliable and low latency communications,'' \emph{IEEE Transactions on Wireless Communications}, vol.~17, no.~10, pp. 6693--6706, 2018.

\bibitem{jindal2010unified}
N.~Jindal and A.~Lozano, ``A unified treatment of optimum pilot overhead in multipath fading channels,'' \emph{IEEE Transactions on Communications}, vol.~58, no.~10, pp. 2939--2948, 2010.

\bibitem{lin2021pilot}
X.~Lin, X.~Zhu, Y.~Jiang, and J.~Cao, ``Pilot overhead vs. pilot power: Short packet structure optimization for urllc over continuous fading,'' in \emph{2021 IEEE Global Communications Conference (GLOBECOM)}.\hskip 1em plus 0.5em minus 0.4em\relax IEEE, 2021, pp. 01--06.

\bibitem{ji2019pilot}
H.~Ji, W.~Kim, and B.~Shim, ``Pilot-less sparse vector coding for short packet transmission,'' \emph{IEEE Wireless Communications Letters}, 2019.

\bibitem{jiang2021road}
W.~Jiang, B.~Han, M.~A. Habibi, and H.~D. Schotten, ``The road towards 6g: A comprehensive survey,'' \emph{IEEE Open Journal of the Communications Society}, vol.~2, pp. 334--366, 2021.

\bibitem{sheikh2024comparative}
M.~Sheikh-Hosseini, F.~Rahdari, H.~Ghasemnezhad, S.~Ahmadi, and M.~Uysal, ``A comparative performance evaluation of ofdm, gfdm, and otfs in impulsive noise channels,'' \emph{IEEE Open Journal of the Communications Society}, 2024.

\bibitem{tarboush2022single}
S.~Tarboush, H.~Sarieddeen, M.-S. Alouini, and T.~Y. Al-Naffouri, ``Single-versus multicarrier terahertz-band communications: A comparative study,'' \emph{IEEE Open Journal of the Communications Society}, vol.~3, pp. 1466--1486, 2022.

\bibitem{xiao2024rethinking}
Z.~Xiao, X.~Liu, Y.~Zeng, J.~A. Zhang, S.~Jin, and R.~Zhang, ``Rethinking waveform for 6g: Harnessing delay-doppler alignment modulation,'' \emph{IEEE Communications Magazine}, 2024.

\bibitem{SEldessoki17}
S.~Eldessoki, D.~Wieruch, and B.~Holfeld, ``Impact of waveforms on coexistence of mixed numerologies in 5g urllc networks,'' in \emph{WSA 2017; 21th International ITG Workshop on Smart Antennas}, 2017, pp. 1--6.

\bibitem{zhang2015filtered}
X.~Zhang, M.~Jia, L.~Chen, J.~Ma, and J.~Qiu, ``Filtered-ofdm-enabler for flexible waveform in the 5th generation cellular networks,'' in \emph{2015 IEEE Global Communications Conference (GLOBECOM)}.\hskip 1em plus 0.5em minus 0.4em\relax IEEE, 2015, pp. 1--6.

\bibitem{zhang2018mixed}
X.~Zhang, L.~Zhang, P.~Xiao, D.~Ma, J.~Wei, and Y.~Xin, ``Mixed numerologies interference analysis and inter-numerology interference cancellation for windowed ofdm systems,'' \emph{IEEE Transactions on Vehicular Technology}, vol.~67, no.~8, pp. 7047--7061, 2018.

\bibitem{sybis2016channel}
M.~Sybis, K.~Wesolowski, K.~Jayasinghe, V.~Venkatasubramanian, and V.~Vukadinovic, ``Channel coding for ultra-reliable low-latency communication in {5G} systems,'' in \emph{Vehicular Technology Conference (VTC-Fall), 2016 IEEE 84th}.\hskip 1em plus 0.5em minus 0.4em\relax IEEE, 2016, pp. 1--5.

\bibitem{berrou1993near}
C.~Berrou, A.~Glavieux, and P.~Thitimajshima, ``Near shannon limit error-correcting coding and decoding: Turbo-codes. 1,'' in \emph{Proceedings of ICC'93-IEEE International Conference on Communications}, vol.~2.\hskip 1em plus 0.5em minus 0.4em\relax IEEE, 1993, pp. 1064--1070.

\bibitem{arikan2009channel}
E.~Arikan, ``Channel polarization: A method for constructing capacity-achieving codes for symmetric binary-input memoryless channels,'' \emph{IEEE Transactions on information Theory}, vol.~55, no.~7, pp. 3051--3073, 2009.

\bibitem{tal2015list}
I.~Tal and A.~Vardy, ``List decoding of polar codes,'' \emph{IEEE Transactions on Information Theory}, vol.~61, no.~5, pp. 2213--2226, 2015.

\bibitem{maiya2012low}
S.~V. Maiya, D.~J. Costello, and T.~E. Fuja, ``Low latency coding: Convolutional codes vs. ldpc codes,'' \emph{IEEE Transactions on Communications}, vol.~60, no.~5, pp. 1215--1225, 2012.

\bibitem{gallager1962low}
R.~Gallager, ``Low-density parity-check codes,'' \emph{IRE Transactions on information theory}, vol.~8, no.~1, pp. 21--28, 1962.

\bibitem{wu2018low}
X.~Wu, M.~Jiang, C.~Zhao, L.~Ma, and Y.~Wei, ``Low-rate pbrl-ldpc codes for urllc in 5g,'' \emph{IEEE Wireless Communications Letters}, vol.~7, no.~5, pp. 800--803, 2018.

\bibitem{divsalar2005low}
D.~Divsalar, S.~Dolinar, and C.~Jones, ``Low-rate ldpc codes with simple protograph structure,'' in \emph{Proceedings. International Symposium on Information Theory, 2005. ISIT 2005.}\hskip 1em plus 0.5em minus 0.4em\relax IEEE, 2005, pp. 1622--1626.

\bibitem{KNiu2012CL}
K.~Niu and K.~Chen, ``Crc-aided decoding of polar codes,'' \emph{IEEE communications letters}, vol.~16, no.~10, pp. 1668--1671, 2012.

\bibitem{tang2016improving}
H.~Tang, W.~Zhang, W.~Hardjawana, and B.~Vucetic, ``Improving latency and reliability in {5G} internet-of-things networks,'' in \emph{2016 IEEE International Conference on Smart Grid Communications (SmartGridComm)}.\hskip 1em plus 0.5em minus 0.4em\relax IEEE, 2016, pp. 509--513.

\bibitem{tarneberg2017utilizing}
W.~Tarneberg, M.~Karaca, A.~Robertsson, F.~Tufvesson, and M.~Kihl, ``Utilizing massive mimo for the tactile internet: Advantages and trade-offs,'' in \emph{2017 IEEE International Conference on Sensing, Communication and Networking (SECON Workshops)}.\hskip 1em plus 0.5em minus 0.4em\relax IEEE, 2017, pp. 1--6.

\bibitem{casciano2019enabling}
G.~Casciano, P.~Baracca, and S.~Buzzi, ``Enabling ultra reliable wireless communications for factory automation with distributed mimo,'' \emph{arXiv preprint arXiv:1907.03530}, 2019.

\bibitem{kashima2016large}
T.~Kashima, J.~Qiu, H.~Shen, C.~Tang, T.~Tian, X.~Wang, X.~Hou, H.~Jiang, A.~Benjebbour, Y.~Saito \emph{et~al.}, ``Large scale massive {MIMO} field trial for {5G} mobile communications system,'' in \emph{2016 International Symposium on Antennas and Propagation (ISAP)}.\hskip 1em plus 0.5em minus 0.4em\relax IEEE, 2016, pp. 602--603.

\bibitem{xu20163d}
C.~Xu, J.~Cosmas, Y.~Zhang, P.~Lazaridis, G.~Araniti, and Z.~D. Zaharis, ``{3D MIMO} radio channel modeling of a weighted linear array system of antennas for {5G} cellular systems,'' in \emph{2016 International Conference on Telecommunications and Multimedia (TEMU)}.\hskip 1em plus 0.5em minus 0.4em\relax IEEE, 2016, pp. 1--6.

\bibitem{busari2019terahertz}
S.~A. Busari, K.~M.~S. Huq, S.~Mumtaz, and J.~Rodriguez, ``Terahertz massive {MIMO} for beyond-{5G} wireless communication,'' in \emph{ICC 2019-2019 IEEE International Conference on Communications (ICC)}.\hskip 1em plus 0.5em minus 0.4em\relax IEEE, 2019, pp. 1--6.

\bibitem{vu2017ultra}
T.~K. Vu, C.-F. Liu, M.~Bennis, M.~Debbah, M.~Latva-Aho, and C.~S. Hong, ``Ultra-reliable and low latency communication in mmwave-enabled massive mimo networks,'' \emph{IEEE Communications Letters}, vol.~21, no.~9, pp. 2041--2044, 2017.

\bibitem{LLu14}
L.~Lu, G.~Y. Li, A.~L. Swindlehurst, A.~Ashikhmin, and R.~Zhang, ``An overview of massive mimo: Benefits and challenges,'' \emph{IEEE journal of selected topics in signal processing}, vol.~8, no.~5, pp. 742--758, 2014.

\bibitem{WTarneberg17}
W.~Tarneberg, M.~Karaca, A.~Robertsson, F.~Tufvesson, and M.~Kihl, ``Utilizing massive mimo for the tactile internet: Advantages and trade-offs,'' in \emph{2017 IEEE International Conference on Sensing, Communication and Networking (SECON Workshops)}.\hskip 1em plus 0.5em minus 0.4em\relax IEEE, 2017, pp. 1--6.

\bibitem{TVu17}
T.~K. {Vu}, C.~{Liu}, M.~{Bennis}, M.~{Debbah}, M.~{Latva-aho}, and C.~S. {Hong}, ``Ultra-reliable and low latency communication in mmwave-enabled massive mimo networks,'' \emph{IEEE Communications Letters}, vol.~21, no.~9, pp. 2041--2044, Sep. 2017.

\bibitem{peng2022resourceUplink}
Q.~Peng, H.~Ren, C.~Pan, N.~Liu, and M.~Elkashlan, ``Resource allocation for uplink cell-free massive mimo enabled urllc in a smart factory,'' \emph{IEEE Transactions on Communications}, vol.~71, no.~1, pp. 553--568, 2022.

\bibitem{peng2023resourceDownlink}
------, ``Resource allocation for cell-free massive mimo-enabled urllc downlink systems,'' \emph{IEEE Transactions on Vehicular Technology}, 2023.

\bibitem{zhang2024performanceRician}
Y.~Zhang, W.~Xia, H.~Zhao, Y.~Zhu, W.~Xu, and W.~Lu, ``Performance analysis of cell-free massive mimo-urllc systems over correlated rician fading channels with phase shifts,'' \emph{IEEE Transactions on Wireless Communications}, 2024.

\bibitem{huang2024performanceSE_EE}
Y.~Huang, Y.~Jiang, F.-C. Zheng, P.~Zhu, D.~Wang, and X.~You, ``Performance of spectral and energy efficiency in cell-free massive mimo-aided urllc system with short blocklength communications,'' \emph{IEEE Transactions on Vehicular Technology}, 2024.

\bibitem{ding2022joint}
C.~Ding, C.~Zeng, C.~Chang, J.-B. Wang, and M.~Lin, ``Joint precoding for mimo radar and urllc in isac systems,'' in \emph{Proceedings of the 1st ACM MobiCom Workshop on Integrated Sensing and Communications Systems}, 2022, pp. 12--18.

\bibitem{behdad2024joint}
Z.~Behdad, {\"O}.~T. Demir, K.~W. Sung, and C.~Cavdar, ``{Joint Processing and Transmission Energy Optimization for ISAC in Cell-Free Massive MIMO with URLLC},'' \emph{arXiv preprint arXiv:2401.10315}, 2024.

\bibitem{nikbakht2025mimo}
H.~Nikbakht, Y.~C. Eldar, and H.~V. Poor, ``A mimo isac system for ultra-reliable and low-latency communications,'' \emph{arXiv preprint arXiv:2501.13025}, 2025.

\bibitem{yuan2023orthogonal}
W.~Yuan, J.~Zou, Y.~Cui, X.~Li, J.~Mu, and K.~Han, ``Orthogonal time frequency space and predictive beamforming-enabled urllc in vehicular networks,'' \emph{IEEE Wireless Communications}, vol.~30, no.~2, pp. 56--62, 2023.

\bibitem{lipp2016variations}
T.~Lipp and S.~Boyd, ``Variations and extension of the convex--concave procedure,'' \emph{Optimization and Engineering}, vol.~17, no.~2, pp. 263--287, 2016.

\bibitem{vu2018ultra}
T.~K. Vu, M.~Bennis, M.~Debbah, M.~Latva-Aho, and C.~S. Hong, ``Ultra-reliable communication in {5G} {mmWave} networks: {A} risk-sensitive approach,'' \emph{IEEE Communications Letters}, vol.~22, no.~4, pp. 708--711, 2018.

\bibitem{ford2017achieving}
R.~Ford, M.~Zhang, M.~Mezzavilla, S.~Dutta, S.~Rangan, and M.~Zorzi, ``Achieving ultra-low latency in {5G} millimeter wave cellular networks,'' \emph{IEEE Communications Magazine}, vol.~55, no.~3, pp. 196--203, 2017.

\bibitem{echigo2021deep}
H.~Echigo, Y.~Cao, M.~Bouazizi, and T.~Ohtsuki, ``A deep learning-based low overhead beam selection in mmwave communications,'' \emph{IEEE Transactions on Vehicular Technology}, vol.~70, no.~1, pp. 682--691, 2021.

\bibitem{ding2021context}
H.~Ding and K.~G. Shin, ``Context-aware beam tracking for 5g mmwave v2i communications,'' \emph{IEEE Transactions on Mobile Computing}, 2021.

\bibitem{levanen2014radio}
T.~Levanen, J.~Pirskanen, and M.~Valkama, ``Radio interface design for ultra-low latency millimeter-wave communications in {5G} era,'' in \emph{2014 IEEE Globecom Workshops (GC Wkshps)}.\hskip 1em plus 0.5em minus 0.4em\relax IEEE, 2014, pp. 1420--1426.

\bibitem{dutta2017frame}
S.~Dutta, M.~Mezzavilla, R.~Ford, M.~Zhang, S.~Rangan, and M.~Zorzi, ``Frame structure design and analysis for millimeter wave cellular systems,'' \emph{IEEE Transactions on Wireless Communications}, vol.~16, no.~3, pp. 1508--1522, 2017.

\bibitem{yoshioka2016field}
S.~Yoshioka, Y.~Inoue, S.~Suyama, Y.~Kishiyama, Y.~Okumura, J.~Kepler, and M.~Cudak, ``Field experimental evaluation of beamtracking and latency performance for {5G} {mmWave} radio access in outdoor mobile environment,'' in \emph{2016 IEEE 27th Annual International Symposium on Personal, Indoor, and Mobile Radio Communications (PIMRC)}.\hskip 1em plus 0.5em minus 0.4em\relax IEEE, 2016, pp. 1--6.

\bibitem{noh2017mmwave}
G.~Noh, J.~Kim, H.~S. Chung, B.~Hui, Y.-M. Choi, and I.~Kim, ``mmwave-based mobile backhaul transceiver for high speed train communication systems,'' in \emph{2017 IEEE Globecom Workshops (GC Wkshps)}.\hskip 1em plus 0.5em minus 0.4em\relax IEEE, 2017, pp. 1--5.

\bibitem{cui2018optimal}
Y.~Cui, X.~Fang, Y.~Fang, and M.~Xiao, ``Optimal nonuniform steady mmwave beamforming for high-speed railway,'' \emph{IEEE Transactions on Vehicular Technology}, vol.~67, no.~5, pp. 4350--4358, 2018.

\bibitem{zhang2021design}
Q.~Zhang, X.~Wang, Z.~Li, and Z.~Wei, ``Design and performance evaluation of joint sensing and communication integrated system for 5g mmwave enabled cavs,'' \emph{IEEE Journal of Selected Topics in Signal Processing}, vol.~15, no.~6, pp. 1500--1514, 2021.

\bibitem{wu2017signal}
Z.~Wu, F.~Zhao, and X.~Liu, ``Signal space diversity aided dynamic multiplexing for embb and urllc traffics,'' in \emph{2017 3rd IEEE International Conference on Computer and Communications (ICCC)}.\hskip 1em plus 0.5em minus 0.4em\relax IEEE, 2017, pp. 1396--1400.

\bibitem{gashema2018spatial}
G.~Gashema, S.~Bhardwaj, A.~Abdukhakimov, D.-S. Kim, and J.-M. Lee, ``Spatial diversity to support urllc through unlicensed spectrum in industrial wireless network systems,'' in \emph{2018 IEEE 3rd International Conference on Communication and Information Systems (ICCIS)}.\hskip 1em plus 0.5em minus 0.4em\relax IEEE, 2018, pp. 141--145.

\bibitem{gao2016improving}
Y.~Gao, T.~Yang, and B.~Hu, ``Improving the transmission reliability in smart factory through spatial diversity with arq,'' in \emph{2016 IEEE/CIC International Conference on Communications in China (ICCC)}.\hskip 1em plus 0.5em minus 0.4em\relax IEEE, 2016, pp. 1--5.

\bibitem{ozyurt2018performance}
S.~{\"O}zyurt and O.~Kucur, ``Performance analysis of maximal ratio combining with transmit antenna selection and signal space diversity under exponential antenna correlation,'' \emph{IET Communications}, vol.~12, no.~5, pp. 612--619, 2018.

\bibitem{panigrahi2017feasibility}
S.~R. Panigrahi, N.~Bjorsell, and M.~Bengtsson, ``Feasibility of large antenna arrays towards low latency ultra reliable communication,'' in \emph{2017 IEEE International Conference on Industrial Technology (ICIT)}.\hskip 1em plus 0.5em minus 0.4em\relax IEEE, 2017, pp. 1289--1294.

\bibitem{kang2022probabilistic}
W.~Kang, ``A probabilistic shaping scheme for mimo systems with signal space diversity,'' in \emph{2022 IEEE Wireless Communications and Networking Conference (WCNC)}.\hskip 1em plus 0.5em minus 0.4em\relax IEEE, 2022, pp. 251--255.

\bibitem{peon2017applying}
P.~G. Pe{\'o}n, E.~Uhlemann, W.~Steiner, and M.~Bjorkman, ``Applying time diversity for improved reliability in a real-time heterogeneous mac protocol,'' in \emph{2017 IEEE 85th Vehicular Technology Conference (VTC Spring)}.\hskip 1em plus 0.5em minus 0.4em\relax IEEE, 2017, pp. 1--7.

\bibitem{kojima2015improved}
T.~Kojima and Y.~Takanashi, ``An improved time diversity combining for helicopter satellite communications,'' in \emph{2015 International Conference on Advanced Technologies for Communications (ATC)}.\hskip 1em plus 0.5em minus 0.4em\relax IEEE, 2015, pp. 6--9.

\bibitem{sato2018accurate}
D.~Sato and T.~Kojima, ``An accurate time diversity combining with a novel channel estimation for helicopter satellite communications,'' in \emph{2018 International Conference on Advanced Technologies for Communications (ATC)}.\hskip 1em plus 0.5em minus 0.4em\relax IEEE, 2018, pp. 89--93.

\bibitem{rao2018packet}
J.~Rao and S.~Vrzic, ``Packet duplication for urllc in 5g: Architectural enhancements and performance analysis,'' \emph{IEEE Network}, vol.~32, no.~2, pp. 32--40, 2018.

\bibitem{centenaro2019system}
M.~Centenaro, D.~Laselva, J.~Steiner, K.~Pedersen, and P.~Mogensen, ``System-level study of data duplication enhancements for {5G} downlink {URLLC},'' \emph{IEEE Access}, 2019.

\bibitem{michalopoulos2019data}
D.~S. Michalopoulos and V.~Pauli, ``Data duplication for high reliability: A protocol-level simulation assessment,'' in \emph{ICC 2019-2019 IEEE International Conference on Communications (ICC)}.\hskip 1em plus 0.5em minus 0.4em\relax IEEE, 2019, pp. 1--7.

\bibitem{boyd2019non}
C.~Boyd, R.~Kotaba, O.~Tirkkonen, and P.~Popovski, ``Non-orthogonal contention-based access for urllc devices with frequency diversity,'' in \emph{2019 IEEE 20th International Workshop on Signal Processing Advances in Wireless Communications (SPAWC)}.\hskip 1em plus 0.5em minus 0.4em\relax IEEE, 2019, pp. 1--5.

\bibitem{senthil2019improved}
B.~Senthil, ``An improved frequency diversity scheme for ofdm systems with frequency offset,'' in \emph{Innovations in Electronics and Communication Engineering}.\hskip 1em plus 0.5em minus 0.4em\relax Springer, 2019, pp. 287--297.

\bibitem{DBLP:journals/corr/SheYQ16a}
\BIBentryALTinterwordspacing
C.~She, C.~Yang, and T.~Q.~S. Quek, ``Uplink transmission design with massive machine type devices in tactile internet,'' \emph{CoRR}, vol. abs/1610.02816, 2016. [Online]. Available: \url{http://arxiv.org/abs/1610.02816}
\BIBentrySTDinterwordspacing

\bibitem{zhao2019joint}
S.~Zhao, Y.~Wang, Y.~Xie, and D.~Xu, ``Joint time-frequency diversity based uplink grant-free transmission scheme for urllc,'' in \emph{2019 11th International Conference on Wireless Communications and Signal Processing (WCSP)}.\hskip 1em plus 0.5em minus 0.4em\relax IEEE, 2019, pp. 1--6.

\bibitem{wu2019urllc}
M.~Wu, Y.~Zhong, G.~Wang, C.~She, X.~Ge, and H.-C. Chao, ``Urllc in large-scale wireless networks with time and frequency diversities,'' in \emph{2019 IEEE Global Communications Conference (GLOBECOM)}.\hskip 1em plus 0.5em minus 0.4em\relax IEEE, 2019, pp. 1--6.

\bibitem{JNielsen2}
J.~J. Nielsen, R.~Liu, and P.~Popovski, ``Ultra-reliable low latency communication using interface diversity,'' \emph{IEEE Transactions on Communications}, vol.~66, no.~3, pp. 1322--1334, March 2018.

\bibitem{desai2020wideband}
A.~Desai, ``Wideband antennas with polarization diversity for wireless applications,'' Ph.D. dissertation, Colorado School of Mines, 2020.

\bibitem{kurma2022urllc}
S.~Kurma, P.~K. Sharma, K.~Singh, S.~Mumtaz, and C.-P. Li, ``Urllc based cooperative industrial iot networks with non-linear energy harvesting,'' \emph{IEEE Transactions on Industrial Informatics}, 2022.

\bibitem{ADestounis}
A.~Destounis, G.~S. Paschos, J.~Arnau, and M.~Kountouris, ``Scheduling urllc users with reliable latency guarantees,'' in \emph{2018 16th International Symposium on Modeling and Optimization in Mobile, Ad Hoc, and Wireless Networks (WiOpt)}.\hskip 1em plus 0.5em minus 0.4em\relax IEEE, 2018, pp. 1--8.

\bibitem{anand2020jointJournal}
A.~Anand, G.~De~Veciana, and S.~Shakkottai, ``Joint scheduling of urllc and embb traffic in 5g wireless networks,'' \emph{IEEE/ACM Transactions On Networking}, vol.~28, no.~2, pp. 477--490, 2020.

\bibitem{GPocovi}
G.~Pocovi, K.~I. Pedersen, and P.~Mogensen, ``Joint link adaptation and scheduling for {5G} ultra-reliable low-latency communications,'' \emph{Ieee Access}, vol.~6, pp. 28\,912--28\,922, 2018.

\bibitem{esswie2018opportunistic}
A.~A. Esswie and K.~I. Pedersen, ``Opportunistic spatial preemptive scheduling for urllc and embb coexistence in multi-user {5G} networks,'' \emph{Ieee Access}, vol.~6, pp. 38\,451--38\,463, 2018.

\bibitem{she2017radio}
C.~She, C.~Yang, and T.~Q. Quek, ``Radio resource management for ultra-reliable and low-latency communications,'' \emph{IEEE Communications Magazine}, vol.~55, no.~6, pp. 72--78, 2017.

\bibitem{shariatmadari2016optimized}
H.~Shariatmadari, S.~Iraji, Z.~Li, M.~A. Uusitalo, and R.~J{\"a}ntti, ``Optimized transmission and resource allocation strategies for ultra-reliable communications,'' in \emph{2016 IEEE 27th Annual International Symposium on Personal, Indoor, and Mobile Radio Communications (PIMRC)}.\hskip 1em plus 0.5em minus 0.4em\relax IEEE, 2016, pp. 1--6.

\bibitem{sun2017energy}
C.~Sun, C.~She, and C.~Yang, ``Energy-efficient resource allocation for ultra-reliable and low-latency communications,'' in \emph{GLOBECOM 2017-2017 IEEE Global Communications Conference}.\hskip 1em plus 0.5em minus 0.4em\relax IEEE, 2017, pp. 1--6.

\bibitem{hytonen2017coordinated}
V.~Hyt{\"o}nen, Z.~Li, B.~Soret, and V.~Nurmela, ``Coordinated multi-cell resource allocation for {5G} ultra-reliable low latency communications,'' in \emph{2017 European Conference on Networks and Communications (EuCNC)}.\hskip 1em plus 0.5em minus 0.4em\relax IEEE, 2017, pp. 1--5.

\bibitem{anand2018resource}
A.~Anand and G.~de~Veciana, ``Resource allocation and harq optimization for urllc traffic in {5G} wireless networks,'' \emph{IEEE Journal on Selected Areas in Communications}, vol.~36, no.~11, pp. 2411--2421, 2018.

\bibitem{AAzari19}
A.~Azari, M.~Ozger, and C.~Cavdar, ``Risk-aware resource allocation for urllc: Challenges and strategies with machine learning,'' \emph{IEEE Communications Magazine}, vol.~57, no.~3, pp. 42--48, 2019.

\bibitem{cheng2020robust}
J.~Cheng, C.~Shen, and S.~Xia, ``Robust urllc packet scheduling of ofdm systems,'' in \emph{2020 IEEE Wireless Communications and Networking Conference (WCNC)}.\hskip 1em plus 0.5em minus 0.4em\relax IEEE, 2020, pp. 1--6.

\bibitem{zhu2019priority}
L.~Zhu, L.~Feng, Z.~Yang, W.~Li, and Q.~Ou, ``Priority-based urllc uplink resource scheduling for smart grid neighborhood area network,'' in \emph{2019 IEEE International Conference on Energy Internet (ICEI)}.\hskip 1em plus 0.5em minus 0.4em\relax IEEE, 2019, pp. 510--515.

\bibitem{wu2020uplink}
S.~Wu, Z.~Wang, Z.~Li, Z.~WeiJun, W.~Shao, B.~Ma, S.~Yao, and Y.~Wang, ``Uplink resource allocation based on short block-length regime in heterogeneous cellular networks for smart grid,'' in \emph{International Conference on Innovative Mobile and Internet Services in Ubiquitous Computing}.\hskip 1em plus 0.5em minus 0.4em\relax Springer, 2020, pp. 213--224.

\bibitem{yu2020joint}
M.~Yu, A.~Tang, X.~Wang, and C.~Han, ``Joint scheduling and power allocation for 6g terahertz mesh networks,'' in \emph{2020 International Conference on Computing, Networking and Communications (ICNC)}.\hskip 1em plus 0.5em minus 0.4em\relax IEEE, 2020, pp. 631--635.

\bibitem{Awolf}
A.~{Wolf}, P.~{Schulz}, M.~{Dörpinghaus}, J.~C.~S. {Santos Filho}, and G.~{Fettweis}, ``How reliable and capable is multi-connectivity?'' \emph{IEEE Transactions on Communications}, vol.~67, no.~2, pp. 1506--1520, Feb 2019.

\bibitem{mahmood01}
N.~H. Mahmood, A.~Karimi, G.~Berardinelli, K.~I. Pedersen, and D.~Laselva, ``On the resource utilization of multi-connectivity transmission for {URLLC} services in {5G} new radio,'' \emph{arXiv preprint arXiv:1904.07963}, 2019.

\bibitem{THobler}
T.~H\"o{\ss}ler, M.~Simsek, and G.~P. Fettweis, ``Mission reliability for urllc in wireless networks,'' \emph{IEEE Communications Letters}, vol.~22, no.~11, pp. 2350--2353, Nov 2018.

\bibitem{nhmahmood1}
N.~H. {Mahmood}, M.~{Lopez}, D.~{Laselva}, K.~{Pedersen}, and G.~{Berardinelli}, ``Reliability oriented dual connectivity for urllc services in {5G} new radio,'' in \emph{2018 15th International Symposium on Wireless Communication Systems (ISWCS)}, Aug 2018, pp. 1--6.

\bibitem{CShe}
C.~{She}, Z.~{Chen}, C.~{Yang}, T.~Q.~S. {Quek}, Y.~{Li}, and B.~{Vucetic}, ``Improving network availability of ultra-reliable and low-latency communications with multi-connectivity,'' \emph{IEEE Transactions on Communications}, vol.~66, no.~11, pp. 5482--5496, Nov 2018.

\bibitem{THolber2}
T.~H\"o{\ss}ler, M.~{Simsek}, and G.~P. {Fettweis}, ``Joint analysis of channel availability and time-based reliability metrics for wireless urllc,'' in \emph{2018 IEEE Global Communications Conference (GLOBECOM)}, Dec 2018.

\bibitem{xue2024cooperativeDeep}
J.~Xue, K.~Yu, T.~Zhang, H.~Zhou, L.~Zhao, and X.~Shen, ``Cooperative deep reinforcement learning enabled power allocation for packet duplication urllc in multi-connectivity vehicular networks,'' \emph{IEEE Transactions on Mobile Computing}, 2024.

\bibitem{andrews2001providing}
M.~Andrews, K.~Kumaran, K.~Ramanan, A.~Stolyar, P.~Whiting, and R.~Vijayakumar, ``Providing quality of service over a shared wireless link,'' \emph{IEEE Communications magazine}, vol.~39, no.~2, pp. 150--154, 2001.

\bibitem{basukala2009performance}
R.~Basukala, H.~M. Ramli, and K.~Sandrasegaran, ``Performance analysis of exp/pf and m-lwdf in downlink 3gpp lte system,'' in \emph{2009 First Asian Himalayas International Conference on Internet}.\hskip 1em plus 0.5em minus 0.4em\relax IEEE, 2009, pp. 1--5.

\bibitem{piro2011two}
G.~Piro, L.~A. Grieco, G.~Boggia, R.~Fortuna, and P.~Camarda, ``Two-level downlink scheduling for real-time multimedia services in lte networks,'' \emph{IEEE Transactions on Multimedia}, vol.~13, no.~5, pp. 1052--1065, 2011.

\bibitem{MIHossain19}
M.~I. Husain, M.~E. Haque, and F.~Tariq, ``An efficient packet scheduling algorithm for urllc systems,'' in \emph{2020 International Conference on UK-China Emerging Technologies (UCET)}, 2020, pp. 1--4.

\bibitem{amjad2018latency}
Z.~Amjad, A.~Sikora, J.-P. Lauffenburger, and B.~Hilt, ``Latency reduction in narrowband 4g lte networks,'' in \emph{2018 15th International Symposium on Wireless Communication Systems (ISWCS)}.\hskip 1em plus 0.5em minus 0.4em\relax IEEE, 2018, pp. 1--5.

\bibitem{arnjad2018latency}
Z.~Arnjad, A.~Sikora, B.~Hilt, and J.-P. Lauffenburger, ``Latency reduction for narrowband lte with semi-persistent scheduling,'' in \emph{2018 IEEE 4th International Symposium on Wireless Systems within the International Conferences on Intelligent Data Acquisition and Advanced Computing Systems (IDAACS-SWS)}.\hskip 1em plus 0.5em minus 0.4em\relax IEEE, 2018, pp. 196--198.

\bibitem{IAshraf2016ETFA}
S.~A. Ashraf, I.~Aktas, E.~Eriksson, K.~W. Helmersson, and J.~Ansari, ``Ultra-reliable and low-latency communication for wireless factory automation: From lte to 5g,'' in \emph{2016 IEEE 21st International Conference on Emerging Technologies and Factory Automation (ETFA)}.\hskip 1em plus 0.5em minus 0.4em\relax IEEE, 2016, pp. 1--8.

\bibitem{haque2023survey}
M.~E. Haque, F.~Tariq, M.~R. Khandaker, K.-K. Wong, and Y.~Zhang, ``A survey of scheduling in 5g urllc and outlook for emerging 6g systems,'' \emph{IEEE Access}, 2023.

\bibitem{zhao2024jointBeam}
X.~Zhao and Y.-J.~A. Zhang, ``Joint beamforming and scheduling for integrated sensing and communication systems in urllc: A pomdp approach,'' \emph{IEEE Transactions on Communications}, 2024.

\bibitem{serror2015channel}
M.~Serror, C.~Dombrowski, K.~Wehrle, and J.~Gross, ``Channel coding versus cooperative arq: Reducing outage probability in ultra-low latency wireless communications,'' in \emph{2015 IEEE Globecom Workshops (GC Wkshps)}.\hskip 1em plus 0.5em minus 0.4em\relax IEEE, 2015, pp. 1--6.

\bibitem{shariatmadari2015analysis}
H.~Shariatmadari, S.~Iraji, and R.~J{\"a}ntti, ``Analysis of transmission methods for ultra-reliable communications,'' in \emph{2015 IEEE 26th Annual International Symposium on Personal, Indoor, and Mobile Radio Communications (PIMRC)}.\hskip 1em plus 0.5em minus 0.4em\relax IEEE, 2015, pp. 2303--2308.

\bibitem{hwang2017adaptive}
J.~Hwang, H.~Saki, and M.~Shikh-Bahaei, ``Adaptive modulation and coding and cooperative arq in a cognitive radio system,'' in \emph{2017 International Conference on Advances in Computing, Communications and Informatics (ICACCI)}.\hskip 1em plus 0.5em minus 0.4em\relax IEEE, 2017, pp. 310--315.

\bibitem{chiu2018cross}
H.-L. Chiu and S.-H. Wu, ``Cross-layer performance analysis of cooperative arq with opportunistic multi-point relaying in mobile networks,'' \emph{IEEE Transactions on Wireless Communications}, vol.~17, no.~6, pp. 4191--4205, 2018.

\bibitem{serror2016performance}
M.~Serror, Y.~Hu, C.~Dombrowski, K.~Wehrle, and J.~Gross, ``Performance analysis of cooperative arq systems for wireless industrial networks,'' in \emph{2016 IEEE 17th International Symposium on A World of Wireless, Mobile and Multimedia Networks (WoWMoM)}.\hskip 1em plus 0.5em minus 0.4em\relax IEEE, 2016, pp. 1--4.

\bibitem{dianati2006node}
M.~Dianati, X.~Ling, K.~Naik, and X.~Shen, ``A node-cooperative arq scheme for wireless ad hoc networks,'' \emph{IEEE Transactions on Vehicular Technology}, vol.~55, no.~3, pp. 1032--1044, 2006.

\bibitem{yu2006cooperative}
G.~Yu, Z.~Zhang, and P.~Qiu, ``Cooperative arq in wireless networks: Protocols description and performance analysis,'' in \emph{2006 IEEE International Conference on Communications}, vol.~8.\hskip 1em plus 0.5em minus 0.4em\relax IEEE, 2006, pp. 3608--3614.

\bibitem{selmi2016efficient}
A.~Selmi, M.~Siala, and H.~Boujemaa, ``Efficient cross-layer design for mimo-harq wireless systems,'' in \emph{2016 International Symposium on Signal, Image, Video and Communications (ISIVC)}.\hskip 1em plus 0.5em minus 0.4em\relax IEEE, 2016, pp. 337--342.

\bibitem{chen2018cross}
Y.-F. Chen, S.-M. Tseng, C.-H. Shen, and M.-S. He, ``Cross layer 1, 2 and 5 resource allocation in uplink turbo-coded harq based ofdma video transmission systems,'' \emph{Wireless Personal Communications}, vol.~98, no.~2, pp. 1997--2008, 2018.

\bibitem{avranas2019throughput}
A.~Avranas, M.~Kountouris, and P.~Ciblat, ``Throughput maximization and ir-harq optimization for urllc traffic in {5G} systems,'' in \emph{ICC 2019-2019 IEEE International Conference on Communications (ICC)}.\hskip 1em plus 0.5em minus 0.4em\relax IEEE, 2019, pp. 1--6.

\bibitem{deghel2018joint}
M.~Deghel, S.~E. Elayoubi, A.~Galindo-Serrano, and R.~Visoz, ``Joint optimization of link adaptation and harq retransmissions for urllc services,'' in \emph{2018 25th International Conference on Telecommunications (ICT)}.\hskip 1em plus 0.5em minus 0.4em\relax IEEE, 2018, pp. 21--26.

\bibitem{imamura2017low}
Y.~Imamura, D.~Muramatsu, Y.~Kishiyama, and K.~Higuchi, ``Low latency hybrid arq method using channel state information before channel decoding,'' in \emph{2017 23rd Asia-Pacific Conference on Communications (APCC)}.\hskip 1em plus 0.5em minus 0.4em\relax IEEE, 2017, pp. 1--6.

\bibitem{she2016cross}
C.~She, C.~Yang, and T.~Q. Quek, ``Cross-layer transmission design for tactile internet,'' in \emph{2016 IEEE Global Communications Conference (GLOBECOM)}.\hskip 1em plus 0.5em minus 0.4em\relax IEEE, 2016, pp. 1--6.

\bibitem{devassy2019reliable}
R.~Devassy, G.~Durisi, G.~C. Ferrante, O.~Simeone, and E.~Uysal, ``Reliable transmission of short packets through queues and noisy channels under latency and peak-age violation guarantees,'' \emph{IEEE Journal on Selected Areas in Communications}, vol.~37, no.~4, pp. 721--734, 2019.

\bibitem{avranas2018energy}
A.~Avranas, M.~Kountouris, and P.~Ciblat, ``Energy-latency tradeoff in ultra-reliable low-latency communication with retransmissions,'' \emph{IEEE Journal on Selected Areas in Communications}, vol.~36, no.~11, pp. 2475--2485, 2018.

\bibitem{vora2018effective}
A.~Vora and K.-D. Kang, ``Effective {5G} wireless downlink scheduling and resource allocation in cyber-physical systems,'' \emph{Technologies}, vol.~6, no.~4, p. 105, 2018.

\bibitem{zhang2019statistical}
X.~Zhang, J.~Wang, and H.~V. Poor, ``Statistical qos-driven energy-efficiency optimization for urllc over {5G} mobile wireless networks in the finite blocklength regime,'' in \emph{2019 53rd Annual Conference on Information Sciences and Systems (CISS)}.\hskip 1em plus 0.5em minus 0.4em\relax IEEE, 2019, pp. 1--6.

\bibitem{femenias2017downlink}
G.~Femenias, F.~Riera-Palou, X.~Mestre, and J.~J. Olmos, ``Downlink scheduling and resource allocation for 5g mimo-multicarrier: Ofdm vs fbmc/oqam,'' \emph{IEEE access}, vol.~5, pp. 13\,770--13\,786, 2017.

\bibitem{JHamamreh1}
J.~M. Hamamreh, Z.~E. Ankarali, and H.~Arslan, ``Cp-less ofdm with alignment signals for enhancing spectral efficiency, reducing latency, and improving {PHY} security of {5G} services,'' \emph{IEEE Access}, vol.~6, pp. 63\,649--63\,663, 2018.

\bibitem{pradhan2023blocklength}
A.~Pradhan, S.~Das, and M.~J. Piran, ``Blocklength optimization and power allocation for energy-efficient and secure urllc in industrial iot,'' \emph{IEEE Internet of Things Journal}, 2023.

\bibitem{xiang2020secure}
Z.~Xiang, W.~Yang, Y.~Cai, Z.~Ding, and Y.~Song, ``Secure transmission design in harq assisted cognitive noma networks,'' \emph{IEEE Transactions on Information Forensics and Security}, vol.~15, pp. 2528--2541, 2020.

\bibitem{dizdar2022rate}
O.~Dizdar and B.~Clerckx, ``Rate-splitting multiple access for communications and jamming in multi-antenna multi-carrier cognitive radio systems,'' \emph{IEEE Transactions on Information Forensics and Security}, vol.~17, pp. 628--643, 2022.

\bibitem{xu2021quantum}
D.~Xu and P.~Ren, ``Quantum learning based nonrandom superimposed coding for secure wireless access in 5g urllc,'' \emph{IEEE Transactions on Information Forensics and Security}, vol.~16, pp. 2429--2444, 2021.

\bibitem{dos2021rate}
E.~J. Dos~Santos, R.~D. Souza, and J.~L. Rebelatto, ``Rate-splitting multiple access for urllc uplink in physical layer network slicing with embb,'' \emph{IEEE Access}, vol.~9, pp. 163\,178--163\,187, 2021.

\bibitem{liu2023networkslicing}
Y.~Liu, B.~Clerckx, and P.~Popovski, ``Network slicing for embb, urllc, and mmtc: An uplink rate-splitting multiple access approach,'' \emph{IEEE Transactions on Wireless Communications}, 2023.

\bibitem{yang2023networkslicing}
P.~Yang, X.~Xi, T.~Q. Quek, J.~Chen, C.~Xianbin, and W.~Dapeng, ``Network slicing for urllc,'' \emph{Ultra-Reliable and Low-Latency Communications (URLLC) Theory and Practice: Advances in 5G and Beyond}, pp. 215--239, 2023.

\bibitem{liu2018massive}
L.~Liu and W.~Yu, ``Massive connectivity with massive mimo—part i: Device activity detection and channel estimation,'' \emph{IEEE Transactions on Signal Processing}, vol.~66, no.~11, pp. 2933--2946, 2018.

\bibitem{seok2019secure}
B.~Seok, J.~C.~S. Sicato, T.~Erzhena, C.~Xuan, Y.~Pan, and J.~H. Park, ``Secure d2d communication for 5g iot network based on lightweight cryptography,'' \emph{Applied Sciences}, vol.~10, no.~1, p. 217, 2019.

\bibitem{mustafa2020lightweight}
I.~Mustafa and C.~Y. Huang, ``Lightweight cryptographic urllc for 5g-v2x,'' in \emph{2020 International Symposium on Networks, Computers and Communications (ISNCC)}.\hskip 1em plus 0.5em minus 0.4em\relax IEEE, 2020, pp. 1--6.

\bibitem{qu2023adversarial}
A.~Qu, Y.~Tang, and W.~Ma, ``Adversarial attacks on deep reinforcement learning-based traffic signal control systems with colluding vehicles,'' \emph{ACM Transactions on Intelligent Systems and Technology}, vol.~14, no.~6, pp. 1--22, 2023.

\bibitem{xu2023environment}
H.~Xu, ``Environment poisoning in reinforcement learning: attacks and resilience,'' Ph.D. dissertation, Nanyang Technological University, 2023.

\bibitem{qiaoben2024understanding}
Y.~Qiaoben, C.~Ying, X.~Zhou, H.~Su, J.~Zhu, and B.~Zhang, ``Understanding adversarial attacks on observations in deep reinforcement learning,'' \emph{Science China Information Sciences}, vol.~67, no.~5, pp. 1--15, 2024.

\bibitem{huang2024federated}
W.~Huang, M.~Ye, Z.~Shi, G.~Wan, H.~Li, B.~Du, and Q.~Yang, ``Federated learning for generalization, robustness, fairness: A survey and benchmark,'' \emph{IEEE Transactions on Pattern Analysis and Machine Intelligence}, 2024.

\bibitem{kumar2023impact}
K.~N. Kumar, C.~K. Mohan, and L.~R. Cenkeramaddi, ``The impact of adversarial attacks on federated learning: A survey,'' \emph{IEEE Transactions on Pattern Analysis and Machine Intelligence}, 2023.

\bibitem{strodthoff2019enhanced}
N.~Strodthoff, B.~G{\"o}ktepe, T.~Schierl, C.~Hellge, and W.~Samek, ``Enhanced machine learning techniques for early harq feedback prediction in 5g,'' \emph{IEEE Journal on Selected Areas in Communications}, vol.~37, no.~11, pp. 2573--2587, 2019.

\bibitem{zhang2019machine}
J.~Zhang, X.~Xu, K.~Zhang, B.~Zhang, X.~Tao, and P.~Zhang, ``Machine learning based flexible transmission time interval scheduling for embb and urllc coexistence scenario,'' \emph{IEEE Access}, vol.~7, pp. 65\,811--65\,820, 2019.

\bibitem{esswie2020online}
A.~A. Esswie, K.~I. Pedersen, and P.~E. Mogensen, ``Online radio pattern optimization based on dual reinforcement-learning approach for 5g urllc networks,'' \emph{IEEE Access}, vol.~8, pp. 132\,922--132\,936, 2020.

\bibitem{weinand2019supervised}
A.~Weinand, R.~Sattiraju, M.~Karrenbauer, and H.~D. Schotten, ``Supervised learning for physical layer based message authentication in urllc scenarios,'' in \emph{2019 IEEE 90th vehicular technology conference (VTC2019-fall)}.\hskip 1em plus 0.5em minus 0.4em\relax IEEE, 2019, pp. 1--7.

\bibitem{zhu2019supervised}
G.~Zhu, J.~Zan, Y.~Yang, and X.~Qi, ``A supervised learning based qos assurance architecture for 5g networks,'' \emph{IEEE Access}, vol.~7, pp. 43\,598--43\,606, 2019.

\bibitem{abdelsadek2020resource}
M.~Y. Abdelsadek, Y.~Gadallah, and M.~H. Ahmed, ``Resource allocation of urllc and embb mixed traffic in 5g networks: A deep learning approach,'' in \emph{GLOBECOM 2020-2020 IEEE Global Communications Conference}.\hskip 1em plus 0.5em minus 0.4em\relax IEEE, 2020, pp. 1--6.

\bibitem{zhao2023energy}
H.~Zhao, B.~Xu, H.~Huang, Q.~Wang, C.~Zhu, and G.~Gui, ``Energy efficient power allocation for ultra-reliable and low-latency communications via unsupervised learning,'' \emph{IET Communications}, 2023.

\bibitem{sun2023unsupervised}
C.~Sun, C.~She, and C.~Yang, ``Unsupervised deep learning for optimizing wireless systems with instantaneous and statistic constraints,'' \emph{Ultra-Reliable and Low-Latency Communications (URLLC) Theory and Practice: Advances in 5G and Beyond}, pp. 85--117, 2023.

\bibitem{sun2019unsupervised}
C.~Sun and C.~Yang, ``Unsupervised deep learning for ultra-reliable and low-latency communications,'' in \emph{2019 IEEE Global Communications Conference (GLOBECOM)}.\hskip 1em plus 0.5em minus 0.4em\relax IEEE, 2019, pp. 1--6.

\bibitem{orim2019cluster}
P.~Orim, N.~Ventura, and J.~Mwangama, ``Cluster-based random access scheme for 5g urllc,'' in \emph{2019 IEEE International Conference on Advanced Networks and Telecommunications Systems (ANTS)}.\hskip 1em plus 0.5em minus 0.4em\relax IEEE, 2019, pp. 1--6.

\bibitem{li2020deep}
J.~Li and X.~Zhang, ``Deep reinforcement learning-based joint scheduling of embb and urllc in 5g networks,'' \emph{IEEE Wireless Communications Letters}, vol.~9, no.~9, pp. 1543--1546, 2020.

\bibitem{liu2023machine}
Y.~Liu, Y.~Deng, A.~Nallanathan, and J.~Yuan, ``Machine learning for 6g enhanced ultra-reliable and low-latency services,'' \emph{IEEE Wireless Communications}, vol.~30, no.~2, pp. 48--54, 2023.

\bibitem{huang2020deep}
Y.~Huang, S.~Li, C.~Li, Y.~T. Hou, and W.~Lou, ``A deep-reinforcement-learning-based approach to dynamic embb/urllc multiplexing in 5g nr,'' \emph{IEEE Internet of Things Journal}, vol.~7, no.~7, pp. 6439--6456, 2020.

\bibitem{ali2021federated}
R.~Ali, Y.~B. Zikria, S.~Garg, A.~K. Bashir, M.~S. Obaidat, and H.~S. Kim, ``A federated reinforcement learning framework for incumbent technologies in beyond 5g networks,'' \emph{IEEE Network}, vol.~35, no.~4, pp. 152--159, 2021.

\bibitem{pan2022asynchronous}
C.~Pan, Z.~Wang, H.~Liao, Z.~Zhou, X.~Wang, M.~Tariq, and S.~Al-Otaibi, ``Asynchronous federated deep reinforcement learning-based urllc-aware computation offloading in space-assisted vehicular networks,'' \emph{IEEE Transactions on Intelligent Transportation Systems}, 2022.

\bibitem{zhang2022federated}
H.~Zhang, H.~Zhou, and M.~Erol-Kantarci, ``Federated deep reinforcement learning for resource allocation in o-ran slicing,'' in \emph{GLOBECOM 2022-2022 IEEE Global Communications Conference}.\hskip 1em plus 0.5em minus 0.4em\relax IEEE, 2022, pp. 958--963.

\bibitem{almarshed2023swift}
S.~Almarshed, D.~Triantafyllopoulou, and K.~Moessner, ``Swift harq based on machine learning for latency minimization in urllc,'' \emph{IEEE access}, vol.~11, pp. 113\,422--113\,436, 2023.

\bibitem{hendaoui2024dynamic}
S.~Hendaoui, F.~Hendaoui, and N.~Zangar, ``Dynamic proactive-reactive scheduling for urllc in 5g: Leveraging xgboost and network virtualization,'' \emph{Physical Communication}, p. 102553, 2024.

\bibitem{ishteyaq2024unleashing}
I.~Ishteyaq, K.~Muzaffar, N.~Shafi, and M.~A. Alathbah, ``Unleashing the power of tomorrow: Exploration of next frontier with 6g networks and cutting edge technologies,'' \emph{IEEE Access}, 2024.

\bibitem{robaglia2024deep}
B.-M. Robaglia, M.~Coupechoux, and D.~Tsilimantos, ``Deep reinforcement learning for uplink scheduling in noma-urllc networks,'' \emph{IEEE Transactions on Machine Learning in Communications and Networking}, 2024.

\bibitem{alsenwi2024distributed}
M.~Alsenwi, E.~Lagunas, and S.~Chatzinotas, ``Distributed learning framework for embb-urllc multiplexing in open radio access networks,'' \emph{IEEE Transactions on Network and Service Management}, 2024.

\bibitem{hasan2024federated}
M.~K. Hasan, A.~A. Habib, S.~Islam, N.~Safie, T.~M. Ghazal, M.~A. Khan, A.~I. Alzahrani, N.~Alalwan, S.~Kadry, and A.~Masood, ``Federated learning enables 6 g communication technology: Requirements, applications, and integrated with intelligence framework,'' \emph{Alexandria Engineering Journal}, vol.~91, pp. 658--668, 2024.

\bibitem{bonati2021intelligence}
L.~Bonati, S.~D'Oro, M.~Polese, S.~Basagni, and T.~Melodia, ``Intelligence and learning in o-ran for data-driven nextg cellular networks,'' \emph{IEEE Communications Magazine}, vol.~59, no.~10, pp. 21--27, 2021.

\bibitem{JPark2020}
J.~Park, S.~Samarakoon, H.~Shiri, M.~K. Abdel-Aziz, T.~Nishio, A.~Elgabli, and M.~Bennis, ``Extreme urllc: Vision, challenges, and key enablers,'' \emph{arXiv preprint arXiv:2001.09683}, 2020.

\bibitem{zhang2023holographic}
H.~Zhang, H.~Zhang, B.~Di, and L.~Song, ``Holographic integrated sensing and communications: Principles, technology, and implementation,'' \emph{IEEE Communications Magazine}, vol.~61, no.~5, pp. 83--89, 2023.

\bibitem{shafie2022terahertz}
A.~Shafie, N.~Yang, C.~Han, J.~M. Jornet, M.~Juntti, and T.~K{\"u}rner, ``Terahertz communications for 6g and beyond wireless networks: Challenges, key advancements, and opportunities,'' \emph{IEEE Network}, vol.~37, no.~3, pp. 162--169, 2022.

\bibitem{chataut20246g}
R.~Chataut, M.~Nankya, and R.~Akl, ``6g networks and the ai revolution—exploring technologies, applications, and emerging challenges,'' \emph{Sensors}, vol.~24, no.~6, p. 1888, 2024.

\bibitem{khan2021blockchain}
A.~H. Khan, N.~U. Hassan, C.~Yuen, J.~Zhao, D.~Niyato, Y.~Zhang, and H.~V. Poor, ``Blockchain and 6g: The future of secure and ubiquitous communication,'' \emph{IEEE Wireless Communications}, vol.~29, no.~1, pp. 194--201, 2021.

\bibitem{hasan2024blockchain}
K.~M.~B. Hasan, M.~Sajid, M.~A. Lapina, M.~Shahid, and K.~Kotecha, ``Blockchain technology meets 6 g wireless networks: A systematic survey,'' \emph{Alexandria Engineering Journal}, vol.~92, pp. 199--220, 2024.

\bibitem{liu2024jointcoop}
Z.~Liu, J.~Zhang, Z.~Liu, D.~W.~K. Ng, and B.~Ai, ``Joint cooperative clustering and power control for energy-efficient cell-free xl-mimo with multi-agent reinforcement learning,'' \emph{IEEE Transactions on Communications}, 2024.

\bibitem{shen2023towardimmersive}
X.~Shen, J.~Gao, M.~Li, C.~Zhou, S.~Hu, M.~He, and W.~Zhuang, ``Toward immersive communications in 6g,'' \emph{Frontiers in Computer Science}, vol.~4, p. 1068478, 2023.

\bibitem{khan2022digitaltwin}
L.~U. Khan, W.~Saad, D.~Niyato, Z.~Han, and C.~S. Hong, ``Digital-twin-enabled 6g: Vision, architectural trends, and future directions,'' \emph{IEEE Communications Magazine}, vol.~60, no.~1, pp. 74--80, 2022.

\bibitem{baccour2023zero}
E.~Baccour, M.~S. Allahham, A.~Erbad, A.~Mohamed, A.~R. Hussein, and M.~Hamdi, ``Zero touch realization of pervasive artificial intelligence as a service in 6g networks,'' \emph{IEEE Communications Magazine}, vol.~61, no.~2, pp. 110--116, 2023.

\bibitem{PYang6G}
P.~Yang, Y.~Xiao, M.~Xiao, and S.~Li, ``6g wireless communications: Vision and potential techniques,'' \emph{IEEE Network}, vol.~33, no.~4, pp. 70--75, 2019.

\bibitem{ZZhang6g}
Z.~Zhang, Y.~Xiao, Z.~Ma, M.~Xiao, Z.~Ding, X.~Lei, G.~K. Karagiannidis, and P.~Fan, ``6g wireless networks: Vision, requirements, architecture, and key technologies,'' \emph{IEEE Vehicular Technology Magazine}, vol.~14, no.~3, pp. 28--41, 2019.

\bibitem{WSaad6g}
W.~Saad, M.~Bennis, and M.~Chen, ``A vision of 6g wireless systems: Applications, trends, technologies, and open research problems,'' \emph{IEEE Network}, vol.~34, no.~3, pp. 134--142, 2020.

\bibitem{MGiordani6g}
M.~Giordani, M.~Polese, M.~Mezzavilla, S.~Rangan, and M.~Zorzi, ``Toward 6g networks: Use cases and technologies,'' \emph{IEEE Communications Magazine}, vol.~58, no.~3, pp. 55--61, 2020.

\bibitem{EStrinati6g}
E.~Calvanese~Strinati, S.~Barbarossa, J.~L. Gonzalez-Jimenez, D.~Ktenas, N.~Cassiau, L.~Maret, and C.~Dehos, ``6g: The next frontier: From holographic messaging to artificial intelligence using subterahertz and visible light communication,'' \emph{IEEE Vehicular Technology Magazine}, vol.~14, no.~3, pp. 42--50, 2019.

\bibitem{tariq20236g}
F.~Tariq, M.~Khandaker, and I.~S. Ansari, \emph{6G wireless: the communication paradigm beyond 2030}.\hskip 1em plus 0.5em minus 0.4em\relax CRC Press, 2023.

\bibitem{xu2021faster}
T.~Xu and I.~Darwazeh, ``Faster urllc: Deep learning waveform fingerprinting,'' in \emph{2021 IEEE International Conference on Communications Workshops (ICC Workshops)}.\hskip 1em plus 0.5em minus 0.4em\relax IEEE, 2021, pp. 1--6.

\bibitem{she2021tutorial}
C.~She, C.~Sun, Z.~Gu, Y.~Li, C.~Yang, H.~V. Poor, and B.~Vucetic, ``A tutorial on ultrareliable and low-latency communications in 6g: Integrating domain knowledge into deep learning,'' \emph{Proceedings of the IEEE}, vol. 109, no.~3, pp. 204--246, 2021.

\bibitem{pourkabirian2024vision}
A.~Pourkabirian, M.~S. Kordafshari, A.~Jindal, and M.~H. Anisi, ``A vision of 6g urllc: Physical-layer technologies and enablers,'' \emph{IEEE Communications Standards Magazine}, vol.~8, no.~2, pp. 20--27, 2024.

\bibitem{GZhao2018ComMag}
G.~{Zhao}, M.~A. {Imran}, Z.~{Pang}, Z.~{Chen}, and L.~{Li}, ``Toward real-time control in future wireless networks: Communication-control co-design,'' \emph{IEEE Communications Magazine}, vol.~57, no.~2, pp. 138--144, February 2019.

\bibitem{TZeng2018ICC}
T.~Zeng, O.~Semiari, W.~Saad, and M.~Bennis, ``Integrated communications and control co-design for wireless vehicular platoon systems,'' in \emph{2018 IEEE International Conference on Communications (ICC)}.\hskip 1em plus 0.5em minus 0.4em\relax IEEE, 2018, pp. 1--6.

\bibitem{aldababsa2024survey}
M.~Aldababsa, S.~{\"O}zyurt, O.~Kucur \emph{et~al.}, ``A survey on orthogonal time frequency space modulation,'' \emph{IEEE Open Journal of the Communications Society}, 2024.

\bibitem{haif2024novelOCDM}
H.~Haif, S.~E. Zegrar, and H.~Arslan, ``Novel ocdm transceiver design for doubly-dispersive channels,'' \emph{IEEE Transactions on Vehicular Technology}, 2024.

\bibitem{thaj2021orthogonaltime}
T.~Thaj and E.~Viterbo, ``Orthogonal time sequency multiplexing modulation,'' in \emph{2021 IEEE Wireless Communications and Networking Conference (WCNC)}.\hskip 1em plus 0.5em minus 0.4em\relax IEEE, 2021, pp. 1--7.

\bibitem{bemani2023affine}
A.~Bemani, N.~Ksairi, and M.~Kountouris, ``Affine frequency division multiplexing for next generation wireless communications,'' \emph{IEEE Transactions on Wireless Communications}, 2023.

\bibitem{zaman2023comprehensive}
S.~Zaman, F.~Tariq, M.~Khandaker, and R.~T. Khan, ``A comprehensive overview of security and privacy in the 6g era,'' \emph{6G Wireless}, pp. 203--258, 2023.

\bibitem{wu2023split}
W.~Wu, M.~Li, K.~Qu, C.~Zhou, X.~Shen, W.~Zhuang, X.~Li, and W.~Shi, ``Split learning over wireless networks: Parallel design and resource management,'' \emph{IEEE Journal on Selected Areas in Communications}, vol.~41, no.~4, pp. 1051--1066, 2023.

\bibitem{lin2024efficient}
Z.~Lin, G.~Zhu, Y.~Deng, X.~Chen, Y.~Gao, K.~Huang, and Y.~Fang, ``Efficient parallel split learning over resource-constrained wireless edge networks,'' \emph{IEEE Transactions on Mobile Computing}, 2024.

\bibitem{kim2023bargaining}
M.~Kim, A.~DeRieux, and W.~Saad, ``A bargaining game for personalized, energy efficient split learning over wireless networks,'' in \emph{2023 IEEE Wireless Communications and Networking Conference (WCNC)}.\hskip 1em plus 0.5em minus 0.4em\relax IEEE, 2023, pp. 1--6.

\bibitem{lin2024split}
Z.~Lin, G.~Qu, X.~Chen, and K.~Huang, ``Split learning in 6g edge networks,'' \emph{IEEE Wireless Communications}, 2024.

\bibitem{khan2023resource}
L.~U. Khan, M.~Guizani, and C.~S. Hong, ``Resource optimized hierarchical split federated learning for wireless networks,'' in \emph{Proceedings of Cyber-Physical Systems and Internet of Things Week 2023}, 2023, pp. 254--259.

\bibitem{khan2023joint}
L.~U. Khan, M.~Guizani, A.~Al-Fuqaha, C.~S. Hong, D.~Niyato, and Z.~Han, ``A joint communication and learning framework for hierarchical split federated learning,'' \emph{IEEE Internet of Things Journal}, 2023.

\bibitem{xu2023accelerating}
C.~Xu, J.~Li, Y.~Liu, Y.~Ling, and M.~Wen, ``Accelerating split federated learning over wireless communication networks,'' \emph{IEEE Transactions on Wireless Communications}, 2023.

\bibitem{behdad2024interplay}
Z.~Behdad, {\"O}.~T. Demir, K.~W. Sung, and C.~Cavdar, ``{Interplay between sensing and communication in cell-free massive MIMO with URLLC users},'' \emph{arXiv preprint arXiv:2401.10133}, 2024.

\bibitem{lyu2024rate-splitting}
X.~Lyu, S.~Aditya, J.~Kim, and B.~Clerckx, ``{Rate-Splitting Multiple Access: The First Prototype and Experimental Validation of its Superiority over SDMA and NOMA},'' \emph{IEEE Transactions on Wireless Communications}, 2024.

\end{thebibliography}

\begin{IEEEbiography}{Md. Emdadul Haque} is a Professor in the Department of Information and Communication Engineering, University of Rajshahi, Bangladesh. He has served as the chairman of the department from 2019 to 2022. He completed his BSc from Khulna University and MSc from Bangladesh University of Engineering and Technology (BUET) and PhD from Tokyo Institute of Technology, Japan. His research interest includes wireless communication and sensing, healthcare technologies and low latency communications.
\end{IEEEbiography}

\begin{IEEEbiography}{Faisal Tariq} is a Senior Lecturer at the James Watt School of Engineering, University of Glasgow, United Kingdom. Prior to this, worked at the Queen Mary University of London. He received his PhD from The Open University, UK. His research interests include 56/6G wireless communications, physical layer security and machine learning applications. He received best paper award in IEEE WPMC 2013. He is serving as an Associate Editor for IEEE Wireless Communications Letters and Editor  for Elsevier Journal of Networks and Computer Applications.
\end{IEEEbiography}

\begin{IEEEbiography}{Md. Sakir Hossain} (S'15-M'18-SM'24) received Ph.D. degree in information and computer science from Saitama University, Japan. He currently works as an Associate Professor at BRAC University, Bangladesh. He worked as a Postdoctoral
Researcher at Czech Technical University, Prague, Czech Republic. He works on the development of solutions for UAV-assisted wireless networks, waveform design, biomedical signal processing, and cyber security.
\end{IEEEbiography}

\begin{IEEEbiography}{Muhammad R A Khandaker} (S'10-M'13-SM'18)
is currently working with Main Roads, Western Australia. Previously, he was an Assistant Professor in the School of Engineering and Physical Sciences at Heriot-Watt University. Before joining Heriot-Watt, he worked as a Postdoctoral Research Fellow at University College London, UK, (July 2013 - June 2018). He served as an Associate Editor for the IEEE Wireless Communications Letters, IEEE Communications Letters, IEEE ACCESS and the EURASIP Journal On Wireless Communications and Networking.
\end{IEEEbiography}

\begin{IEEEbiography}{Muhammad Ali Imran} received his M.Sc. (Distinction) and Ph.D. degrees from Imperial College London, UK, in 2002 and 2007, respectively. He is a Professor in Communication Systems in the University of Glasgow, Dean of graduate studies, Head of Communications Sensing and Imaging (CSI) research group and Di- rector of Glasgow UESTC Centre of Educational Development and Innovation. He has led a number of multimillion-funded international research projects and the ”new physical layer” work area for 5G innovation centre at Surrey. He has a global collaborative research network spanning both academia and key industrial players in the field of wireless communications. He has supervised 50+ successful PhD graduates and published over 600 peer-reviewed research papers including more than 100 IEEE Transaction papers. Prof. Imran is a Fellow of IEEE.
\end{IEEEbiography}

\begin{IEEEbiography}{Kai-Kit Wong} received the BEng, the MPhil, and the PhD degrees, all in Electrical and Electronic Engineering, from the Hong Kong University of Science and Technology, Hong Kong, in 1996, 1998, and 2001, respectively. After graduation, he took up academic and research positions at the University of Hong Kong, Lucent Technologies, Bell-Labs, Holmdel, the Smart Antennas Research Group of Stanford University, and the University of Hull, UK. He is Chair in Wireless Communications at the Department of Electronic and Electrical Engineering, University College London, UK. His current research centers around 5G and beyond mobile communications. He is a co-recipient of the 2013 IEEE Signal Processing Letters Best Paper Award and the 2000 IEEE VTS Japan Chapter Award at the IEEE Vehicular Technology Conference in Japan in 2000, and a few other international best paper awards. He is Fellow of IEEE and IET and is also on the editorial board of several international journals. He was the Editor-in-Chief for IEEE Wireless Communications Letters from 2020 to 2023.
\end{IEEEbiography}

%\begin{IEEEbiography}{MS} is a ..
%\end{IEEEbiography}

%[{\includegraphics[width=1in,height=1.25in,clip,keepaspectratio]{picture}}]

\end{document}